\PassOptionsToPackage{capitalize,noabbrev,nameinlink}{cleveref}
\PassOptionsToPackage{usenames,dvipsnames}{color}
\PassOptionsToPackage{usenames,table}{xcolor}
\PassOptionsToPackage{final}{microtype}
\PassOptionsToPackage{scaled}{inconsolata}

\documentclass[sigconf]{acmart}

\newcommand{\paperTitle}{Order-Preserving Key Compression for
\texorpdfstring{\\}{}In-Memory Search Trees}
\newcommand{\paperKeywords}{}
\newcommand{\paperAuthors}{}

\copyrightyear{2020}
\acmYear{2020}
\setcopyright{acmlicensed}
\acmConference[SIGMOD'20]{Proceedings of the 2020 ACM SIGMOD International Conference on Management of Data}{June 14--19, 2020}{Portland, OR, USA}
\acmBooktitle{Proceedings of the 2020 ACM SIGMOD International Conference on Management of Data (SIGMOD'20), June 14--19, 2020, Portland, OR, USA}
\acmPrice{15.00}
\acmDOI{10.1145/3318464.3380583}
\acmISBN{978-1-4503-6735-6/20/06}


\hypersetup{
    pdfauthor = {\paperAuthors},
    pdftitle = {\paperTitle},
    pdfkeywords = {\paperKeywords},
    pdfborder={ 0 0 0 }
}

\usepackage{amsmath}
\usepackage{amssymb}
\usepackage{boxedminipage}
\usepackage{xspace}
\usepackage{tabularx}
\usepackage{balance}  % for  \balance command ON LAST PAGE  (only there!)
\usepackage[hyphenbreaks]{breakurl}
\usepackage[font={small}]{caption}
\usepackage{graphicx}
\usepackage{subfig}
\usepackage{cleveref}
\usepackage{enumitem}
\usepackage{boldline}
\usepackage{multirow}

\usepackage[T1]{fontenc}
\usepackage{graphicx}
\usepackage{soul}
\usepackage{pifont}

\usepackage{booktabs}
\usepackage[end]{algpseudocode}
\usepackage{algorithm}

\usepackage{xspace}
\usepackage{adjustbox}

\newcommand{\ope}{HOPE\xspace}
\newcommand{\bplustree}{B+tree\xspace}
\newcommand{\bplustrees}{B+trees\xspace}
\newcommand{\pbtree}{Prefix B+tree\xspace}
\newcommand{\pbtrees}{Prefix B+trees\xspace}
\newcommand{\art}{ART\xspace}
\newcommand{\hot}{HOT\xspace}
\newcommand{\surf}{SuRF\xspace}
\newcommand{\ttree}{T-Tree\xspace}
\newcommand{\tlx}{TLX B+tree\xspace}
\newcommand{\compSelector}{Symbol Selector\xspace}
\newcommand{\compAssigner}{Code Assigner\xspace}
\newcommand{\compDictionary}{Dictionary\xspace}
\newcommand{\compEncoder}{Encoder\xspace}

\newcommand{\encodingFIFC}{FIFC\xspace}
\newcommand{\encodingFIVC}{FIVC\xspace}
\newcommand{\encodingVIVC}{VIVC\xspace}
\newcommand{\encodingVIFC}{VIFC\xspace}

\newcommand{\schemeSingleChar}{Single-Char\xspace}
\newcommand{\schemeDoubleChar}{Double-Char\xspace}
\newcommand{\schemeALM}{ALM\xspace}
\newcommand{\schemeALMImproved}{ALM-Improved\xspace}
\newcommand{\schemeThreeGrams}{3-Grams\xspace}
\newcommand{\schemeFourGrams}{4-Grams\xspace}

\definecolor{todo-color}{rgb}{1,0,0}

\definecolor{comment-color}{rgb}{0.25,0.25,0.25}

\definecolor{rev-color}{rgb}{0,0,0}
\newcommand{\rev}[1]{\textnormal{\color{rev-color}{#1}}\unskip}

\begin{document}

\title{\paperTitle}

%% ==================================================================
%% AUTHORS
%% ==================================================================


\author{Huanchen Zhang}
\affiliation{\institution{Carnegie Mellon University}}
\email{huanche1@cs.cmu.edu}

\author{Xiaoxuan Liu}
\affiliation{\institution{Carnegie Mellon University}}
\email{xiaoxual@andrew.cmu.edu}

\author{David G. Andersen}
\affiliation{\institution{Carnegie Mellon University}}
\email{dga@cs.cmu.edu}

\author{Michael Kaminsky}
\affiliation{\institution{BrdgAI}}
\email{kaminsky@cs.cmu.edu}

\author{Kimberly Keeton}
\affiliation{\institution{Hewlett Packard Labs}}
\email{kimberly.keeton@hpe.com}

\author{Andrew Pavlo}
\affiliation{\institution{Carnegie Mellon University}}
\email{pavlo@cs.cmu.edu}

\renewcommand{\shortauthors}{Huanchen Zhang et al.}

%% ==================================================================
%% ABSTRACT
%% ==================================================================
\begin{abstract}
We present the \emph{High-speed Order-Preserving Encoder} (\ope)
for in-memory search trees.
\ope is a fast dictionary-based compressor that encodes
arbitrary keys while preserving their order.
\ope's approach is to identify common key patterns at a fine granularity
and exploit the entropy to achieve high compression rates with a small dictionary.
%To validate \ope,
We first develop a theoretical model to reason about order-preserving dictionary designs.
We then select six representative compression schemes using this model and implement
them in \ope.
These schemes make different trade-offs between compression rate and encoding speed.
%We evaluate \ope on four data structures used
%in databases: \surf, \art, \hot, and \bplustree.
\rev{We evaluate \ope on five data structures used
in databases: \surf, \art, \hot, \bplustree, and \pbtree.}
Our experiments show that using \ope allows the search trees
to achieve lower query latency (up to 40\% lower)
and better memory efficiency (up to 30\% smaller)
simultaneously for most string key workloads.

%Unlike the dictionary compression techniques in existing databases,
%\ope can encode an arbitrary input key without modifying the dictionary.
\end{abstract}

%% \begin{CCSXML}
%% <ccs2012>
%%  <concept>
%%   <concept_id>10010520.10010553.10010562</concept_id>
%%   <concept_desc>Computer systems organization~Embedded systems</concept_desc>
%%   <concept_significance>500</concept_significance>
%%  </concept>
%%  <concept>
%%   <concept_id>10010520.10010575.10010755</concept_id>
%%   <concept_desc>Computer systems organization~Redundancy</concept_desc>
%%   <concept_significance>300</concept_significance>
%%  </concept>
%%  <concept>
%%   <concept_id>10010520.10010553.10010554</concept_id>
%%   <concept_desc>Computer systems organization~Robotics</concept_desc>
%%   <concept_significance>100</concept_significance>
%%  </concept>
%%  <concept>
%%   <concept_id>10003033.10003083.10003095</concept_id>
%%   <concept_desc>Networks~Network reliability</concept_desc>
%%   <concept_significance>100</concept_significance>
%%  </concept>
%% </ccs2012>
%% \end{CCSXML}

%% \ccsdesc[500]{Computer systems organization~Embedded systems}
%% \ccsdesc[300]{Computer systems organization~Redundancy}
%% \ccsdesc{Computer systems organization~Robotics}
%% \ccsdesc[100]{Networks~Network reliability}

%% \keywords{datasets, neural networks, gaze detection, text tagging}

\maketitle

%% ==================================================================
%% SECTIONS
%% ==================================================================

\section{Introduction}
\label{sec:intro}

%Main memory remains a limited resource in software systems. For example,
SSDs have increased the performance demand on main-memory data structures;
the growing cost gap between DRAM and SSD storage together with increasing
database sizes means that main-memory structures must be both compact \emph{and} fast.
The \$/GB ratio of DRAM versus SSDs increased from 
10$\times$ in 2013 to 40$\times$ in 2018~\cite{www-memory-price}.
Together with growing database sizes, database management
systems (DBMSs) now operate with a lower memory to storage size ratio than before.
DBMS developers in turn are changing how they implement their systems' 
architectures. For example, a major Internet company's engineering team
% Facebook RocksDB~\cite{www-rocksdb}
assumes a 1:100 memory to storage ratio to guide their future system designs~\cite{personal-comm}.

One challenge with assuming that memory is severely limited is that modern online 
transaction processing (OLTP) applications demand that most if not all transactions complete in 
milliseconds or less~\cite{stonebraker2007}. To meet this latency requirement, applications
use many \textit{search trees} (i.e., indexes, filters) in memory to 
minimize the number of I/Os on storage devices. But these search trees consume a large portion of 
the total memory available to the DBMS~\cite{bhattacharjee2009,zhang2016, lasch2019}.

Compression is an obvious way to reduce the memory cost of a DBMS's search trees.
Compression improves the cache performance of the search tree
and allows the DBMS to retain more %non-index
data in memory to further reduce I/Os.
A system must balance these performance gains with the additional 
computational overhead of the compression algorithms.
%which is non-trivial in many scenarios.

Existing search tree compression techniques fall into two categories.
The first is \emph{block-oriented compression} of tree pages using algorithms
such as Snappy~\cite{www-snappy} and LZ4~\cite{www-lz4}.
This approach is beneficial to disk-based trees because it minimizes
data movement between disk and memory.
For in-memory search trees, however, block compression algorithms impose
too much computational overhead because the DBMS is unable to 
operate directly on the search tree data without having to decompress it first~\cite{zhang2016}.
The second approach is to design a \emph{memory-efficient data structure} that avoids
storing unnecessary key information and internal meta-data 
(e.g., pointers)~\cite{leis2013,zhang2016,zhang2018,binna2018,masker2019}.
Although these new designs are smaller than previous implementations,
they still are a major source of DMBSs' memory footprints.
%Although these new designs have smaller memory footprints compared to previous
%implementations, they still are a major source of memory overhead.
%storage overhead, especially with long string keys.
%This is because they store keys in an uncompressed form.

An orthogonal approach is to compress the individual input keys
before inserting them into the search tree.
%This is essential to improving the memory efficiency of the data structures
Key compression is important for reducing index memory consumption 
because real-world databases contain many 
variable-length string attributes~\cite{muller2014} whose size dominates the
data structure's internal overheads.
A common application of string compression is in
columnar DBMSs~\cite{abadi2006}, which often use dictionary compression to replace string values
in a column with fixed-length integers.
Traditional dictionary compression, however, does not work for 
in-memory search trees (e.g., OLTP indexes) for two reasons.  First,
the DBMS must continually grow its dictionary as new keys arrive.
Second, key compression in a search tree must be order-preserving
to support range queries properly.

%\todo{Need to distinguish what you mean by ``key compression'' with ART / 
%HOT / radix trees. One can argue that they are compressing keys by not storing redundant 
%information. They are also able to operate directly on compressed keys, unlike with the 
%blocked-oriented compression.}

%\todo{You need to state what why order-preserving is a desired property somewhere above before you 
%use the term in your description of \ope.}

We, therefore, present
\textbf{\underline{H}igh-speed \underline{O}rder-\underline{P}reserving
\underline{E}ncoder} (\ope),
a dictionary-based key compressor for in-memory search trees
(e.g., B+trees, tries).
\ope includes six entropy encoding schemes that trade
between compression rate and encoding performance.
%a key compressor for in-memory search trees.
%\ope is a dictionary-based entropy-encoding tool for compressing any
%order-preserving data structure (e.g., B+trees, tries).
%We designed \ope as a general-purpose framework for compressing any
%order-preserving data structure (e.g., B+trees, tries).
% It not only reduces the size of the search trees, but also improves the DBMS's query 
% performance in most OLTP workloads.
%\ope is a dictionary-based entropy-encoding tool.
% We developed a theoretical model to guarantee that \ope's dictionaries 
% can encode \emph{any} key in an order-preserving way.
% We implemented six specific compression schemes in \ope,
% trading between compression rate and encoding performance.
When the DBMS creates a tree-based index/filter,
\ope samples the initial bulk-loaded keys and counts the frequencies
of the byte patterns specified by a scheme.
It uses these statistics to generate dictionary symbols
that comply with our theoretical model to preserve key ordering.
\ope then encodes the symbols using either fixed-length codes
or optimal order-preserving prefix codes.
A key insight in \ope is its emphasis on encoding speed
(rather than decoding) because our target search tree queries
need not reconstruct the original keys.
\rev{To evaluate \ope, we applied it to five in-memory search trees: 
\surf~\cite{zhang2018}, \art~\cite{leis2013}, \hot~\cite{binna2018},
\bplustree~\cite{www-tlx-btree}, and \pbtree~\cite{bayer1977}.}
Our experimental results show that \ope reduces their query latency by up to 
40\% and saves their memory consumption by up to 30\%
%\ope improves both performance and memory use
at the same time for most string key workloads.

We make four primary contributions in this paper.
First, we develop a theoretical model to characterize the properties of dictionary encoding.
% including order preservation.
%Second, we introduce the \ope framework to compress keys for in-memory
%search trees with low overhead.
Second, we introduce \ope
to compress keys for in-memory search trees efficiently.
Third, we implement six compression schemes in \ope
to study the compression rate vs. encoding speed trade-off.
Finally, we apply \ope on five trees
and show that \ope improves their performance and memory-efficiency
%in most cases.
simultaneously.

\section{Background and Related Work}
\label{sec:related}

Modern DBMSs rely on in-memory search trees
(e.g., indexes and filters) to achieve high throughput and low latency.
We divide these search trees into three categories.
The first is the B-tree/\bplustree family, including
Cache Sensitive B+trees (CSB+trees)~\cite{rao2000}
and Bw-Trees~\cite{levandoski2013,wang2018}.
They store keys horizontally side-by-side in the leaf nodes
and have good range query performance.
The second category includes tries and radix trees
\cite{fredkin1960,morrison1968,heinz2002,askitis2007,boehm2011,kissinger2012,leis2013,binna2018}.
They store keys vertically to allow prefix compression.
Recent memory-efficient tries such as \art~\cite{leis2013}
and \hot~\cite{binna2018} are faster than
\bplustree{s} on modern hardware.
The last category is hybrid data structures such as Masstree~\cite{mao2012}
that combine the \bplustree and trie designs in a single data structure.

Existing compression techniques for search trees
leverage general-purpose block compression algorithms such as 
LZ77~\cite{www-lz77}, Snappy~\cite{www-snappy}, and LZ4~\cite{www-lz4}.
For example, InnoDB uses the zlib library~\cite{www-zlib}
to compress its \bplustree pages/nodes before they are written to disk.
Block compression algorithms, however, are too slow
for in-memory search trees:
%% This approach works well for disk-based indexes because it minimizes
%% data movement between disk and memory.
%% The computational cost of compressing and decompressing memory pages
%% is hidden by the reduction in I/Os.
%% For in-memory search trees, however, block compression algorithms are
%% too expensive.
query latencies for in-memory \bplustree{s} and tries range from
100s of nanoseconds to a few microseconds,
while the fastest block compression algorithms can
decompress only a few 4~KB memory pages in that time~\cite{www-lz4}.

Recent work has addressed this size problem through new data structures~\cite{zhang2016, zhang2018, 
binna2018, masker2019}.
For example, the succinct trie in \surf consumes only 10 bits
(close to the theoretical optimum) to represent a node~\cite{zhang2018}.
Compressing input keys using \ope, however, is an orthogonal approach
that one can apply to any of the above search tree categories
to achieve additional space savings and performance gains.
%% There are two limitations to this approach.
%% First, instead of a general plug-in method,
%% this approach requires substantial designing and engineering effort
%% for each target application.
%% Second, while this approach minimizes the structural overhead of
%% a search tree, it leaves the actual content (i.e., the keys) stored
%% in the tree uncompressed.
%% As we observe an increasing number of string columns in real-world
%% databases~\cite{muller2014},
%% being able to compress the keys becomes an essential complement to
%% compressing the data structures alone.

One could apply existing field/table-wise compression schemes
to search tree keys.
Whole-key dictionary compression is the most popular scheme used in DBMSs today.
It replaces the values in a column with smaller fixed-length codes
using a dictionary.
Indexes and filters, therefore, could take advantage of those existing
dictionaries for key compression.
There are several problems with this approach.
%beyond the fact that
%not all applications enable dictionary compression in their databases.
First, the dictionary compression must be order-preserving
to allow range queries on search trees.
%so that the search trees can perform proper range queries on the encoded keys.
Order-preserving dictionaries, however, are difficult to maintain
with changing value domains~\cite{liu2019},
which is often the case for string keys in OLTP applications.
Second, the latency of encoding a key is similar to that of querying
the actual indexes/filters
because most order-preserving dictionaries use the same kind of
search trees themselves~\cite{binnig2009}.
% For example, a state-of-the-art order-preserving string compression
% scheme proposed by Binnig et al. uses B+trees
% as the dictionary data structure.
Finally, dictionary compression only works well for columns
with low/moderate cardinalities.
If most of the values are unique, then the larger dictionary
negates the size reduction in the actual fields.

\rev{Existing order-preserving frequency-based compression schemes, including
the one used in DB2 BLU~\cite{raman2013} and padded encoding~\cite{li2015},
exploit the column value distribution skew by assigning smaller codes to
more frequent values.
%to achieve better compression rate and higher scan performance.
Variable-length codes, however, are inefficient to locate,
decode, and process in parallel. DB2 BLU, thus, only uses up to a few
different code sizes per column and stores the codes of the same size
together to speed up queries. Padded encoding, on the other hand, pads the
variable-length codes with zeros at the end so that all codes are of the
same length (i.e., the maximum length of the variable-length codes) to
facilitate scan queries. DB2 BLU and padded encoding are designed for
column stores where most queries are reads, and updates are often in batches.
%Although both designs consider value frequencies, they
Both designs still use the whole-key dictionary compression discussed above
and therefore, cannot encode new values without extending the dictionary,
which can cause expensive
re-encodes of the column. HOPE, however, can encode arbitrary input values
using the same dictionary while preserving their ordering.
Such property is desirable for write-intensive OLTP indexes.}

The focus of this paper is on compressing string keys.
Numeric keys are already small and can be further compressed
using techniques, such as null suppression and delta encoding~\cite{abadi2006}.
\rev{Prefix compression and suffix truncation are common techniques
  used in \bplustrees.
  \ope can provide additional benefits on top of these compression methods.}
  %as shown in the \pbtree evaluation in \cref{sec:tree-eval}.}
%% Prefix compression is a common technique used in
%% \bplustree{s}, where each node only stores
%% the common prefix of its keys once~\cite{bayer1977}.
%% Prefix compression can achieve at most the same level
%% of reduction as a radix tree.
%% Suffix truncation is another common technique where nodes skip the suffixes after the keys are 
%% uniquely identified in the tree~\cite{leis2013, zhang2018, binna2018}.
%% Suffix truncation is a lossy scheme,
%% and it trades a higher false positive rate for better memory efficiency.

Prior studies considered entropy encoding schemes such as
Huffman~\cite{huffman1952} and arithmetic coding~\cite{witten1987}
too slow for columnar data compression
because their variable-length codes are slow to 
decode~\cite{chen2001, abadi2006, raman2006, binnig2009, bhattacharjee2009, liu2019}.
%For example, DB2 BLU only uses up to a few different code sizes~\cite{raman2013}.
This concern does not apply to search trees
because non-covering index and filter queries do not reconstruct
the original keys\footnote{The search tree can recover the original keys
if needed: entropy encoding is lossless.
Exploring lossy compression on search trees is future work}.
In addition, entropy encoding schemes produce high compression rates
even with small dictionaries
because they exploit common patterns at a fine granularity.
% Small dictionaries are preferable in terms of both performance and space.

%% To efficiently compress arbitrary string keys for in-memory search trees,
%% we argue that a better approach is to use
%% order-preserving entropy encoding.
%% Entropy encoding schemes such as
%% Huffman coding~\cite{huffman1952} and arithmetic coding~\cite{witten1987}
%% are considered slow in previous studies
%% \cite{chen2001, abadi2006, raman2006, binnig2009, bhattacharjee2009, liu2019}
%% because they generate variable-length codes that can hurt decoding speed.
%% This is not a problem when compressing search trees in DMBSs
%% because most index and filter queries do not require decoding the
%% keys\footnote{Even for queries that need to recover the original keys
%% from the search trees (e.g., covering indexes),
%% entropy encoding schemes can satisfy those queries
%% because the compression is lossless unlike suffix truncation.}.
%% In addition, entropy encoding schemes can produce good compression rate
%% even with small dictionaries
%% because they explore common patterns at a fine granularity.
%% Small dictionaries are preferable because they reduce the
%% compression overhead in terms of both performance and space.}

Antoshenkov et al.~\cite{antoshenkov1996, antoshenkov1997}
proposed an order-preserving string compressor
with a string parsing algorithm (ALM)
to guarantee the order of the encoded results.
We introduce our string axis model in the next section,
which is inspired by the ALM method but is more general.
The ALM compressor belongs to a specific category in
our compression model.

%% The order-preserving part of the problem is well studied
%% in previous literature.
%% The Hu-Tucker algorithm~\cite{hu1971} generates optimal
%% prefix codes that also preserve orders.
%% Antoshenkov et al.~\cite{antoshenkov1996, antoshenkov1997}
%% introduced a string parsing algorithm (the ALM method)
%% that guarantees the order of the encoded results
%% as long as the codes and the dictionary symbols follow the same order.
%% We leverage these two techniques in TODO.
%% The goals that we want to achieve with our compressor, however,
%% is more practical.
%% First, the compressor must have competitive compression rate
%% so that it can reduce the overall search tree size.
%% Second, the compressor must be fast so that the performance gained from
%% a smaller search tree is not offset by the compression overhead.
%% Finally, the compressor must be lightweight and easy-to-use
%% for a wide range of search trees.

%beating ALM

%Maybe a transition to the next section.

\section{Compression Model}
\label{sec:model}

Different dictionary encoding schemes,
ranging from Huffman encoding~\cite{huffman1952} to the
\schemeALM-based compressor~\cite{antoshenkov1997},
provide different capabilities and guarantees.
For example, some can encode arbitrary input strings
while others preserve order.
In this section, we introduce a unified model,
called the string axis model, to characterize the properties
of a dictionary encoding scheme.
This model is inspired by the \schemeALM string parsing
algorithm~\cite{antoshenkov1996}, which solves the
order-preserving problem for dictionary-based string compression.
Using the string axis model, we can construct a wide range of
dictionary-based compression schemes that can serve our target application
(i.e., key compression for in-memory search trees).
We divide qualified schemes into four categories, each making
different trade-offs. % between compression rate and performance.
We then briefly describe six representative compression schemes
supported by \ope.

%% ---------------------------------------------------------------
%% The String Axis Model
%% ---------------------------------------------------------------
\subsection{The String Axis Model}
\label{sec:model:axis}

\begin{figure}[t]
    \centering
    \includegraphics[width=0.85\columnwidth]{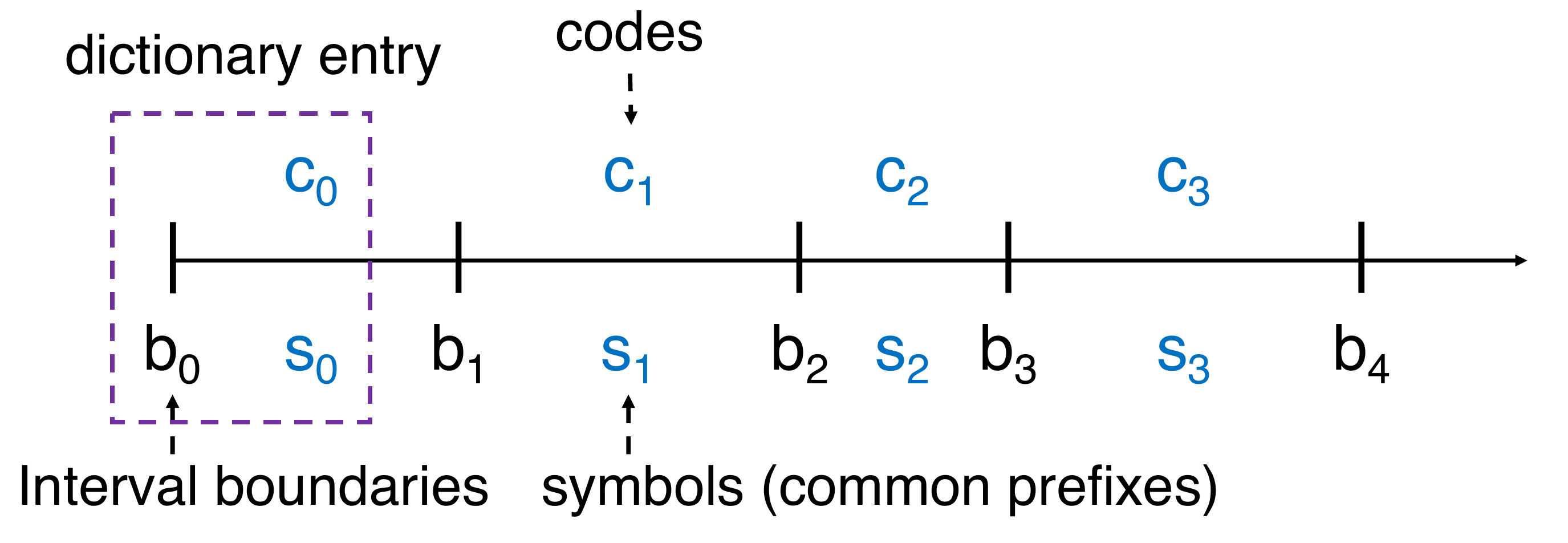}
    \caption{
        \textbf{String Axis Model} --
        All strings are divided into connected intervals in lexicographical order. Source 
        strings in the same interval share a common prefix ($s_{i}$) that maps to code
        ($c_i$).
%         \ope recursively encodes a string by finding its interval and replacing each prefix 
%         with that interval's code.
    }
    \label{fig:string-axis}
\end{figure}

%As shown in \cref{fig:string-axis}, a \emph{string axis} is an axis that
%lays out all possible source strings in lexicographical order.
As shown in \cref{fig:string-axis}, a \emph{string axis} lays out
all possible source strings on a single axis in lexicographical order.
We can represent a dictionary encoding scheme using this 
model and highlight three important properties:
(1) completeness, (2) unique decodability, and (3) order-preserving.

Let $\Sigma$ denote the source string alphabet.
$\Sigma^*$ is the set of all possible finite-length strings over $\Sigma$.
Similarly, let $X$ denote the code alphabet
and $X^*$ be the code space.
Typically, $\Sigma$ is the set of all characters,
and $X = \{0, 1\}$.
%A dictionary $D$ maps a subset of the source strings $S$ to the set of codes $C$:
A dictionary $D: S \rightarrow C, S \in \Sigma^*, C \in X^*$
maps a subset of the source strings $S$ to the set of codes $C$.

%% \vspace*{-0.15in}
%% \begin{equation*}
%%     D: S \rightarrow C, S \in \Sigma^*, C \in X^*
%% \end{equation*}
%% \vspace*{-0.15in}

\begin{figure}[t]
    \centering
    \includegraphics[width=0.82\columnwidth]{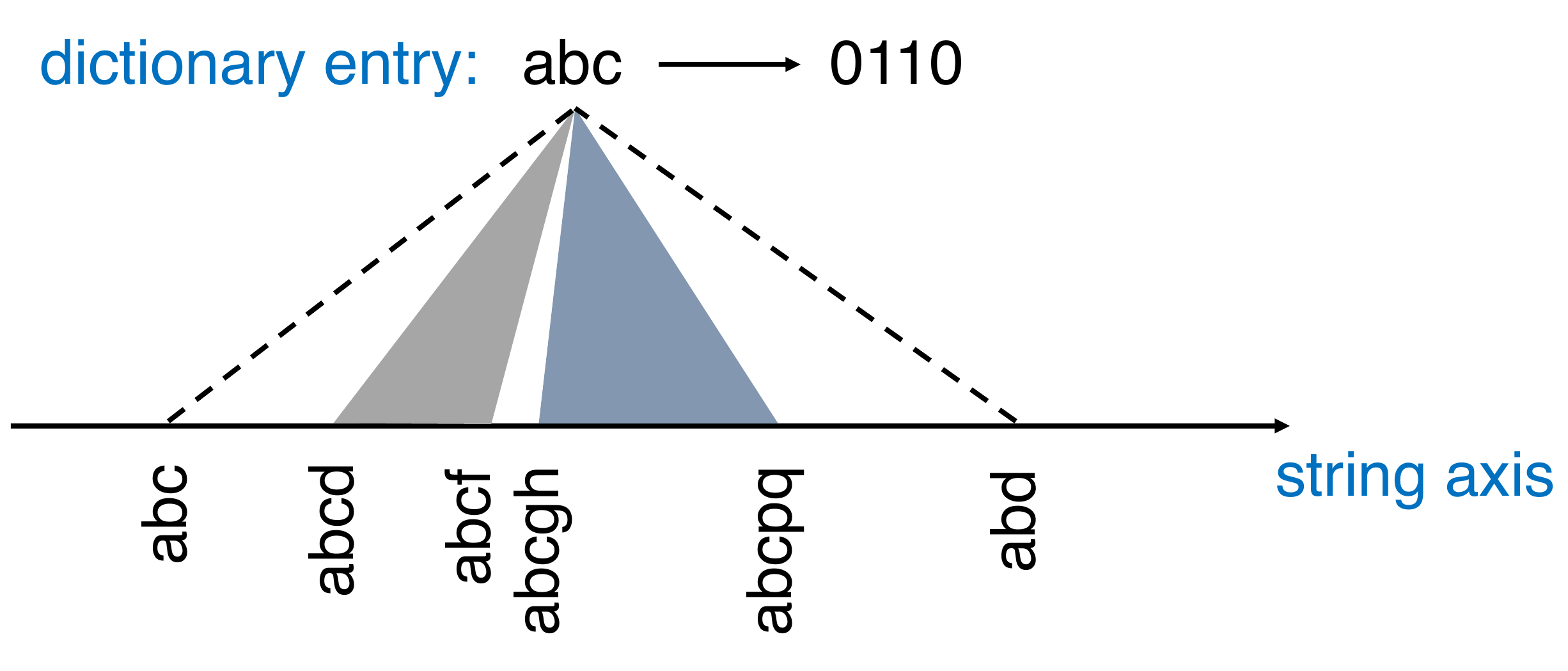}
    \caption{
        \textbf{Dictionary Entry Example} --
        All sub-intervals of \texttt{[abc, abd)} are valid mappings for dictionary entry abc 
        $\longrightarrow$ 0110.
    }
    \label{fig:axis-mapping}
\end{figure}

On the string axis, a dictionary entry $s_i \rightarrow c_i$ is mapped to
an interval $I_i$, where $s_i$ is a prefix of all strings within $I_i$.
The choice of $I_i$ is not unique.
For example, as shown in \cref{fig:axis-mapping},
both $\left[\mbox{\texttt{abcd}}, \mbox{\texttt{abcf}}\right)$ and $\left[\mbox{\texttt{abcgh}}, 
\mbox{\texttt{abcpq}}\right)$ are valid mappings for
dictionary entry \texttt{abc}$\rightarrow$0110.
In fact, any sub-interval of $\left[\mbox{\texttt{abc}}, \mbox{\texttt{abd}}\right)$ is a valid mapping
in this example.
If a source string $src$ falls into the interval $I_i$,
then a dictionary lookup on $src$ returns the corresponding
entry $s_i \rightarrow c_i$.

We can model the dictionary encoding method as a recursive process.
Given a source string $src$,
one can lookup $src$ in the dictionary and obtain an entry
$(s \rightarrow c) \in D, s \in S, c \in C$,
such that $s$ is a prefix of $src$, i.e., $src = s \boldsymbol{\cdot}
src_{suffix}$, where ``$\boldsymbol{\cdot}$'' is the concatenation operation.
We then replace $s$ with $c$ in $src$ and repeat the process\footnote{One can use a different 
dictionary at every step. For performance reasons, we consider a single dictionary
throughout the process in this paper.} using $src_{suffix}$.

To guarantee that encoding always makes progress, we must ensure that every 
dictionary lookup is successful. This means that for any $src$, there must exist a
dictionary entry $s \rightarrow c$ such that $len(s)\!>\!0$ and
$s$ is a prefix of $src$.
In other words, we must consume some prefix from the source string
at every lookup.
We call this property \textbf{dictionary completeness}.
Existing dictionary compression schemes for DBMSs are usually not complete
because they only assign codes to the string values already seen by the DBMS.
%\todo{$\leftarrow$ when are they not? Examples?}
These schemes cannot encode arbitrary strings unless they grow the dictionary,
but growing to accommodate new entries may require the DBMS to re-encode the entire corpus~\cite{binnig2009}.
In the string axis model, a dictionary is complete if and only if
the union of all the intervals (i.e., $\bigcup I_i$)
covers the entire string axis.

A dictionary encoding $Enc: \Sigma^* \rightarrow X^*$
is \textbf{uniquely decodable} if $Enc$ is an \emph{injection} (i.e., there is a one-to-one 
mapping 
from every element of $\Sigma^*$ to an element in $X^*$).
To guarantee unique decodability, we must ensure that
(1) there is only one way to encode a source string and
(2) every encoded result is unique.
Under our string axis model, these requirements are equivalent to
(1) all intervals $I_i$'s are disjoint and
(2) the set of codes $C$ used in the dictionary are uniquely decodable
(we only consider prefix codes in this paper).

With these requirements, we can use the string axis model to construct a dictionary
that is both complete and uniquely decodable.
As shown in \cref{fig:string-axis},
for a given dictionary size of $n$ entries,
we first divide the string axis into $n$ consecutive intervals
$I_0, I_1, \dots, I_{n-1}$, where the max-length common prefix
$s_i$ of all strings in $I_i$ is not empty (i.e., $len(s_i)\!>\!0$)
for each interval.
%We use $b_0, b_1, \dots, b_{n-1}, b_n = maxstring$
We use $b_0, b_1, \dots, b_{n-1}, b_n$
to denote interval boundaries.
That is, $I_i = [b_i, b_{i+1})$ for $i = 0, 1, \dots, n-1$.
We then assign a set of uniquely decodable codes
$c_0, c_1, \dots, c_{n-1}$ to the intervals.
Our dictionary is thus $s_i \rightarrow c_i, i = 0, 1, \dots, n\!-\!1$.
A dictionary lookup maps the source string to
a single interval $I_i$, where $b_i < src < b_{i+1}$.

We can achieve the \textbf{order-preserving} property on top of unique decodability
by assigning monotonically increasing codes
$c_0\!<\!c_1\!<\!\dots\!<\!c_{n-1}$ to the intervals.
This is easy to prove.
Suppose there are two source strings ($src_1$, $src_2$), where $src_1\!<\!src_2$.
If $src_1$ and $src_2$ belong to the same interval $I_i$ in the dictionary,
they must share common prefix $s_i$. Replacing $s_i$ with $c_i$ in
each string does not affect their relative ordering.
If $src_1$ and $src_2$ map to different intervals $I_i$ and $I_j$,
then $Enc(src_1)\!=\!c_i \boldsymbol{\cdot} Enc(src_{1_{suffix}})$,
$Enc(src_2)\!=\!c_j \boldsymbol{\cdot} Enc(src_{2_{suffix}})$.
Since $src_1\!<\!src_2$, $I_i$ must preceed $I_j$ on the string axis.
That means $c_i\!<\!c_j$.
Because $c_i$'s are prefix codes,
$c_i \boldsymbol{\cdot} Enc(src_{1_{suffix}}) < c_j 
\boldsymbol{\cdot} Enc(src_{2_{suffix}})$.
%regardless of what the suffixes are.

For encoding search tree keys,
we prefer schemes that are complete and order-preserving;
unique decodability is implied by the latter property.
Completeness allows the scheme to encode arbitrary keys, while
order-preserving guarantees that the search tree supports
meaningful range queries on the encoded keys.
%\rev{For search tree applications that do not require unique decodability,
%a lossy compression scheme might be acceptable (or even preferable).
%Exploring lossy compression is out of the scope of this paper.}
%and we defer it to future work.}

%% ---------------------------------------------------------------
%% Exploiting Entropy
%% ---------------------------------------------------------------
\subsection{Exploiting Entropy}
\label{sec:model:entropy}

For a dictionary encoding scheme to reduce the size of the corpus,
its emitted codes must be shorter than the source strings.
Given a complete, order-preserving dictionary 
$D: s_i \rightarrow c_i, i = 0, 1, \dots, n\!-\!1$,
let $p_i$ denote the probability that a dictionary entry is accessed
at each step during the encoding of an arbitrary source string.
%\todo{$\leftarrow$ by who? what do you mean by ``accessed''?}
Because the dictionary is complete and uniquely decodable
(implied by order-preserving), $\sum_{i=0}^{n-1}p_i = 1$.
The encoding scheme achieves the best compression when
the compression rate
$CPR = {\sum_{i=0}^{n-1} len(s_i)p_i} \big{/} {\sum_{i=0}^{n-1} len(c_i)p_i}$
is maximized.

%% The compression rate is:

%% \vspace*{-0.12in}
%% \begin{equation*}
%%     %$CPR = \sum_{i=0}^{n-1} len(s_i)p_i / \sum_{i=0}^{n-1} len(c_i)p_i$.
%%     CPR = \frac{\sum_{i=0}^{n-1} len(s_i)p_i}{\sum_{i=0}^{n-1} len(c_i)p_i}
%% \end{equation*}

%% The dictionary encoding scheme achieves the best compression
%% when $CPR$ is maximized.

\begin{figure}[t]
    \centering
    \includegraphics[width=0.95\columnwidth]{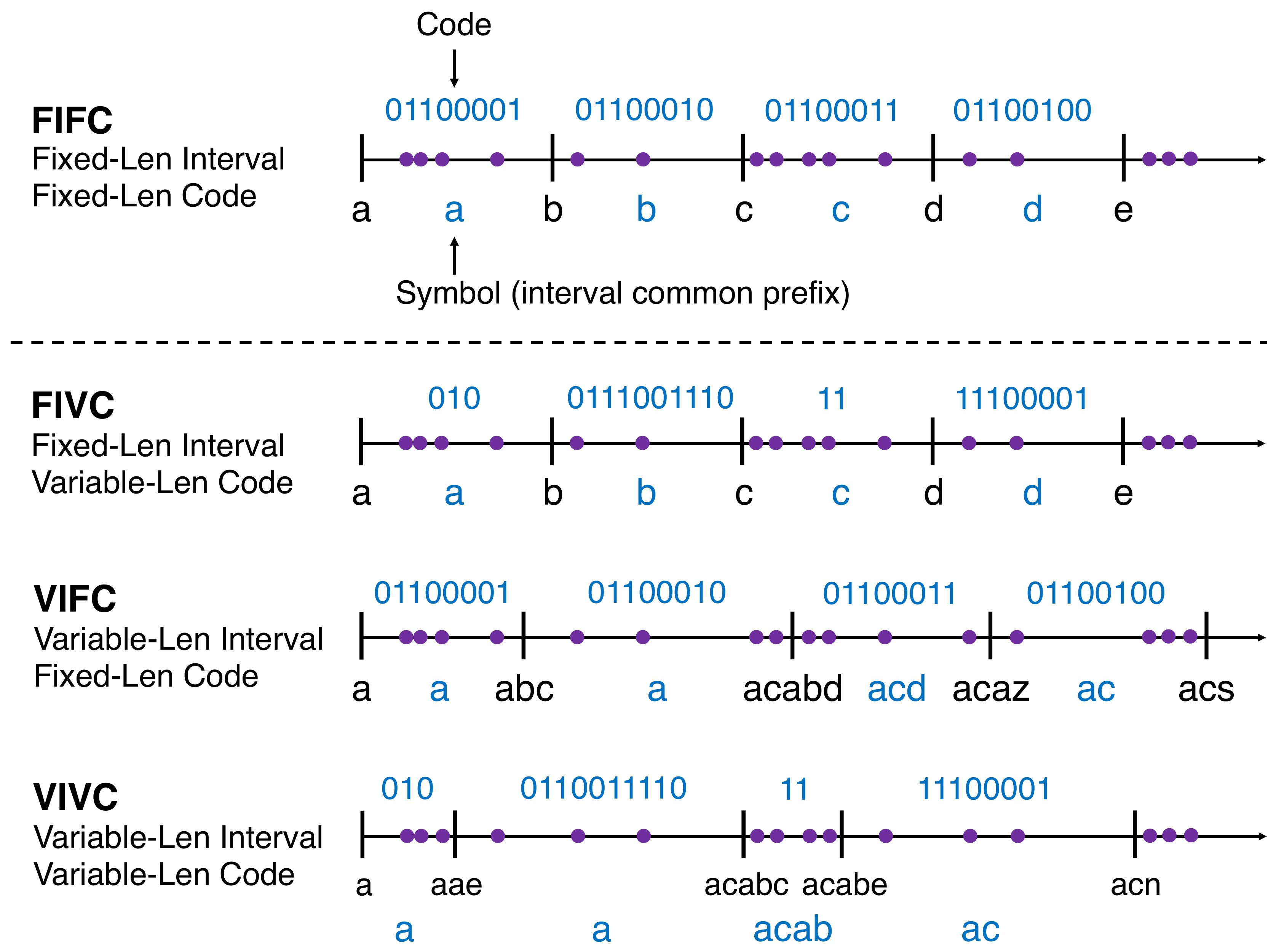}
    \caption{
        \textbf{Compression Models --}
        Four categories of complete and order-preserving dictionary encoding schemes.
%         based on interval length and code length.
    }
    \label{fig:cpr-model}
\end{figure}

According to the string axis model,
we can characterize a dictionary encoding scheme in two aspects:
(1) how to divide intervals and (2) what code to assign to each interval.
Interval division determines the symbol lengths ($len(s_i)$)
and the access probability distribution ($p_i$) in a dictionary.
Code assignment exploits the entropy %of those intervals
in $p_i$'s
by using shorter codes ($c_i$) for more frequently-accessed intervals.

We consider two interval-division strategies:
fixed-length intervals and variable-length intervals.
For code assignment, we consider two types of prefix codes:
fixed-length codes and optimal variable-length codes.
We, therefore, divide all complete and order-preserving dictionary
encoding schemes into four categories, as shown in \cref{fig:cpr-model}.
\\ \vspace{-0.1in}

%% -----------------------
%% \encodingFIFC
%% -----------------------
\noindent \textbf{Fixed-length Interval, Fixed-length Code (\encodingFIFC): }
% \encodingFIFC stands for Fixed-length Interval, Fixed-length Code.
This is the baseline scheme because ASCII encodes characters in this way.
We do not consider this category for compression.
\\ \vspace{-0.1in}

%% -----------------------
%% FIVC
%% -----------------------
\noindent \textbf{Fixed-length Interval, Variable-length Code (\encodingFIVC):}
% \encodingFIVC stands for Fixed-length Interval, Variable-length Code.
This category is the classic Hu-Tucker encoding~\cite{hu1971}.
If order-preserving is not required,
both Huffman encoding~\cite{huffman1952} and arithmetic
encoding~\cite{witten1987} also belong to this category\footnote{Arithmetic encoding does not 
operate the same way as a typical dictionary encoder. But its underlying principle matches this 
category.}.
Although intervals have a fixed length,
access probabilities are not evenly distributed among the intervals.
Using optimal (prefix) codes, thus, maximizes the compression rate.
\\ \vspace{-0.1in}

%% -----------------------
%% VIFC
%% -----------------------
\noindent \textbf{Variable-length Interval, Fixed-length Code (\encodingVIFC):}
% \encodingVIFC stands for Variable-length Interval, Fixed-length Code.
This category is represented by the ALM
string compression algorithm proposed by Antoshenkov~\cite{antoshenkov1997}.
%(we refer to the algorithm as \schemeALM in the rest of the paper).
Because the code lengths are fixed (i.e., $len(c_i) = L$),
the compression rate $CPR = \frac{1}{L} \sum_{i=0}^{n-1} len(s_i)p_i$.
\schemeALM applied the ``equalizing'' heuristic of letting
$len(s_0)p_0 = len(s_1)p_1 = \dots = len(s_n)p_n$
%to try to achieve optimal compression (i.e., maximize $CPR$).
to maximize $CPR$.
We note that the example in \cref{fig:cpr-model} has two intervals with
the same dictionary symbol.
This is allowed because only one of the intervals will contain a specific source string,
%which uniquely determines the result of a dictionary lookup.
%We only use the dictionary symbols (more precisely the lengths of the symbols)
%to decide the number of characters from the source string that we have encoded
%for each step.
Also, by using variable-length intervals, we no longer
have the ``concatenation property'' for the encoded results
(e.g., $Code(ab) \ne Code(a) \boldsymbol{\cdot} Code(b)$).
This property, however, is not a requirement for our target application.
\\ \vspace{-0.1in}

%% -----------------------
%% VIVC
%% -----------------------
\noindent \textbf{\mbox{Variable-length Interval, Variable-length Code (\encodingVIVC):}}
% \encodingVIVC stands for Variable-length Interval, Variable-length Code.
To the best of our knowledge, this category is unexplored by previous work.
Although Antoshenkov suggests that \schemeALM could benefit from
a supplementary variable-length code~\cite{antoshenkov1997},
it is neither implemented nor evaluated.
\encodingVIVC has the most flexibility in building dictionaries
(one can view \encodingFIFC, \encodingFIVC, and \encodingVIFC as special cases of \encodingVIVC),
and it can potentially lead to an optimal compression rate.
We describe the \encodingVIVC schemes in \ope in \cref{sec:model:scheme}.
\\ \vspace{-0.1in}

Although \encodingVIVC schemes can have higher compression rates than the other schemes,
both fixed-length intervals and fixed-length codes have performance
advantages over their variable-length counterparts.
Fixed-length intervals create smaller and faster dictionary structures,
while fixed-length codes are more efficient to decode.
Our objective is to find the best trade-off between compression rate and
encoding performance for in-memory search tree keys.

%% ---------------------------------------------------------------
%% Compression Schemes
%% ---------------------------------------------------------------
% \vspace*{-0.1in}
\subsection{Compression Schemes}
\label{sec:model:scheme}

\begin{figure}[t]
    \centering
    \subfloat[\schemeSingleChar (\encodingFIVC)]{
        \includegraphics[width=0.95\columnwidth]{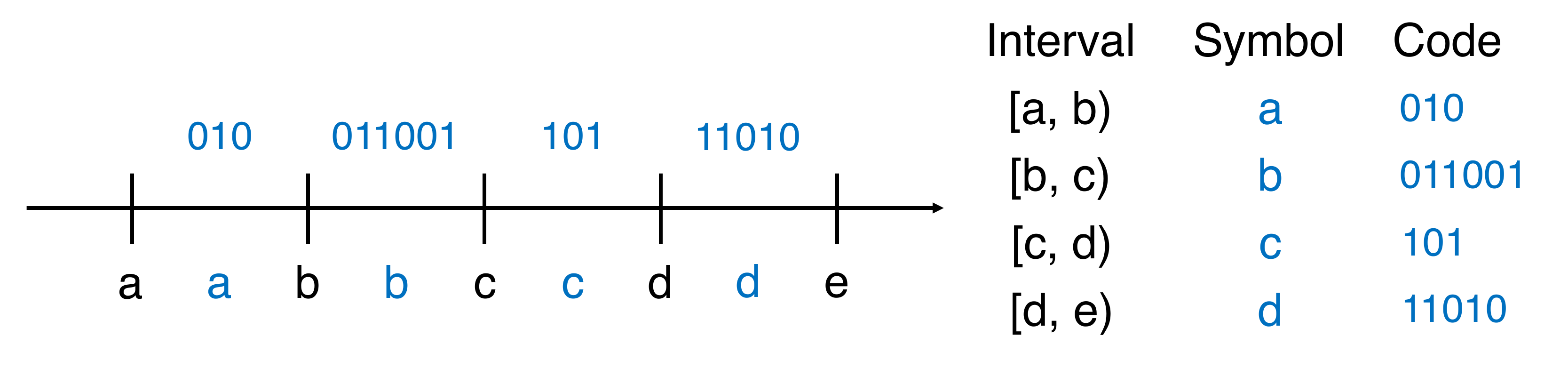}
        \label{fig:single-char}
    }
    \\
    \subfloat[\schemeDoubleChar (\encodingFIVC)]{
        \includegraphics[width=0.95\columnwidth]{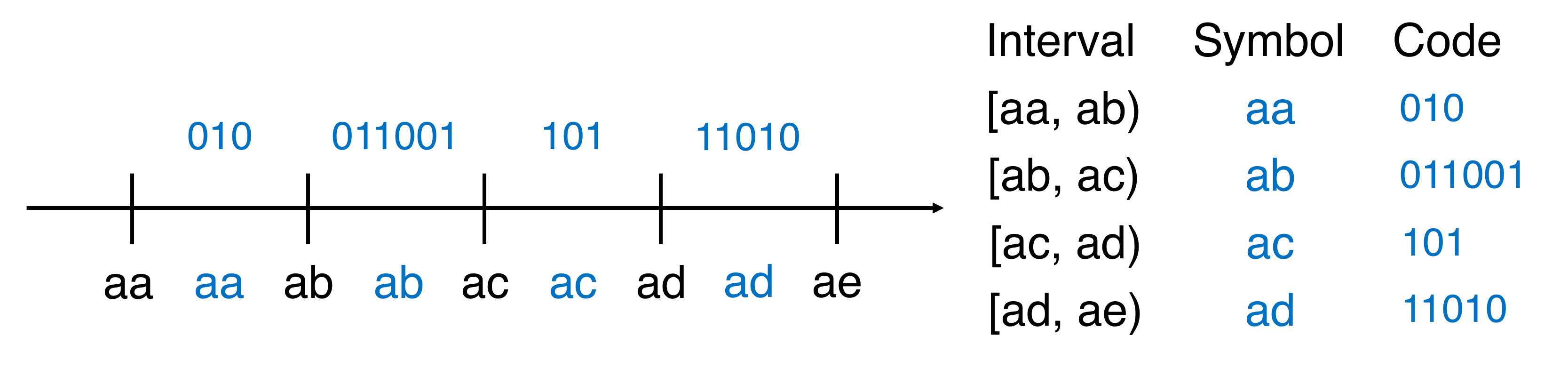}
        \label{fig:double-char}
    }
    \\
    \subfloat[\schemeALM (\encodingVIFC)]{
        \includegraphics[width=0.95\columnwidth]{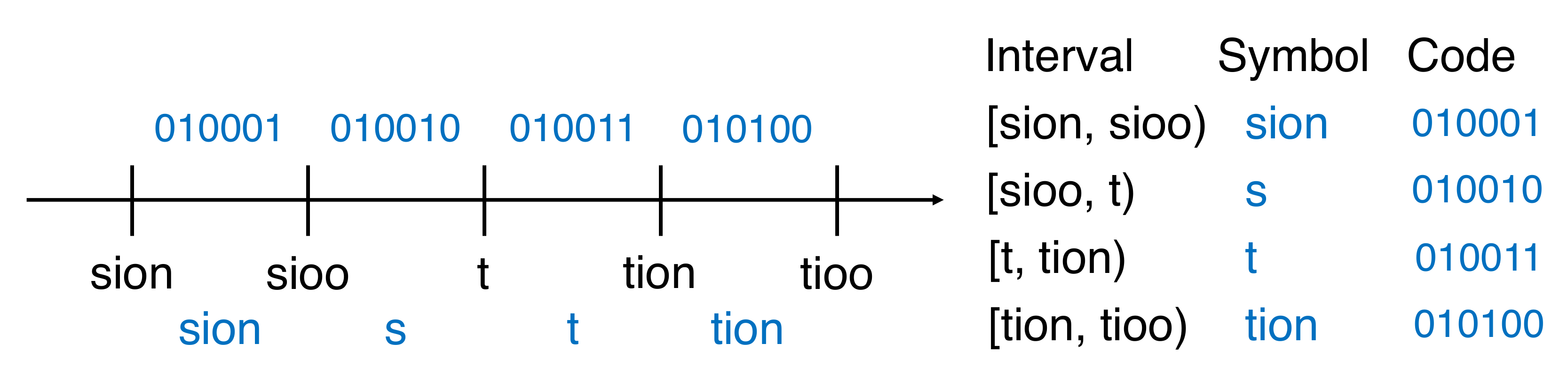}
        \label{fig:alm}
    }
    \\
    \subfloat[\schemeThreeGrams (\encodingVIVC)]{
        \includegraphics[width=0.95\columnwidth]{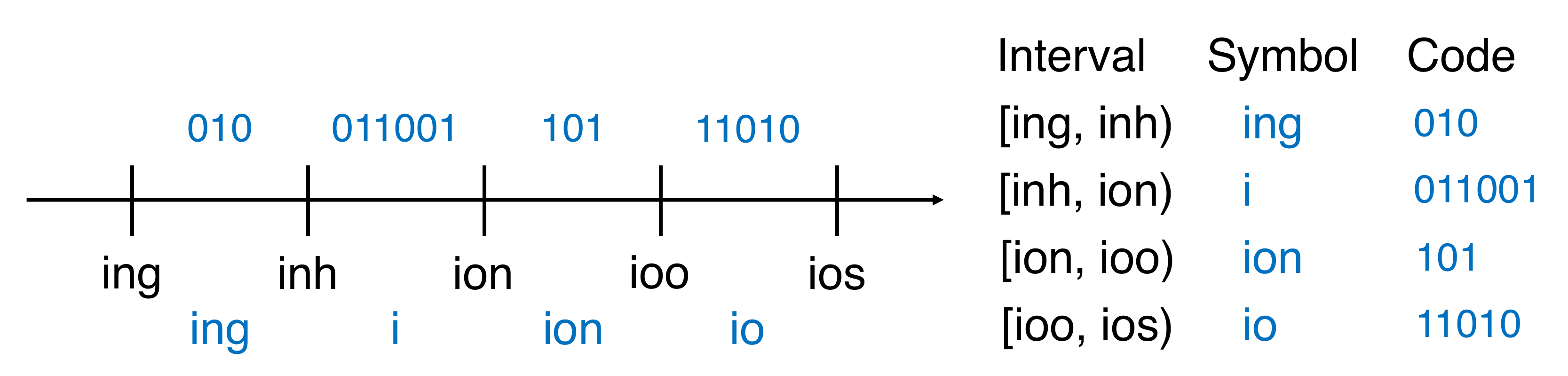}
        \label{fig:3-grams}
    }
    \\
    \subfloat[\schemeFourGrams (\encodingVIVC)]{
        \includegraphics[width=0.95\columnwidth]{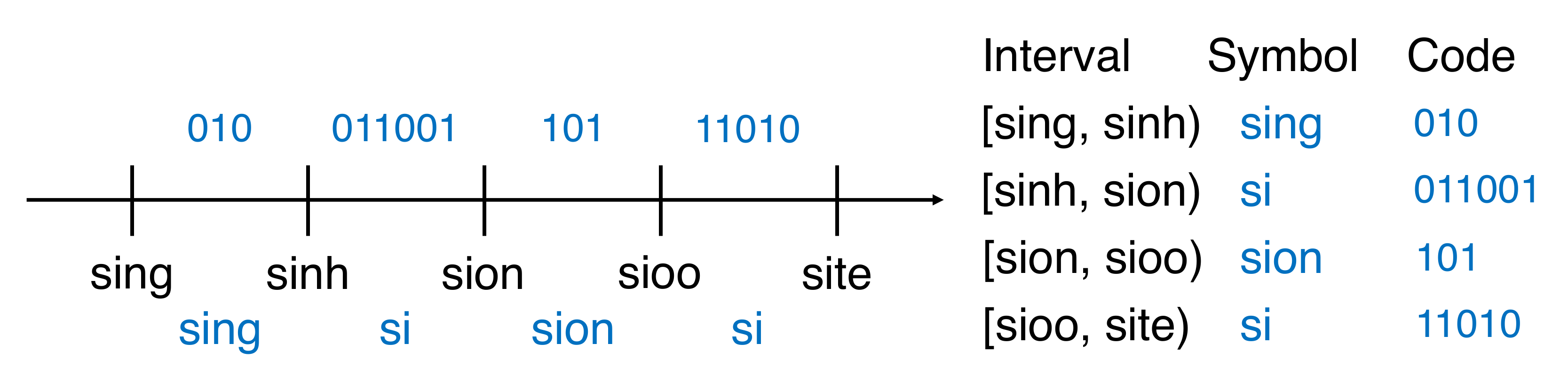}
        \label{fig:4-grams}
    }
    \\
    \subfloat[\schemeALMImproved (\encodingVIVC)]{
        \includegraphics[width=0.95\columnwidth]{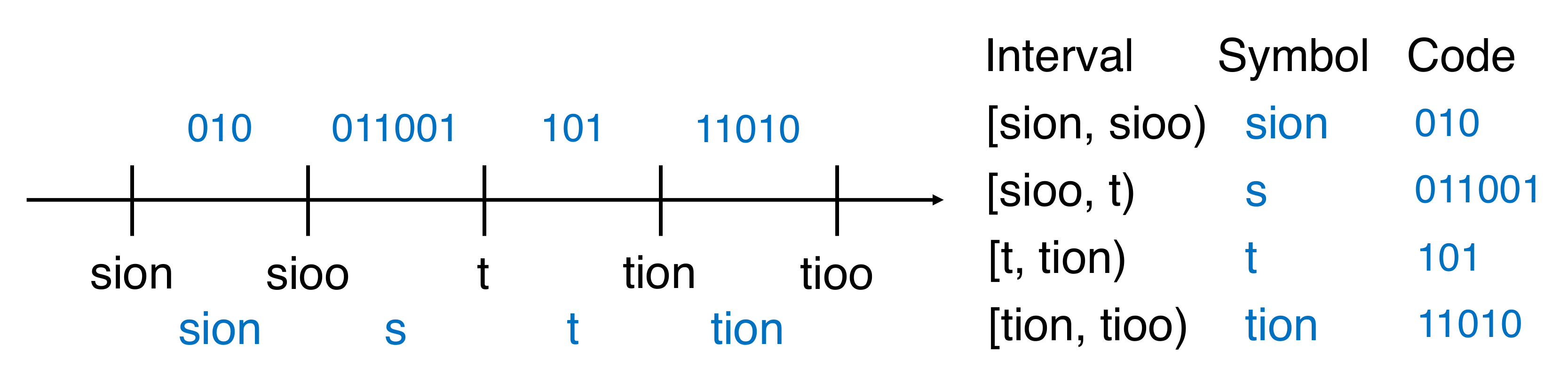}
        \label{fig:alm-improved}
    }
    \caption{
        \textbf{Compression Schemes} --
        Example dictionary segments.
    }
    \label{fig:cpr-schemes}
\end{figure}

Based on the above dictionary encoding models,
we next introduce six compression schemes implemented in \ope,
as shown in \cref{fig:cpr-schemes},
%within the \ope framework.
% These are the built-in schemes in \ope.
We select these schemes from the three viable categories (\encodingFIVC, \encodingVIFC, 
and \encodingVIVC).
Each scheme makes different trade-offs between compression rate and
encoding performance.
We first describe them at a high level and then provide their
implementation details in~\cref{sec:impl}.
\\ \vspace{-0.1in}

%% -----------------------
%% Single-Char
%% -----------------------
\noindent \textbf{\schemeSingleChar}
is the \encodingFIVC compression algorithm used in
Huffman encoding and arithmetic encoding.
The fixed-length intervals have consecutive single characters
as the boundaries (e.g., $\left[\mbox{\texttt{a}}, \mbox{\texttt{b}}\right)$, 
$\left[\mbox{\texttt{b}}, \mbox{\texttt{c}}\right)$).
The dictionary symbols are 8-bit ASCII characters,
and the dictionary has a fixed 256 entries.
The codes assigned to the symbols are Hu-Tucker codes.
Hu-Tucker codes are optimal order-preserving prefix codes
%(they are essentially order-preserving Huffman codes).
\cref{fig:single-char} shows an example dictionary segment.
\\ \vspace{-0.1in}

%% -----------------------
%% Double-Char
%% -----------------------
\noindent \textbf{\schemeDoubleChar}
is a \encodingFIVC compression algorithm that is similar to \schemeDoubleChar,
except that the interval boundaries are consecutive double characters (e.g.,
$\left[\mbox{\texttt{aa}}, \mbox{\texttt{ab}}\right)$,
$\left[\mbox{\texttt{ab}}, \mbox{\texttt{ac}}\right)$).
To make the dictionary complete, we introduce a terminator
character $\varnothing$ before all ASCII characters to fill the interval gaps
between $\left[\mbox{\texttt{a`\textbackslash255'}}, \mbox{\texttt{b}}\right)$ and 
$\left[\mbox{\texttt{b`\textbackslash0'}}, \mbox{\texttt{b`\textbackslash1'}}\right)$,
for example, with interval 
$\left[\mbox{\texttt{b}}\varnothing, \mbox{\texttt{b`\textbackslash0'}}\right)$.
% The dictionary uses Hu-Tucker codes.
\cref{fig:double-char} shows an example dictionary.
This scheme should achieve better compression than \schemeSingleChar
because it exploits the first-order entropy of the source strings
instead of the zeroth-order entropy.
\\ \vspace{-0.1in}

%% -----------------------
%% ALM
%% -----------------------
\noindent \textbf{\schemeALM}
is a state-of-the-art \encodingVIFC string compression algorithm.
To determine the interval boundaries from a set of sample source strings (e.g., initial 
keys for an index), \schemeALM first selects substring patterns that are long and frequent.
Specifically, for a substring pattern $s$, it computes
$len(s) \times freq(s)$, where $freq(s)$ represents the number of
occurrence of $s$ in the sample set.
\schemeALM includes $s$ in its dictionary if the product is greater than
a threshold $W$.
%It then fills in each gap between the selected symbols with one or more intervals.
It then creates one or more intervals between the adjacent selected symbols.
The goal of the algorithm is to make the above product
(i.e., length of common prefix $\times$ access frequency)
for each interval as equal as possible.
The detailed algorithm is described in~\cite{antoshenkov1997}.

\schemeALM uses monotonically increasing fixed-length codes.
\cref{fig:alm} shows an example dictionary segment.
The dictionary size for \schemeALM depends on the threshold $W$.
One must binary search on $W$'s to obtain a desired dictionary size.
\\ \vspace{-0.1in}

%% -----------------------
%% 3-Grams
%% -----------------------
\noindent \textbf{\schemeThreeGrams}
is a \encodingVIVC compression algorithm where the interval boundaries are
%the first \encodingVIVC compression algorithm that we support in \ope.
%As the name suggests, the interval boundaries in this scheme are
3-character strings.
Given a set of sample source strings and a dictionary size limit $n$,
the scheme first selects the top $n/2$ most frequent 3-character patterns
and adds them to the dictionary.
For each interval gap between the selected 3-character patterns,
\schemeThreeGrams creates a dictionary entry to cover the gap.
For example, in \cref{fig:3-grams},
``\texttt{ing}'' and ``\texttt{ion}'' are selected frequent patterns
from the first step.
``\texttt{ing}'' and ``\texttt{ion}'' represent intervals
$\left[\mbox{\texttt{ing}}, \mbox{\texttt{inh}}\right)$ and
$\left[\mbox{\texttt{ion}}, \mbox{\texttt{ioo}}\right)$
on the string axis.
Their gap interval $\left[\mbox{\texttt{inh}}, \mbox{\texttt{ion}}\right)$
is also included as a dictionary entry.
3-Grams uses Hu-Tucker codes.
%using Hu-Tucker codes.
%\todo{it's not clear how the new 3-gram dividing point inh is selected, and why the ``gap 
%interval'' in the previous sentence isn't [ing, ion)}
\\ \vspace{-0.1in}

%% -----------------------
%% 4-Grams:
%% -----------------------
\noindent \textbf{\schemeFourGrams}
is a \encodingVIVC compression algorithm similar to \schemeThreeGrams with 4-character string 
boundaries.
\cref{fig:4-grams} shows an example.
%Note that although two dictionary entries in the example
%have the same symbol, there is no ambiguity because
%a dictionary lookup only returns the entry whose
%interval on the string axis contains the source string.
%\todo{this needs more explanation, and earlier in the \
%  text, as it happens in Fig. 3, too}
Compared to \schemeThreeGrams, \schemeFourGrams exploits higher-order entropy; but whether 
it provides a better compression rate over \schemeThreeGrams depends on the dictionary size.
\\ \vspace{-0.1in}

%% -----------------------
%% ALM-Improved
%% -----------------------
\noindent \textbf{\schemeALMImproved}
improves the \schemeALM scheme in two ways.
First, as shown in \cref{fig:alm-improved},
we replace the fixed-length codes in \schemeALM with Hu-Tucker
codes because we observe access skew among the intervals
despite \schemeALM's ``equalizing'' algorithm.
%\todo{not clear; ``although'' -> ``even though''?}
Second, the original \schemeALM counts the frequency for every
substring (of any length) in the sample set,
which is slow and memory-consuming.
In \schemeALMImproved, we simplify this by only collecting
statistics for substrings that are suffixes of the sample source strings.
%\todo{should give intuition for why this still works}
Our evaluation in \cref{sec:micro-eval} shows that
using Hu-Tucker codes improves \schemeALM's compression rate
while counting the frequencies of string suffixes reduces
\schemeALM's build time
without compromising the compression rate.

%A discussion on optimality --> easily extensible

\section{\ope}
\label{sec:impl}

We now present the design and implementation of \ope.
%the \ope framework.
There are two goals in \ope's architecture.
First, \ope must minimize its performance overhead so that
it does not negate the benefits of storing shorter keys.
Second, \ope must be extensible.
From our discussion in \cref{sec:model}, there are many choices in
constructing an order-preserving dictionary encoding scheme.
Although we support six representative schemes in the current version
of \ope, one could, for example, devise better heuristics in
generating dictionary entries to achieve a higher compression rate,
or invent more efficient dictionary data structures to further
reduce encoding latency.
\ope can be easily extended to include such improvements
through its modularized design.

%\todo{Need to define the two phases first: build, encode.}

%% ---------------------------------------------------------------
%% Overview
%% ---------------------------------------------------------------
\subsection{Overview}
\label{sec:impl:overview}

\begin{figure}[t]
    \centering
    \includegraphics[width=0.95\columnwidth]{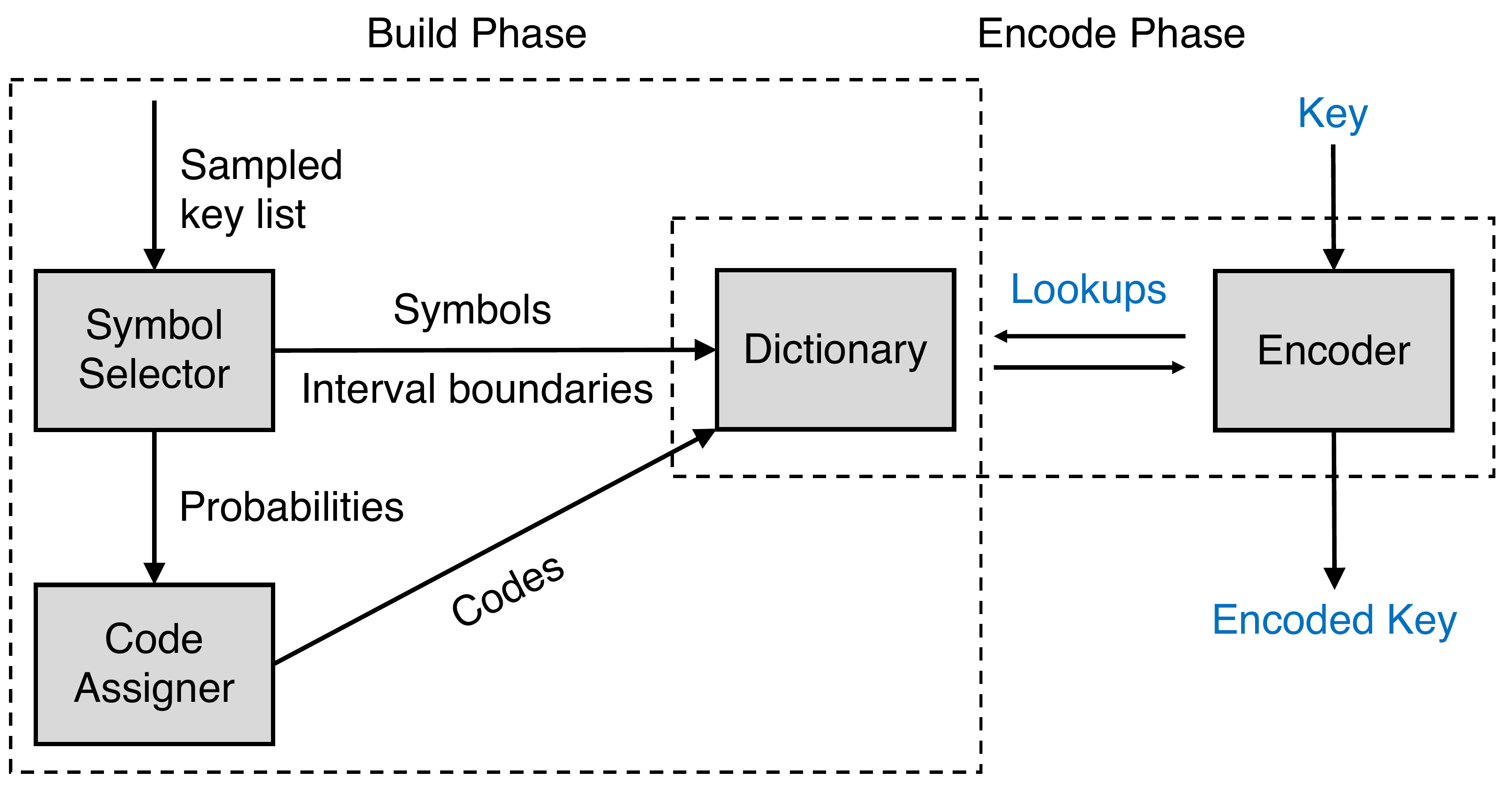}
    \caption{
        \textbf{The \ope Framework} --
        An overview of \ope's modules and their interactions with each other in the two phases.
%         The Build Phase uses the sampled keys to construct the symbol-code dictionary. 
%         The Encode Phase is when a query uses the \compEncoder to convert a key into their 
%         corresponding dictionary code.
    }
    \label{fig:framework}
\end{figure}

As shown in \cref{fig:framework},
\ope executes in two phases (i.e., Build, Encode) and has four modules:
%the \ope framework executes in two phases (Build, Encode) and has four modules:
\textbf{\compSelector}, \textbf{\compAssigner}, \textbf{\compDictionary},
and \textbf{\compEncoder}.
A DBMS provides \ope with a list of sample keys from the search tree.
\ope then produces a \compDictionary and an \compEncoder as its output.
\rev{We note that the size and representativeness of the sampled key list
  only affect the compression rate. The correctness of \ope's compression
  algorithm is guaranteed by the dictionary completeness and
  order-preserving properties discussed in~\cref{sec:model:axis}.
  In other words, any \ope dictionary can both encode arbitrary input keys
  and preserve the original key ordering.}

%In the first step of the build phase, the \compSelector counts the frequencies 
%of string patterns in the sampled key list as specified by the target compression scheme.
%It then divides the string axis into intervals based on
%the collected statistics.
In the first step of the build phase, the \compSelector counts the
frequencies of the specified string patterns in the sampled key list
and then divides the string axis into intervals
based on the heuristics given by the target compression scheme.
The \compSelector generates three outputs for each interval:
(1) dictionary symbol (i.e., the common prefix of the interval),
(2) interval boundaries, and
(3) probability that a source string falls within that interval.

The framework then gives the symbols and interval boundaries to
the \compDictionary module.
Meanwhile, it sends the probabilities to the 
\compAssigner to generate codes for the dictionary symbols.
If the scheme uses fixed-length codes,
the \compAssigner only considers the dictionary size.
If the scheme uses variable-length codes,
the \compAssigner examines the probability distribution
to generate optimal order-preserving prefix codes
(i.e., Hu-Tucker codes).

When the \compDictionary module receives the symbols,
the interval boundaries, and the codes,
it selects an appropriate and fast dictionary data structure
to store the mappings.  The string lengths of the interval boundaries
inform the decision; available data structures range from fixed-length arrays
to general-purpose tries.
The dictionary size is a tunable parameter
for \encodingVIFC and \encodingVIVC schemes.
Using a larger dictionary trades performance for a better compression rate.
% The build phase completes once the dictionary is built.

The encode phase uses only the \compDictionary and \compEncoder modules.
On receiving an uncompressed key, the \compEncoder performs multiple
lookups in the dictionary. Each lookup translates a part of the original
key to some code as described in \cref{sec:model:axis}.
The \compEncoder then concatenates the codes in order and outputs the result.
This encoding process is sequential
for variable-length interval schemes (i.e., \encodingVIFC and \encodingVIVC)
because the remaining source string to be encoded depends on the results 
of earlier dictionary lookups.

% The framework executes the build phase only once unless the
% key distribution changes significantly.
%We contend that this is rare for a database column in most workloads.
%\todo{Can you cite something and get rid of contend}
%A shift in the key distribution, however, does not affect
%\ope's correctness; it only reduces the compression rate.
% The encode phase, on the other hand, is performance critical
% because every insert/query key goes through the encoder
% before the search tree processes the query.

% Given this high-level overview of \ope, 
We next describe the implementation details for each module.
Building a decoder %and its corresponding dictionary
is optional because our target workload for search trees
does not require reconstructing the original keys.
%We leave designing an optimized decoder as future work.

%% ---------------------------------------------------------------
%% Implementation
%% ---------------------------------------------------------------
\subsection{Implementation}
\label{sec:impl:impl}

\begin{table}[t!]
    \centering
%     {\scriptsize \input{tables/modules.tex} }
    \includegraphics[width=\columnwidth]{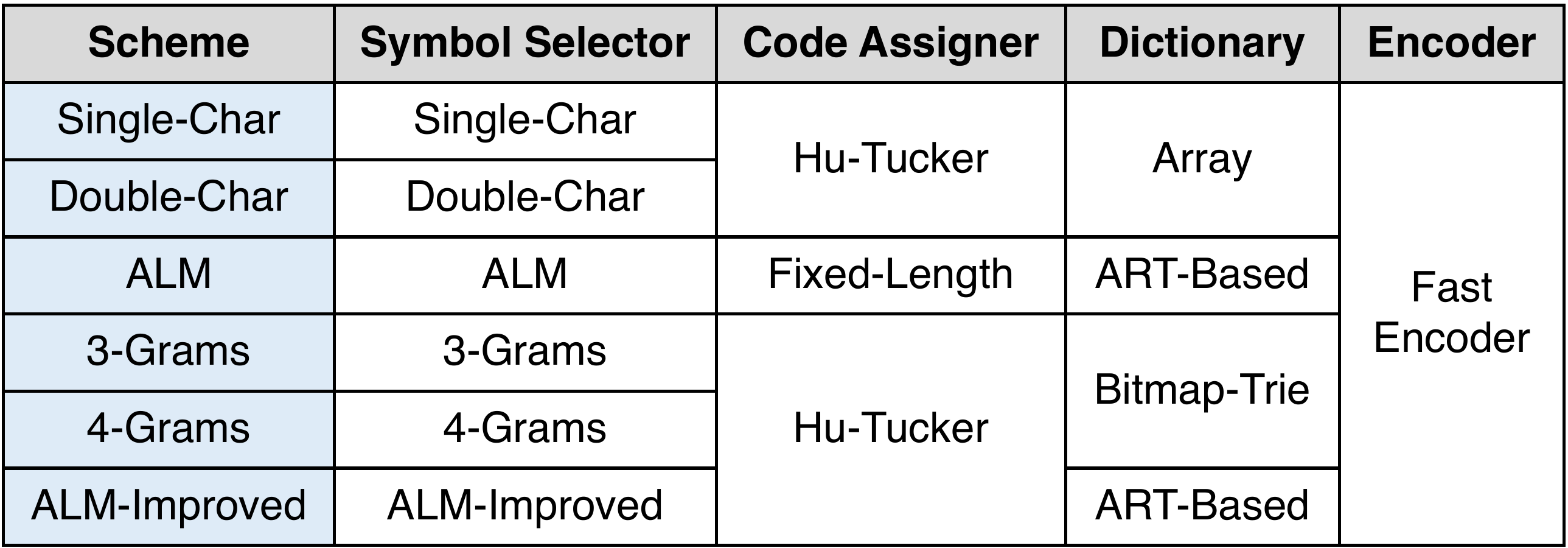}
    \caption{
        \textbf{Module Implementations} --
        The configuration of \ope's six compression schemes.
    }
    \label{tbl:modules}
\end{table}

% We define clear interfaces for the modules in \cref{fig:framework}.
% A module can have multiple implementations.
\ope users can create new compression schemes by combining different module implementations.
%As shown in \cref{tbl:modules},
\ope currently supports the six compression schemes
described in \cref{sec:model:scheme}.
For \compSelector and \compAssigner,
the goal is to generate a dictionary that leads to the maximum compression rate.
%These two modules are garbage-collected after the build phase completes
We no longer need these two modules after the build phase.
%\todo{$\leftarrow$ You did not describe what the garbage is}.
%\todo{Do you just mean their memory is freed; maybe just say that}
We spend extra effort optimizing the \compDictionary and \compEncoder modules 
because they are on the critical path of every search tree query.
\\ \vspace{-0.1in}

%% -----------------------
%% Selector
%% -----------------------
\noindent \textbf{\compSelector:}
% Each compression scheme in \ope has its own \compSelector.
It first counts the occurrences of substring patterns
in the sampled keys using a hash table.
For example, \schemeThreeGrams considers all three-character substrings,
while the \schemeALM examines substrings of all lengths.
%while the \schemeALMImproved \compSelector only analyzes
%``suffix substrings'' \todo{$\leftarrow$ what does this mean?}.
% 
For \schemeSingleChar and \schemeDoubleChar, the interval boundaries are implied
because they are fixed-length-interval schemes.
%(i.e., \encodingFIVC).
For the remaining schemes, the \compSelector{s} divide intervals
using the algorithms described in \cref{sec:model:scheme}:
first identify the most frequent
symbols and then fill the gaps with new intervals.

The \schemeALM and \schemeALMImproved \compSelector{s} require an extra \emph{blending}
step before their interval-division algorithms.
This is because the selected variable-length substrings may not satisfy the \emph{prefix property}
(i.e., a substring can be a prefix of another substring).
For example, ``\texttt{sig}'' and ``\texttt{sigmod}'' may both appear
in the frequency list,
but the interval-division algorithm cannot select both of them
because the two intervals on the string axis are not disjoint:
``\texttt{sigmod}'' is a sub-interval of ``\texttt{sig}''.
A blending algorithm redistributes the occurrence count
of a prefix symbol to its longest extension in the frequency list~\cite{antoshenkov1997}.
We implement this blending algorithm in \ope using a
trie data structure.
%customized trie data structure 
%\todo{$\leftarrow$ what do you mean by ``customized''? what is special about it?}

After the \compSelector decides the intervals,
it performs a test encoding of the sample keys using the intervals
as if the code for each interval has been assigned.
The purpose of this step is to obtain the probability that a source string
(or its remaining suffix) %after certain encoding steps)
falls into each interval
so that the \compAssigner can generate codes based on those probabilities
to maximize compression.
For variable-length-interval schemes, the probabilities
are weighted by the symbol lengths of the intervals.
\\ \vspace{-0.1in}

%% -----------------------
%% Assigner
%% -----------------------
\noindent \textbf{\compAssigner:}
Assume that the \compAssigner receives $N$ probabilities.
%from the \compSelector.
To assign fixed-length codes to them,
the \compAssigner outputs monotonically increasing
integers $0, 1, 2, \cdots, N-1$,
each using $\left \lceil{\log_{2}N}\right \rceil$ bits.
For variable-length codes,
\ope uses the Hu-Tucker algorithm to generate optimal
order-preserving prefix codes.
One could use an alternative method, such as \emph{Range Encoding}~\cite{martin1979}
(i.e., the integer version of Arithmetic Encoding).
Range Encoding, however, requires more bits than Hu-Tucker to ensure that codes are exactly on 
range boundaries to guarantee order-preserving.

% \todo{MK: I would remove the entire next paragraph and take the sentences 
% in the following one as your reference to the algorithm}
The Hu-Tucker algorithm works in four steps \cite{www-hu-tucker}.
%\footnote{\rev{\cite{www-hu-tucker}
%provides a detailed and intuitive description of the Hu-Tucker algorithm.}}.
First, it creates a leaf node for each probability
%received from the \compSelector
and then lists the leaf nodes in interval order.
Second, it recursively merges the two least-frequent nodes
where there are no leaf nodes between them (unlike Huffman).
%it selects the two least-frequent nodes and merges them
%to create a new internal node.
%This new node takes the place of the existing left node.
%Unlike the Huffman algorithm, Hu-Tucker allows two nodes to merge
%only if there are no leaf nodes between them.
This is where the algorithm guarantees order.
%After constructing this tree structure,
After constructing this probability tree,
it next computes the depth of each leaf node to derive the lengths of the codes.
Finally, the algorithm constructs a tree by adding these leaf nodes level-by-level starting
from the deepest and then connecting adjacent nodes at the same level in pairs.
\ope uses this
%new data structure as a Huffman tree
Huffman-tree-like structure to extract the final codes.
Our Hu-Tucker implementation in the \compAssigner uses an improved
algorithm~\cite{yohe1972} that runs in $\mathcal{O}(N^2)$ time
instead of $\mathcal{O}(N^3)$ as in the original paper~\cite{hu1971}.
\\ \vspace{-0.1in}

%% -----------------------
%% Dictionary
%% -----------------------
\noindent \textbf{\compDictionary:}
A dictionary in \ope maps an interval (and its symbol) to a code.
Because the intervals are connected and disjoint, the dictionary
needs to store only the left boundary of each interval as the key.
A key lookup in the dictionary then is a ``greater than or equal to''
index query.
%on the underlying data structure.
For the values,
we store only the codes along with the lengths of the symbols
to determine the number of characters from
the source string that we have consumed at each step.

We implemented three dictionary data structures in \ope.
The first is an \textit{array} for the \schemeSingleChar and \schemeDoubleChar schemes.
Each dictionary entry includes an 8-bit integer to record
the code length and a 32-bit integer
% \footnote{32-bits is enough to store all the codes in our 
% evaluation. For larger dictionaries, \ope may need a 64-bit field.}
to store the code.
The dictionary symbols and the interval left boundaries
are implied by the array offsets.
For example, the $97^{\mbox{\scriptsize th}}$ entry in \schemeSingleChar has the symbol \texttt{a}, 
while the $24770^{\mbox{\scriptsize th}}$
entry in \schemeDoubleChar corresponds to the symbol \texttt{aa}\footnote{
  $24770 = 96 \times (256 + 1) + 97 + 1$.
  The $+1$'s are for the terminators $\varnothing$
  %The $+1$'s are necessary because of the terminator character $\varnothing$.
  %See \cref{sec:model:scheme}.
}.
A lookup in an array-based dictionary is fast because it requires only a single
memory access and the array fits in CPU cache.

\begin{figure}[t]
    \centering
    \includegraphics[width=\columnwidth]{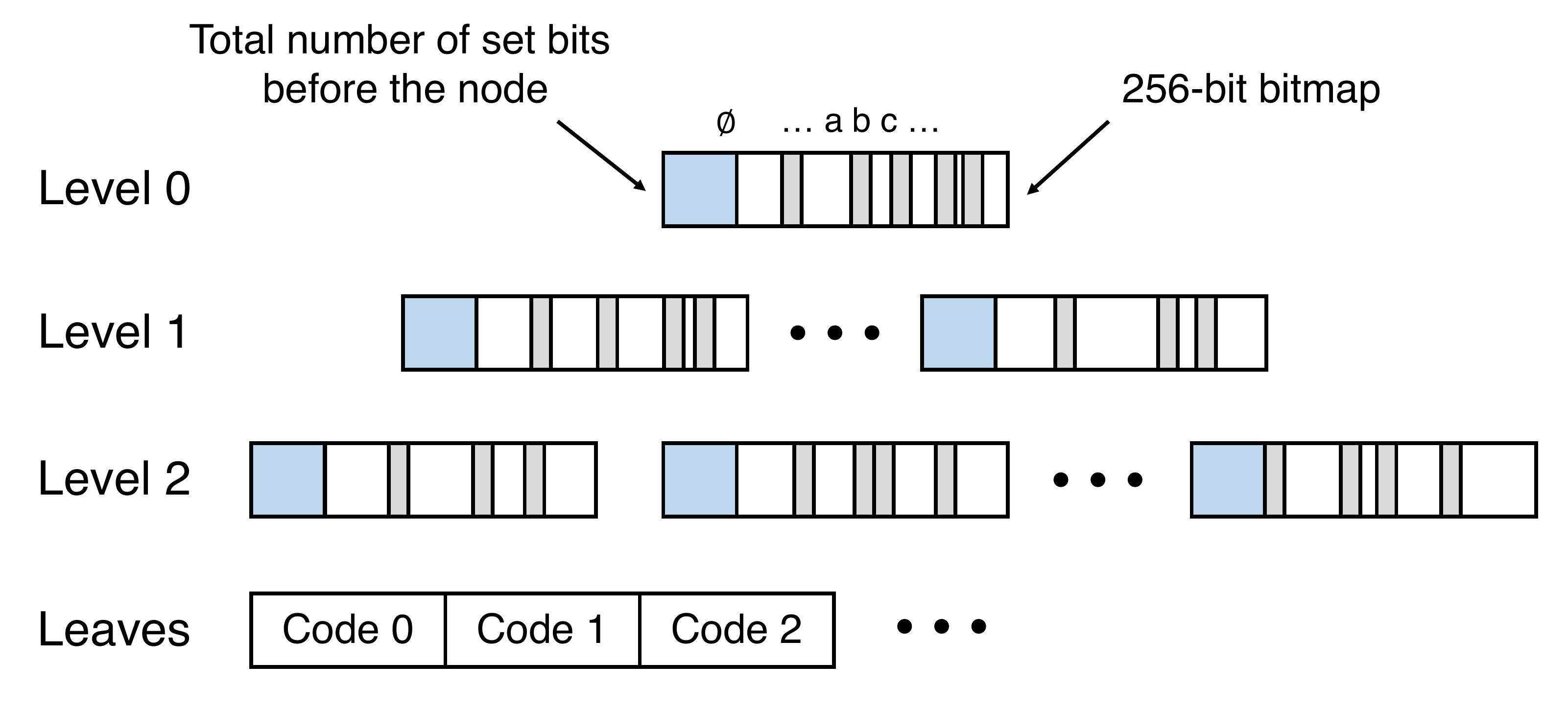}
    \caption{
        \textbf{\schemeThreeGrams Bitmap-Trie \compDictionary} --
        Each node consists of a 256-bit bitmap and a counter.
        The former records the branches of the node and the latter represents the total number of 
        set bits in the bitmaps of all the preceding nodes. 
    }
      \label{fig:3-gram-trie}
\end{figure}

The second dictionary data structure in \ope is a \textit{bitmap-trie} used by the 
\schemeThreeGrams and 
4-Grams schemes.
\cref{fig:3-gram-trie} depicts the structure of a three-level
bitmap-trie for \schemeThreeGrams.
The nodes are stored in an array in breadth-first order.
Each node consists of a 32-bit integer and a 256-bit bitmap.
The bitmap records all the branches of the node.
For example, if the node has a branch labeled \texttt{a},
the $97^{\mbox{\scriptsize th}}$ bit in the bitmap is set.
The integer at the front stores the total number of set bits
in the bitmaps of all the preceding nodes.
Since the stored interval boundaries can be shorter than three characters, the data structure 
borrows the most significant bit from the 32-bit integer
to denote the termination character $\varnothing$.
%In other words, $\varnothing$ is the first bit of the 257-bit bitmap in a node.
% This seems inconsistent with the diagram, where the MSB of the integer is disjoint from the bitmap.

% Navigating the bitmap-trie is fast.
Given a node ($n$, $bitmap$) 
where $n$ is the count of the preceding set bits,
its child node pointed by label $l$ at position
$n + \textit{popcount}(\textit{bitmap}, l)$\footnote{The \texttt{POPCOUNT} CPU instruction 
counts the set bits in a bit-vector. The function $\textit{popcount}(\textit{bitmap}, l)$ counts the
set bits up to position $l$ in $\textit{bitmap}$.}
in the node array.
Our evaluation shows that looking up a bitmap-trie
is 2.3$\times$ faster than binary-searching the dictionary entries
\rev{because it requires fewer memory probes and has better cache performance.}

Finally, we use an \art-based dictionary to serve the
\schemeALM and \schemeALMImproved schemes.
\art is a radix tree that supports variable-length keys~\cite{leis2013}.
We modified three aspects of \art to make it more suitable as a
dictionary.
First, we added support for \emph{prefix keys} in \art. This is necessary because both \texttt{abc} 
and \texttt{abcd}, for example,
can be valid interval boundaries stored in a dictionary.
We also disabled \art's \emph{optimistic common prefix skipping} that compresses paths 
on single-branch nodes by storing only the first few bytes. If a 
corresponding segment of a query key matches the stored bytes during a lookup, \art assumes 
that the key segment also matches the rest of the common prefix
(a final key verification happens against the full tuple).
\ope's \art-based dictionary, however, stores the full common prefix
for each node since it cannot assume that there is a tuple with the original key.
Lastly, we modified the \art's leaf nodes to store the dictionary entries instead of tuple 
pointers.
\\ \vspace{-0.1in}

%% -----------------------
%% Encoder
%% -----------------------
\noindent \textbf{\compEncoder:}
% As shown in \cref{sec:model:axis},
% the encoding algorithm works as follows.
\ope looks up the source string in the dictionary to find an interval
that contains the string. The dictionary returns
the symbol length $L$ and the code $C$.
\ope then concatenates $C$ to the result buffer
and removes the prefix of length $L$ that matches the symbol
from the source string.
It repeats this process on the remaining string until it is empty.

To make the non-byte-aligned code concatenation fast,
\ope stores codes in 64-bit integer buffers. It adds a new code 
to the result buffer in three steps:
(1) left-shift the result buffer to make room for the new code;
(2) write the new code to the buffer using a bit-wise \texttt{OR} instruction;
(3) split the new code if it spans two 64-bit integers.
This procedure costs only a few CPU cycles per code concatenation.

When encoding a batch of sorted keys,
the \compEncoder optimizes the algorithm by first dividing the batch into blocks,
where each block contains a fixed number of keys.
The \compEncoder then encodes the common prefix of the keys within a block
only once, avoiding redundant work.
When the batch size is two,
%we call this optimization pair-encoding.
we call this pair-encoding.
Compared to encoding keys individually,
pair-encoding reduces key compression overhead
for the closed-range queries in a search tree.
%as shown in~\cref{fig:batch} in~\cref{sec:micro-eval}.
\rev{We evaluate batch encoding in~\cref{sec:batch}.}

\section{Integration}
\label{sec:integration}

\begin{figure}[t]
    \centering
    \includegraphics[width=\columnwidth]{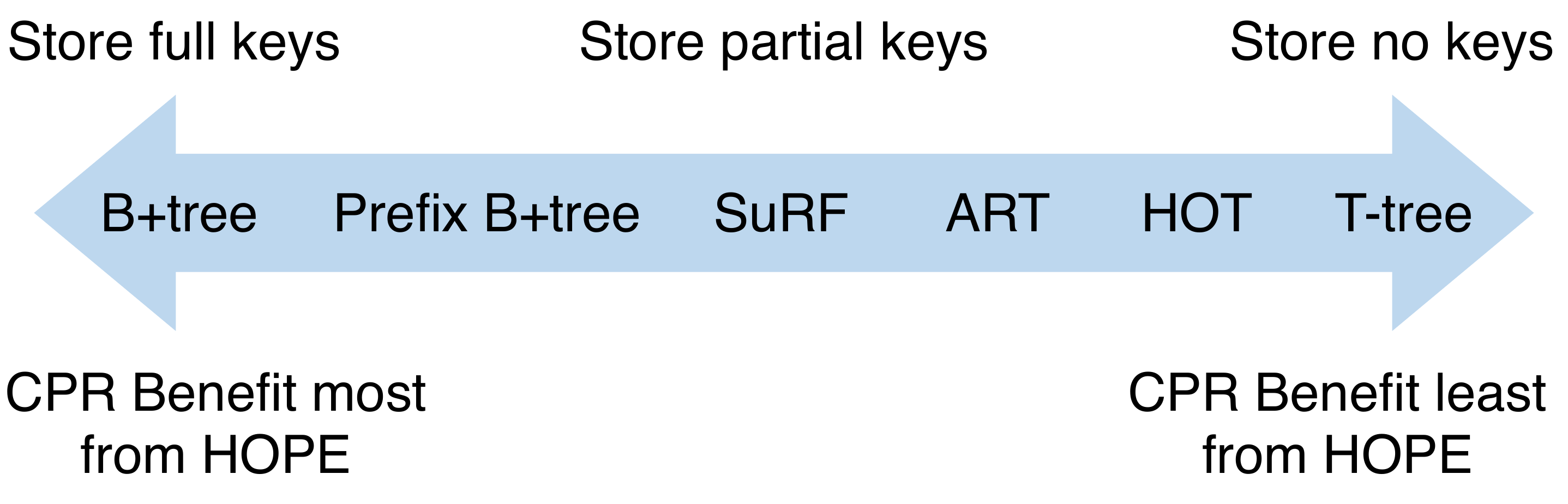}
    \caption{
        \textbf{Search Tree on Key Storage} --
        %\bplustree, \surf, \art, \hot, and \ttree get decreasing
        %benefits from HOPE as the completeness of key storage goes down.
        \rev{%\bplustree, \pbtree, \surf, \art, \hot, and \ttree 
        Search trees get decreasing
        benefits from HOPE, especially in terms of compression rate (CPR),
        as the completeness of key storage goes down.}
    }
    \label{fig:tree-spectrum}
\end{figure}

% \ope targets in-memory search trees.
Integrating \ope in a DBMS is a straightforward process 
because we designed it to be a standalone library that is 
independent of the target search tree data structure.
%and with zero external dependencies.

When the DBMS creates a new search tree,
\ope samples the initial bulk-loaded keys
and constructs the dictionary (i.e., the build phase).
%Once \ope creates the \compDictionary and \compEncoder modules,
After that,
every query for that tree, including the initial bulk-inserts,
must go through the \compEncoder first to compress the keys.
%before it can be processed by the search tree.
If the search tree is initially empty, \ope samples keys as the DBMS inserts them into 
the tree. It then rebuilds the search tree using the compressed
keys once it sees enough samples.
%based on a user-defined threshold.
We use a small sample size because it guarantees fast tree rebuild,
and it does not compromise the compression rate,
as shown in \cref{sec:sample-size}.
%We set this threshold low in our experiments to guarantee a fast tree rebuild.
%\cref{sec:sample-size} shows a sensitivity test on the threshold.
%This threshold is low in our experiments and thus
%this rebuild process is fast because the search tree is still small.
%Our experiments in \cref{sec:tree-eval} show that this threshold is low and thus
%this rebuild process is fast because the search tree is still small.
% \todo{Maybe use the term index in this section intsead of search tree; feels more DB-y}

%As we discussed in \cref{sec:impl:overview}, we typically invoke the \ope
We typically invoke the \ope
framework's Build Phase only once because switching dictionaries
causes the search tree to rebuild, which is particularly
expensive for large trees.
Our assumption is that the value distribution in a database
column is relatively stable, especially at the substring level.
For example, ``\texttt{@gmail.com}'' is likely a common pattern for emails.
%(unless google changes its name to ``The Hulk'').
Because \ope exploits common patterns at relatively fine granularity,
its dictionary remains effective in compressing keys over time.
\rev{We evaluated \ope under a dramatic key distribution change
  in~\cref{sec:updates}
  and observed a compression rate decreases as expected, with simpler
  schemes such as \schemeSingleChar less affected.
  We note that even if a dramatic change in the key distribution happens,
  \ope is not required to rebuild immediately because it still
  guarantees query correctness.
  The system can schedule the reconstruction during maintenance
  to recover the compression rate.}
%\rev{We evaluate \ope under a dramatic key distribution change
%  in~\cref{sec:updates}.}

%We applied \ope to four in-memory search trees used in today's DBMSs:
%\rev{We applied \ope to five in-memory search trees used in today's DBMSs:}
We applied \ope to five in-memory search trees:
\begin{itemize}[leftmargin=*]
\item[$\bullet$] \textbf{\surf}:
  The Succinct Range Filter~\cite{zhang2018}
  is a trie-based data structure that performs
  approximate membership tests for ranges.
  \surf uses succinct data structures
  to achieve an extremely small memory footprint.

\item[$\bullet$] \textbf{\art}:
  The Adaptive Radix Tree~\cite{leis2013, leis2016}
  is the default index structure for HyPer~\cite{kemper2011}.
  \art adaptively selects variable-sized node layouts
  based on fanouts to save space and to improve cache performance.

\item[$\bullet$] \textbf{\hot}:
  The Height Optimized Trie~\cite{binna2018} is a fast and memory-efficient index structure.
  \hot guarantees high node fanouts by combining nodes across trie levels.

\item[$\bullet$] \textbf{\bplustree}:
  We use the cache-optimized
  \tlx~\cite{www-tlx-btree} (formerly known as STX).
  \tlx stores variable-length strings outside the node
  using reference pointers.
  The default node size is 256 bytes, making a fanout of 16
  (8-byte key pointer and 8-byte value pointer per slot).

\item[$\bullet$] \rev{\textbf{\pbtree}}:
  \rev{A \pbtree~\cite{bayer1977} optimizes a plain \bplustree by applying
    prefix and suffix truncation to the nodes~\cite{graefe2011}.
    A \bplustree node with prefix truncation stores the common prefix of
    its keys only once. During a leaf node split, suffix truncation allows
    the parent node to choose the shortest string qualified as a separator
    key. We implemented both techniques on a
    state-of-the-art \bplustree~\cite{leis2019,www-olc-btree}
    other than \tlx for better experimental robustness.}
\end{itemize}

\ope provides the most benefit to search trees that store the full keys.
Many tree indexes for in-memory DBMSs, such as \art and \hot, 
only store partial keys to help the DBMS find the record IDs.
They then verify the results against the full keys after fetching
the in-memory records.
%because the step is as cheap as accessing index nodes.
To understand \ope's interaction with these different search trees,
we arrange them in \cref{fig:tree-spectrum} according to how large
a part of the keys they store.
The \bplustree is at one extreme where the data structure stores full keys.
At the other extreme sits the \ttree~\cite{lehman1985}
(or simply a sorted list of record IDs)
where no keys appear in the data structure.
%\surf, \art, and \hot fall in the middle.
\rev{\pbtree, \surf, \art, and \hot fall in the middle.}
%\ope is more effective towards the \bplustree side.
\rev{\ope is more effective towards the \bplustree side, especially
  in terms of compression rate.}
\rev{The query latency improvement is the difference between
  the speedup due to shorter keys and the overhead of key compression.
  For the \bplustree family, shorter keys means larger fanouts and
  faster string comparisons.
  Although tries only store partial keys,}
%Even though tries only store partial keys,
\ope improves their performance by reducing the tree height.
%The rest of this section analyzes
We next analyze
the latency reduction of using \ope on a trie.

Let $l$ denote the average key length
and $\textit{cpr}$ denote the compression rate
(i.e., uncompressed length / compressed length).
The average height of the original \textit{trie} is $h$.
We use $t_{\textit{trie}}$ to denote the time needed to walk
one level (i.e., one character) down the \textit{trie},
and $t_{\textit{encode}}$ to denote the time needed to compress
one character in \ope.

The average point query latency in the original \textit{trie}
is $h \times t_{\textit{trie}}$,
while this latency in the compressed \textit{trie} is
$l \times t_{\textit{encode}} + \frac{h}{\textit{cpr}} \times t_{\textit{trie}}$,
where $l \times t_{\textit{encode}}$ represents the encoding overhead.
Therefore, the percentage of latency reduction is:
\[
\frac{h \times t_{\textit{trie}} - (l \times t_{\textit{encode}} 
  + \frac{h}{\textit{cpr}} \times t_{\textit{trie}})}{h \times t_{\textit{trie}}}
= 1 - \frac{1}{\textit{cpr}} - \frac{l \times t_{\textit{encode}}}{h \times t_{\textit{trie}}}
\]
If the expression $> 0$, we improve performance.
For example, when evaluating \surf on the email workload
in \cref{sec:tree-eval},
$l = 21.2$ bytes, and $h = 18.2$ levels.
The original \surf has an average point query latency of $1.46 \mu s$.
%the average key length $l$ is $21.2$ bytes.
%The original \surf has the average trie height $h = 18.2$,
%and has an average point query latency of $1.46 \mu s$.
$t_{\textit{trie}}$ is, thus, $\frac{1.46 \mu s}{18.2} = 80.2 ns$.
Our evaluation in \cref{sec:micro-eval} shows that \ope's \schemeDoubleChar scheme achieves
$\textit{cpr} = 1.94$ and $t_{\textit{encode}} = 6.9 ns$ per character.
%a compression rate $\textit{cpr} = 1.94$ and an encoding latency
%per character $t_{\textit{encode}} = 6.9 ns$.
Hence, we estimate that by using \schemeDoubleChar on \surf,
we can reduce the point query latency by 
$1 - \frac{1}{1.94} - \frac{21.2 \times 6.9}{18.2 \times 80.2} = 38\%$.
The real latency reduction is usually higher
($41\%$ in this case as shown in \cref{sec:tree-eval})
because smaller tries also improve %CPU cache locality.
cache performance.

%TODO: Do similar analysis for space and FPR?

\section{\ope Microbenchmarks}
\label{sec:micro-eval}

\begin{figure*}[t]
    \centering
    \begin{adjustbox}{width=1.0\textwidth,totalheight=\textheight,keepaspectratio}

\begin{tabular}{cccc}

  & {\Large \textbf{Email}}
  & {\Large \textbf{Wiki}}
  & {\Large \textbf{URL}} \\

  %% ----------------------------------------------------------------

  {\rotatebox{90}{\hskip 4.8em \Large \textbf{CPR}}}

  & \includegraphics[width=0.33\textwidth]{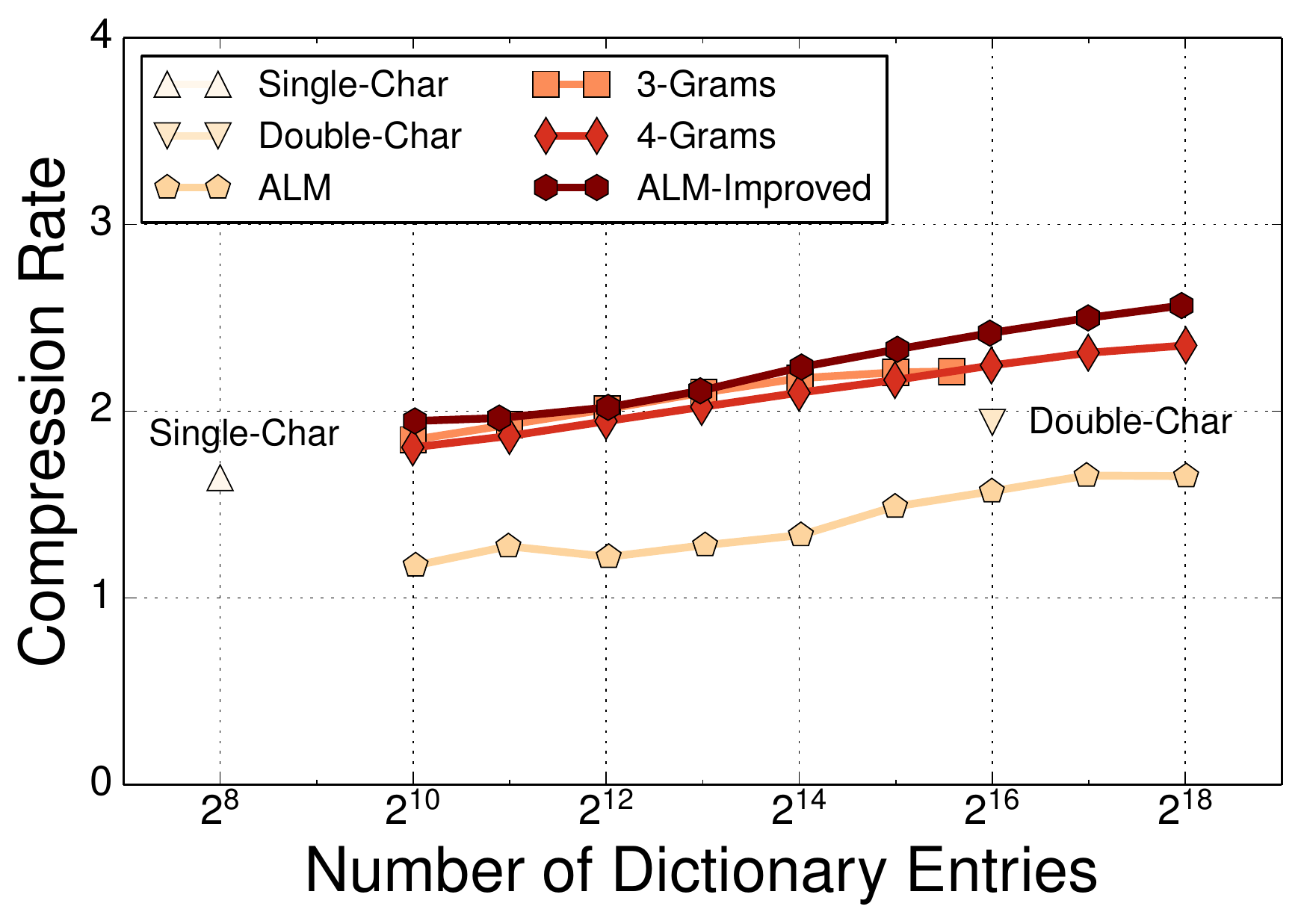}

  & \includegraphics[width=0.33\textwidth]{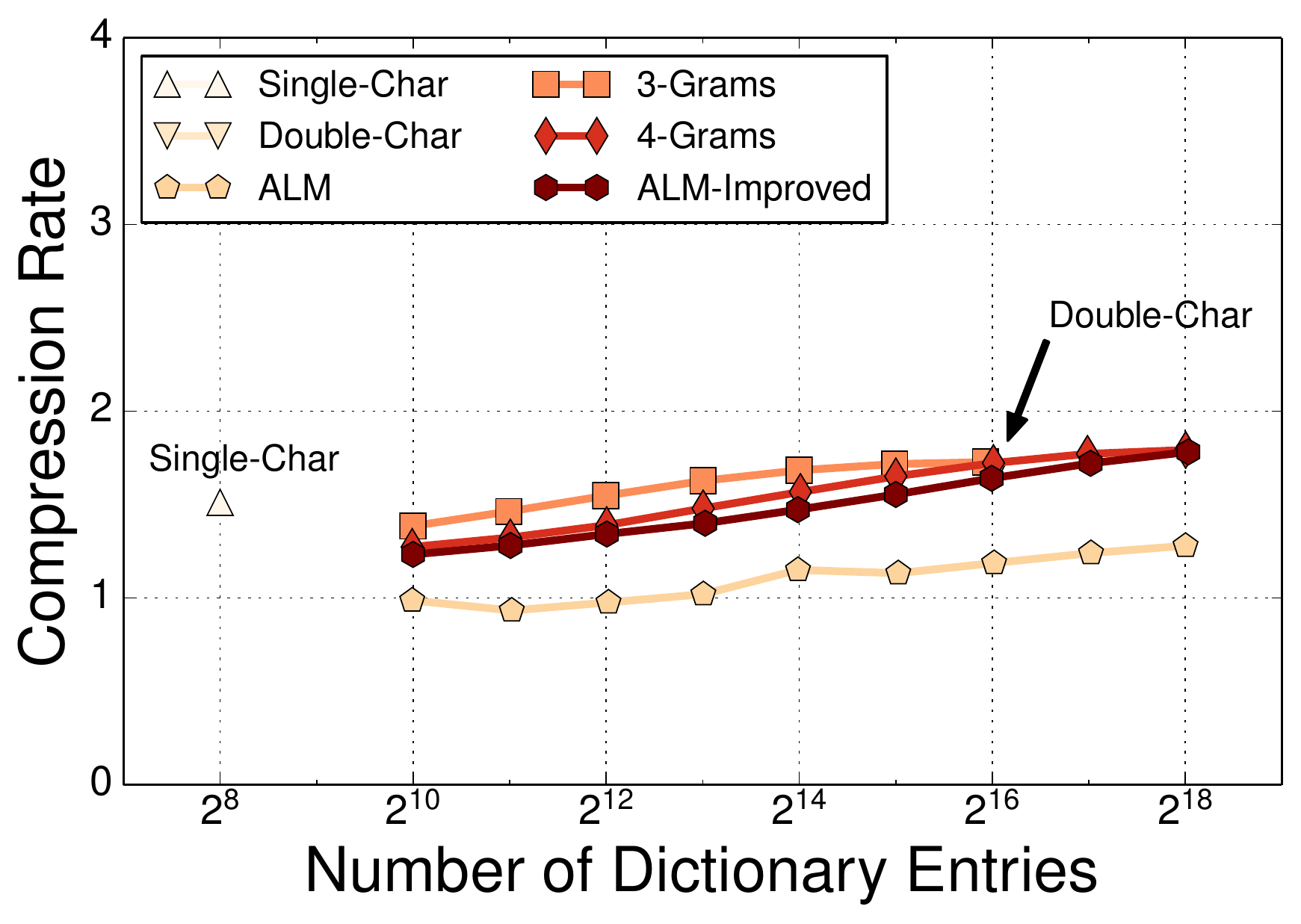}

  & \includegraphics[width=0.33\textwidth]{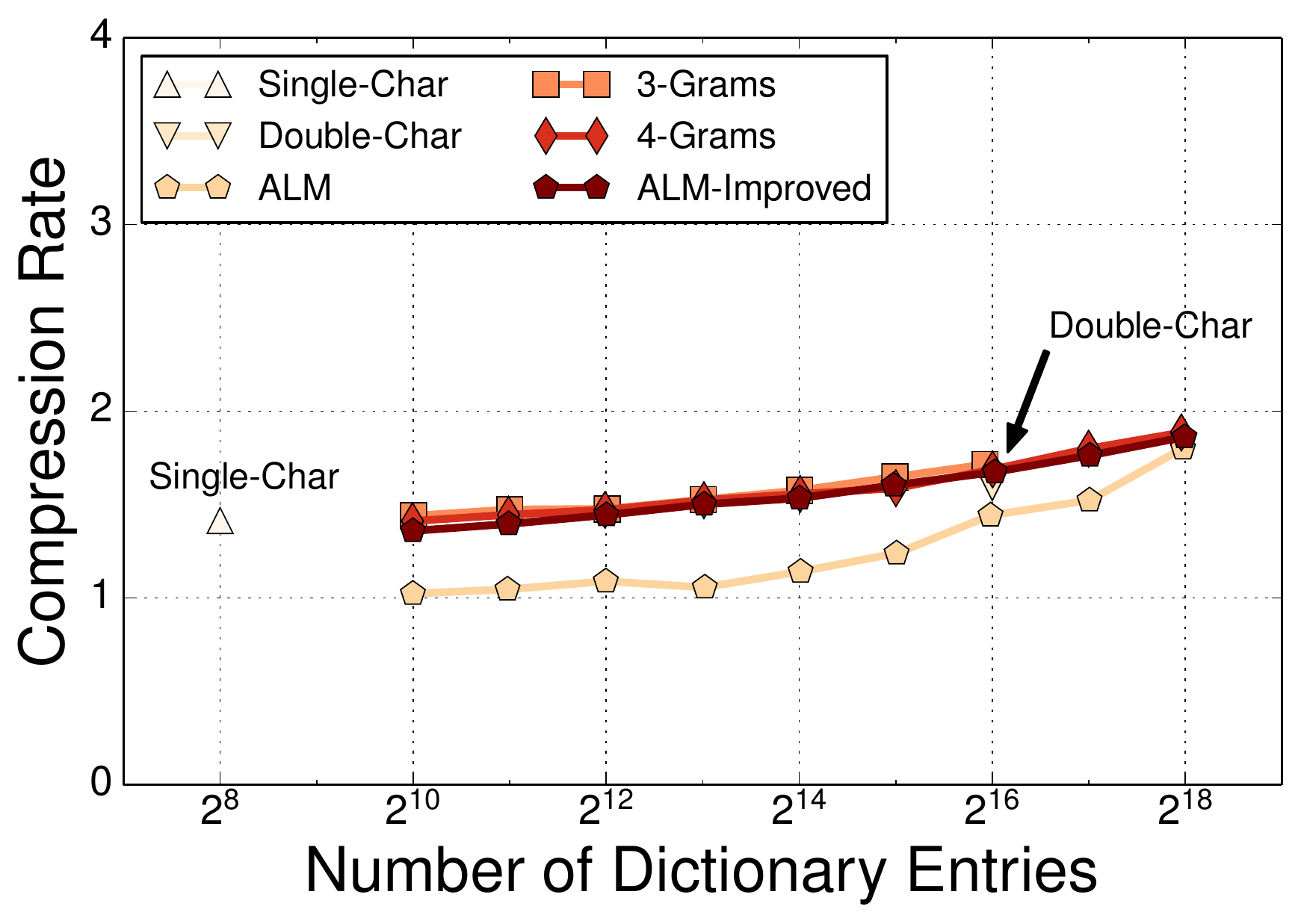} \\

  %% ----------------------------------------------------------------

  {\rotatebox{90}{\hskip 3.6em \Large \textbf{Latency}}}

  & \includegraphics[width=0.33\textwidth]{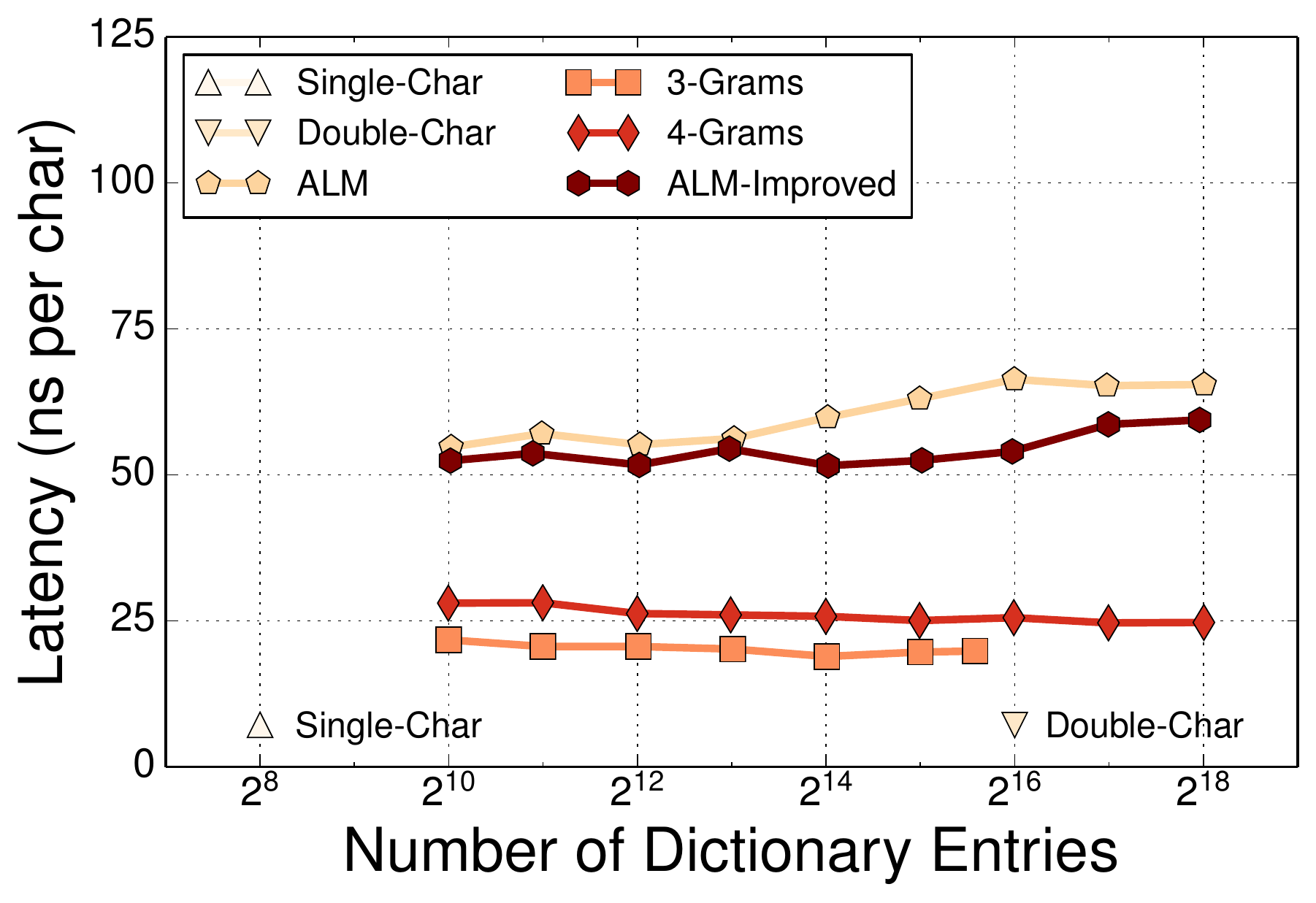}

  & \includegraphics[width=0.33\textwidth]{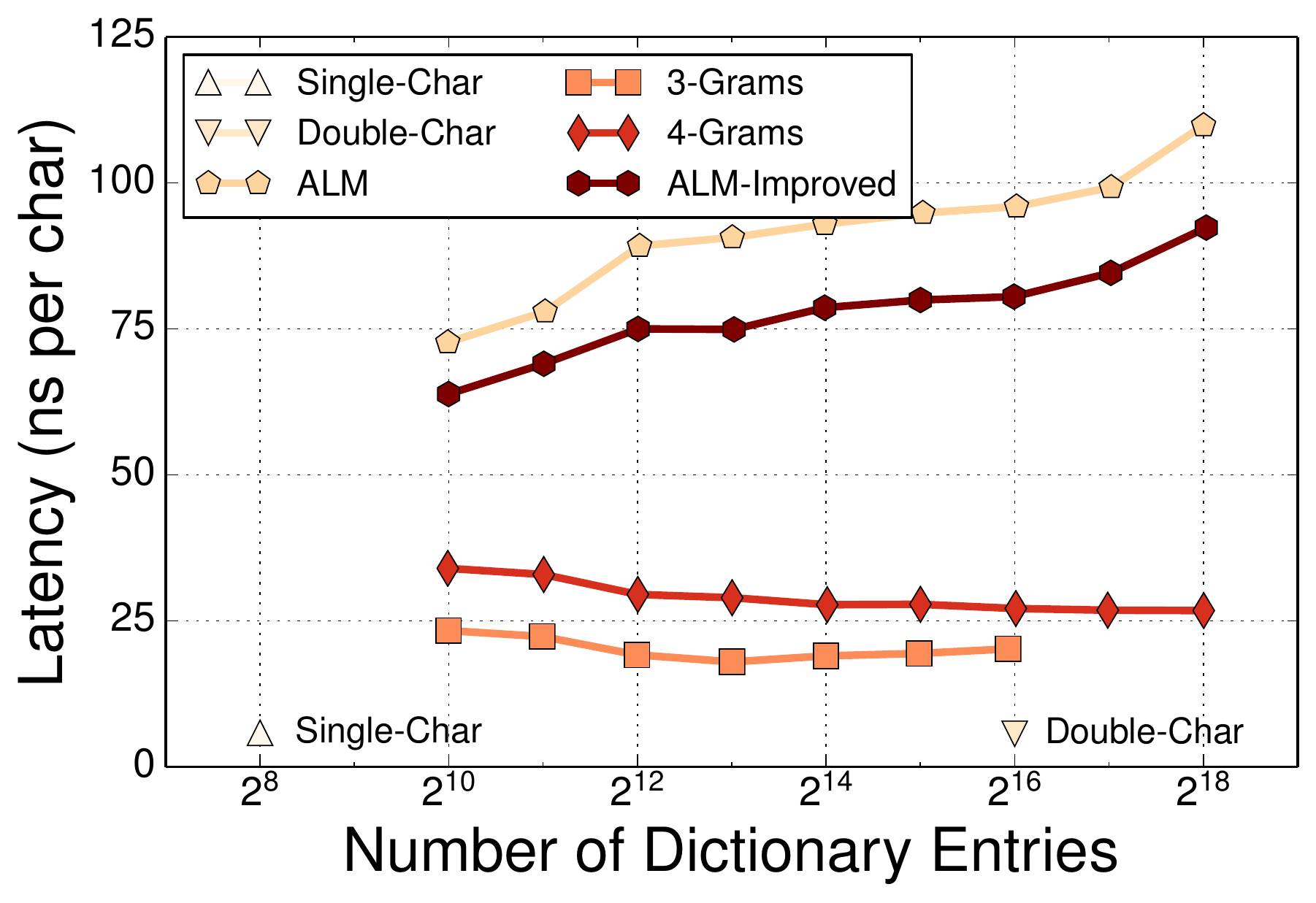}

  & \includegraphics[width=0.33\textwidth]{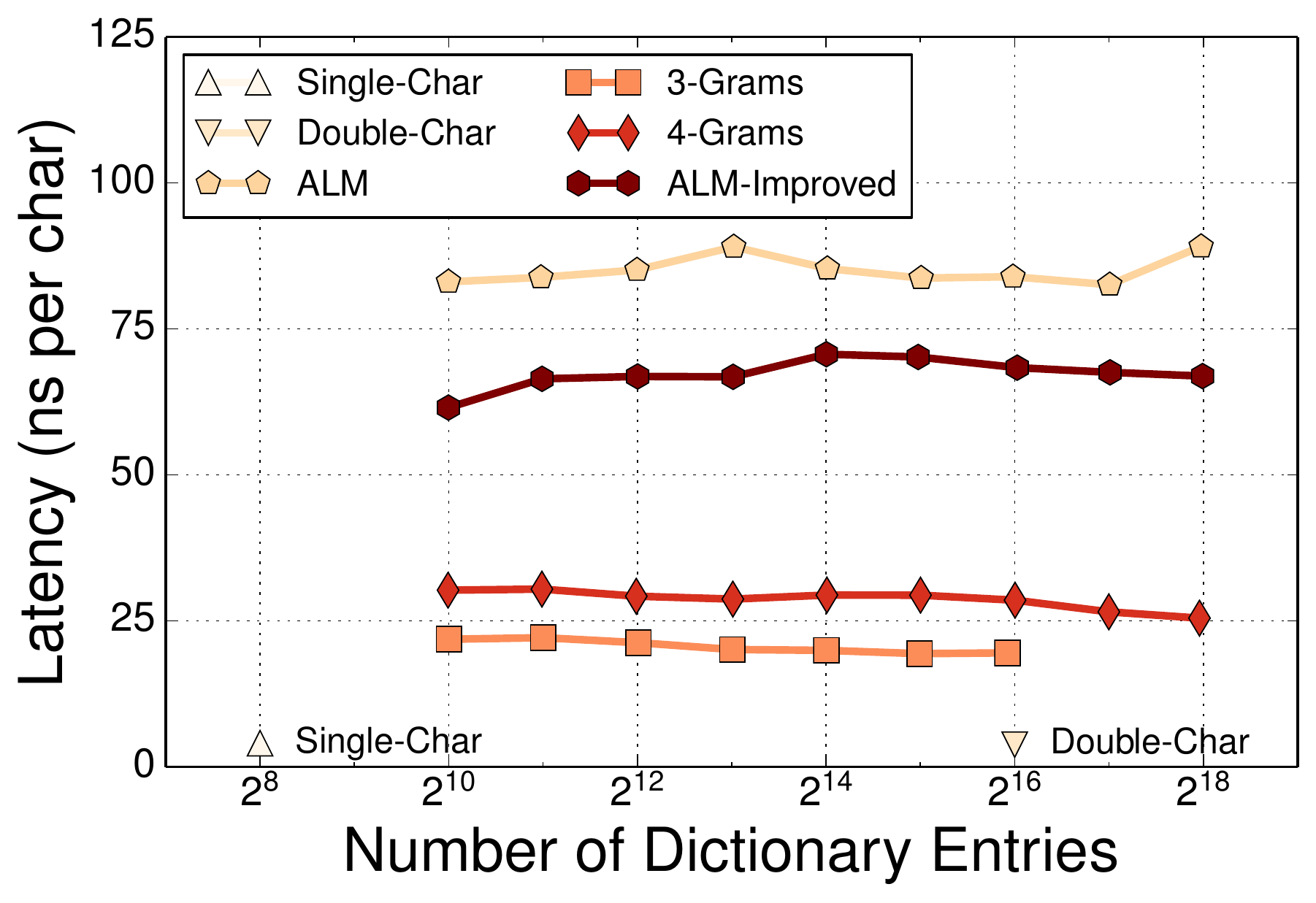} \\

  {\rotatebox{90}{\hskip 3.8em \Large \textbf{Memory}}}

  & \includegraphics[width=0.33\textwidth]{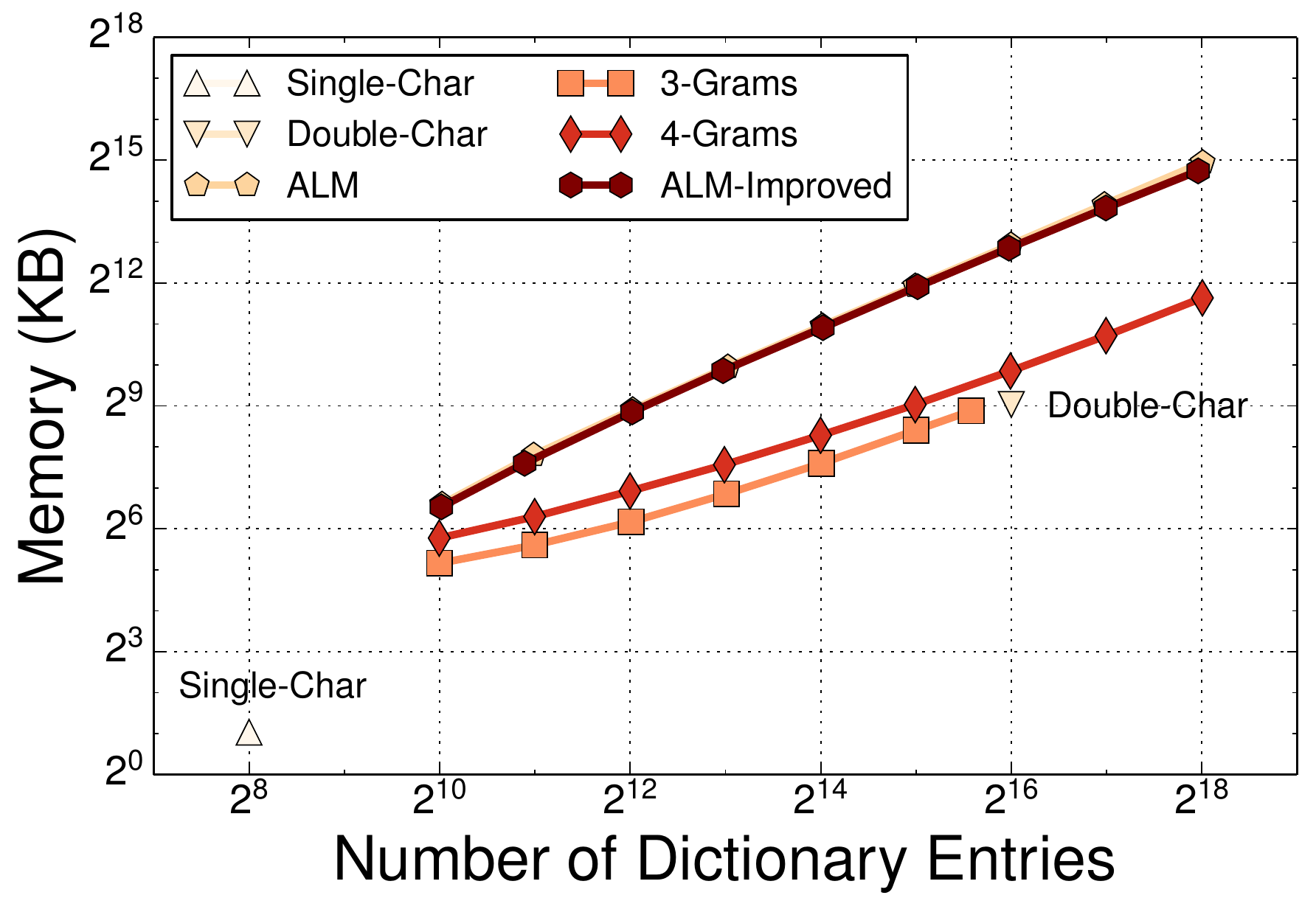}

  & \includegraphics[width=0.33\textwidth]{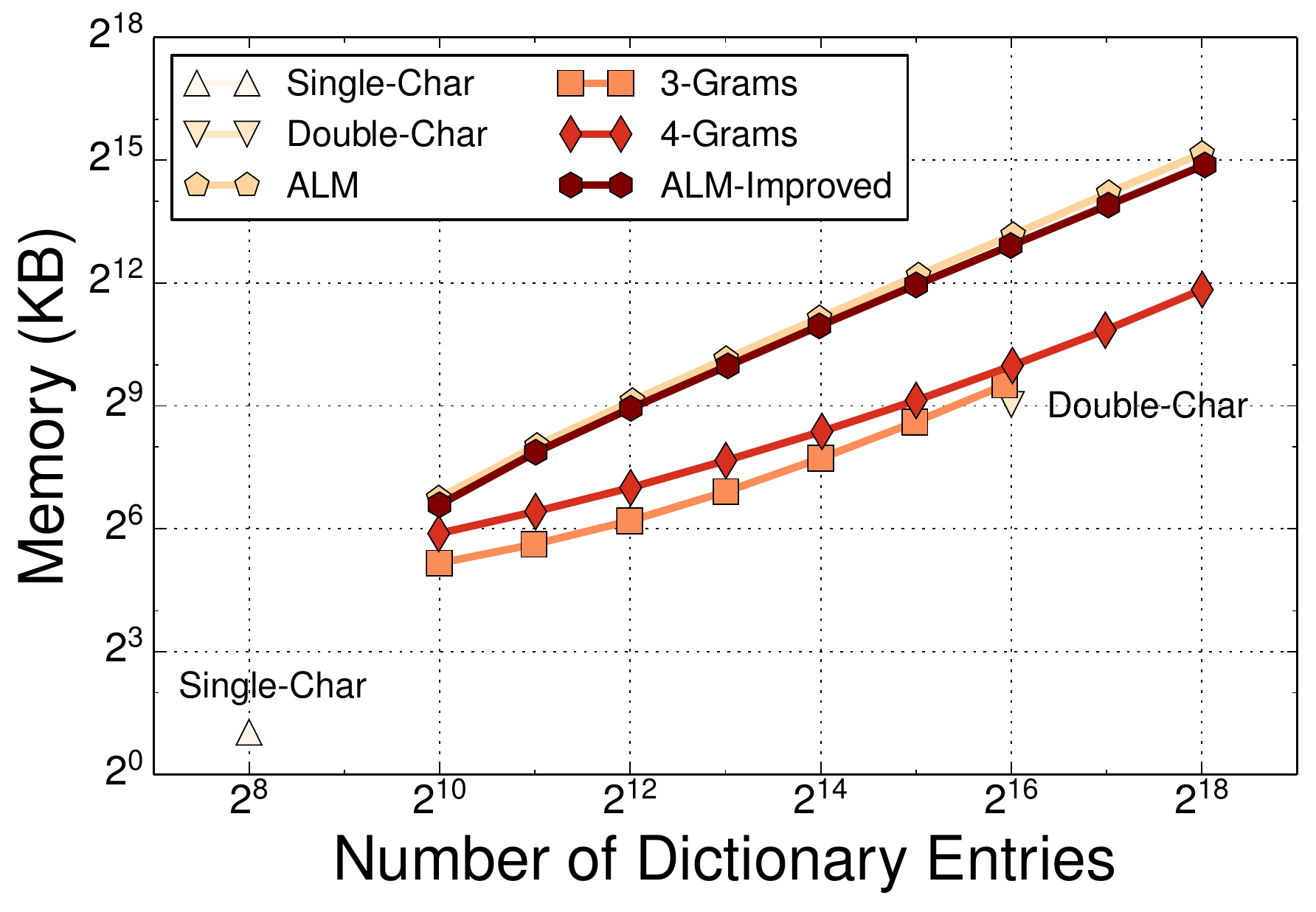}

  & \includegraphics[width=0.33\textwidth]{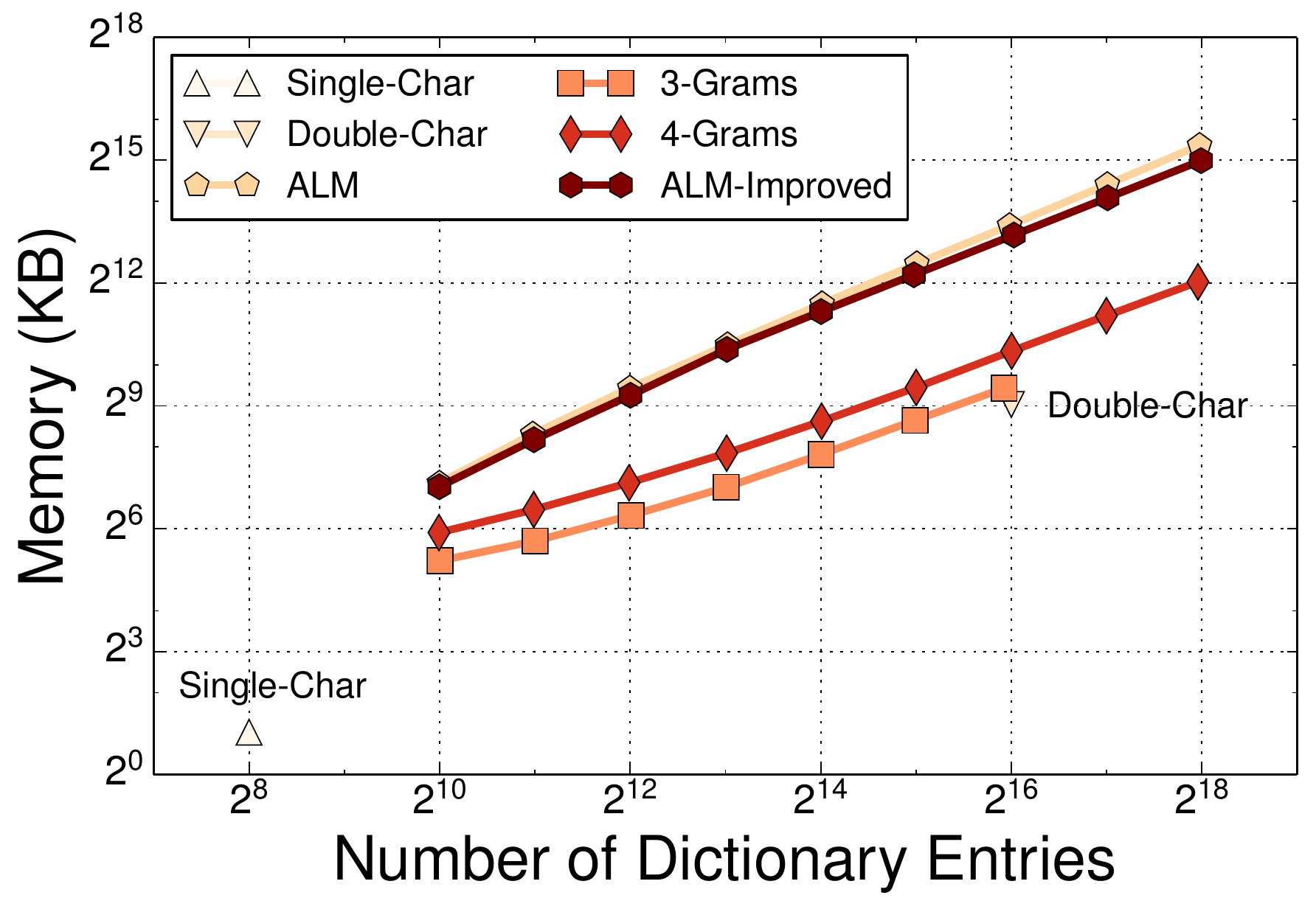} \\

\end{tabular}

\end{adjustbox}

    \caption{
        \textbf{Compression Microbenchmarks} --
        Measurements of \ope's six schemes on the different datasets.
    }
    \label{fig:cpr-lat-mem}
\end{figure*}

%% \begin{figure}[t]
%%     \centering
%%     \includegraphics[width=0.95\columnwidth]{figures/microbench/batch/batch_lat.pdf}
%%     \caption{
%%         \textbf{Batch Encoding} --
%%         Encoding latency measured under varying batch sizes on a pre-sorted 1\% sample of email 
%%         keys. The dictionary size is $2^{16}$ (64K) for \schemeThreeGrams and \schemeFourGrams. 
%%   }
%%   \label{fig:batch}
%% \end{figure}

We evaluate \ope in the next two sections. We first analyze the trade-offs between
compression rate and compression overhead
of different schemes in \ope.
These microbenchmarks help explain the end-to-end
measurements on \ope-integrated search trees
in \cref{sec:tree-eval}.

We use the following datasets for all our experiments:
\begin{itemize}[leftmargin=*]
    \item[$\bullet$] \textbf{Email:}
    25 million email addresses (host reversed -- e.g., ``\texttt{com.gmail@foo}'')
    with an average length of 22 bytes.

    \item[$\bullet$] \textbf{Wiki:}
    14 million article titles from the English version of Wikipedia
    with an average length of 21 bytes~\cite{www-wiki}.
    
    \item[$\bullet$] \textbf{URL:}
    25 million URLs from a 2007 web crawl with an average length of 104 
    bytes~\cite{www-url}.
\end{itemize}

We run our experiments using a machine equipped with
two Intel\textsuperscript{\textregistered}
Xeon\textsuperscript{\textregistered} E5-2630v4 CPUs (2.20GHz,
32~KB L1, 256~KB L2, 25.6~MB L3) and 8$\times$16~GB DDR4 RAM.

In each experiment, we randomly shuffle the target dataset before each trial.
We then select $1\%$ of the entries from the shuffled dataset
as the sampled keys for \ope.
\rev{Our sensitivity test in \cref{sec:sample-size} shows that 1\% is
  large enough for all schemes to reach their maximum compression rates.}
%\rev{We perform a sensitivity test on the sample size in~\cref{sec:sample-size}.}
We repeat each trial three times and report the average result.

%% ---------------------------------------------------------------
%% Performance \& Efficacy
%% ---------------------------------------------------------------
\subsection{Performance \& Efficacy}
\label{sec:micro-eval:perf}
We first evaluate the runtime performance and compression efficacy of 
\ope's six built-in schemes listed in \cref{tbl:modules}. 
\ope compresses the keys one-at-a-time with a single thread.
%\ope performs the compression with a single thread that processes each entry one-at-a-time.
We vary the number of dictionary entries in each trial and measure three facets per scheme:
(1) the  compression rate,
(2) the average encoding latency per character, and
(3) the size of the dictionary.
We compute the compression rate as the uncompressed dataset size 
divided by the compressed dataset size.
We obtain the average encoding latency per character by
dividing the execution time by the total number of bytes
in the uncompressed dataset.

\cref{fig:cpr-lat-mem} shows the experimental results.
%for the \ope's compression schemes.
We vary the number of dictionary entries on the x-axis (log scaled).
The \schemeSingleChar and \schemeDoubleChar schemes
have fixed dictionary sizes of $2^{8}$ and $2^{16}$, respectively.
The \schemeThreeGrams dictionary cannot grow to $2^{18}$
because there are not enough unique three-character patterns
in the sampled keys.
\\ \vspace{-0.1in}

%% -----------------------
%% Compression Rate
%% -----------------------
\noindent \textbf{Compression Rate:}
The first row of \cref{fig:cpr-lat-mem} shows that the \encodingVIVC schemes (\schemeThreeGrams, 
\schemeFourGrams, \schemeALMImproved)
have better compression rates than the others.
This is because \encodingVIVC schemes exploit the source strings' higher-order entropies
to optimize both interval division and code assignment at the same time.
In particular, \schemeALMImproved compresses the keys more than the original \schemeALM
because it uses a better pattern extraction algorithm, and it uses the Hu-Tucker codes
%These Hu-Tucker codes improve compression in \schemeALMImproved because there is still
that leverage the
remaining skew in the dictionary entries' access probabilities. \schemeALM tries to 
equalize these weighted probabilities but our improved version has better efficacy.
\rev{We also note that a larger dictionary produces a better compression rate for the 
variable-length interval schemes.}
\\ \vspace{-0.1in}

%% -----------------------
%% Encoding Latency
%% -----------------------
\noindent \textbf{Encoding Latency:}
The latency results in the second row of \cref{fig:cpr-lat-mem} demonstrate that the simpler schemes 
have lower encoding latency. This is expected because the latency depends largely on 
the dictionary data structures.
\schemeSingleChar and \schemeDoubleChar are the fastest because they use
array dictionaries that are small enough to fit in the CPU's L2 cache.
Our specialized bitmap-tries used by \schemeThreeGrams and \schemeFourGrams
are faster than the general \art-based dictionaries used by
\schemeALM and \schemeALMImproved because
(1) the bitmap speeds up in-node label search;
and (2) the succinct design (without pointers) improves cache performance.

\rev{The figures also show that latency is stable (and even decrease slightly)
in all workloads for \schemeThreeGrams and \schemeFourGrams as
their dictionary sizes increase.}
This is interesting because the cost of performing a lookup in the dictionary increases as 
the dictionary grows in size. The larger dictionaries, however, achieve higher
compression rates such that it reduces lookups:
larger dictionaries have shorter intervals on
the string axis, and shorter intervals usually have longer common prefixes
(i.e., dictionary symbols).
Thus, \ope consumes more bytes from the source string
at each lookup with larger dictionaries,
counteracting the higher per-lookup cost.
\\ \vspace{-0.1in}

%% -----------------------
%% Dictionary Memory
%% -----------------------
\noindent \textbf{Dictionary Memory:}
% (note that the y-axis is also log scaled).
The third row of \cref{fig:cpr-lat-mem} shows that the dictionary sizes for the 
variable-length schemes grow linearly as the number of dictionary entries increases.
Even so, for most dictionaries, the total tree plus dictionary size is still much smaller
than the size of the corresponding uncompressed search tree.
These measurements also show that our bitmap-tries 
for \schemeThreeGrams and \schemeFourGrams are up to %\todo{$\sim$700\%} 
an order of magnitude smaller than the
\art-based dictionaries for all the datasets. The 3-Grams bitmap-trie is only $1.4 \times$
larger than \schemeDoubleChar's fixed-length array of the same size.
\\ \vspace{-0.1in}

%% -----------------------
%% Discussion
%% -----------------------
\noindent \textbf{Discussion:}
Schemes that compress more are slower,
except that the original \schemeALM is strictly worse than the other schemes
in both dimensions.
The latency gaps between schemes are generally
larger than the compression rate gaps.
We evaluate this trade-off in \cref{sec:tree-eval}
by applying the \ope schemes to in-memory search trees.
\\ \vspace{-0.1in}

\begin{figure}[t]
    \centering
    \includegraphics[width=0.795\columnwidth]
        {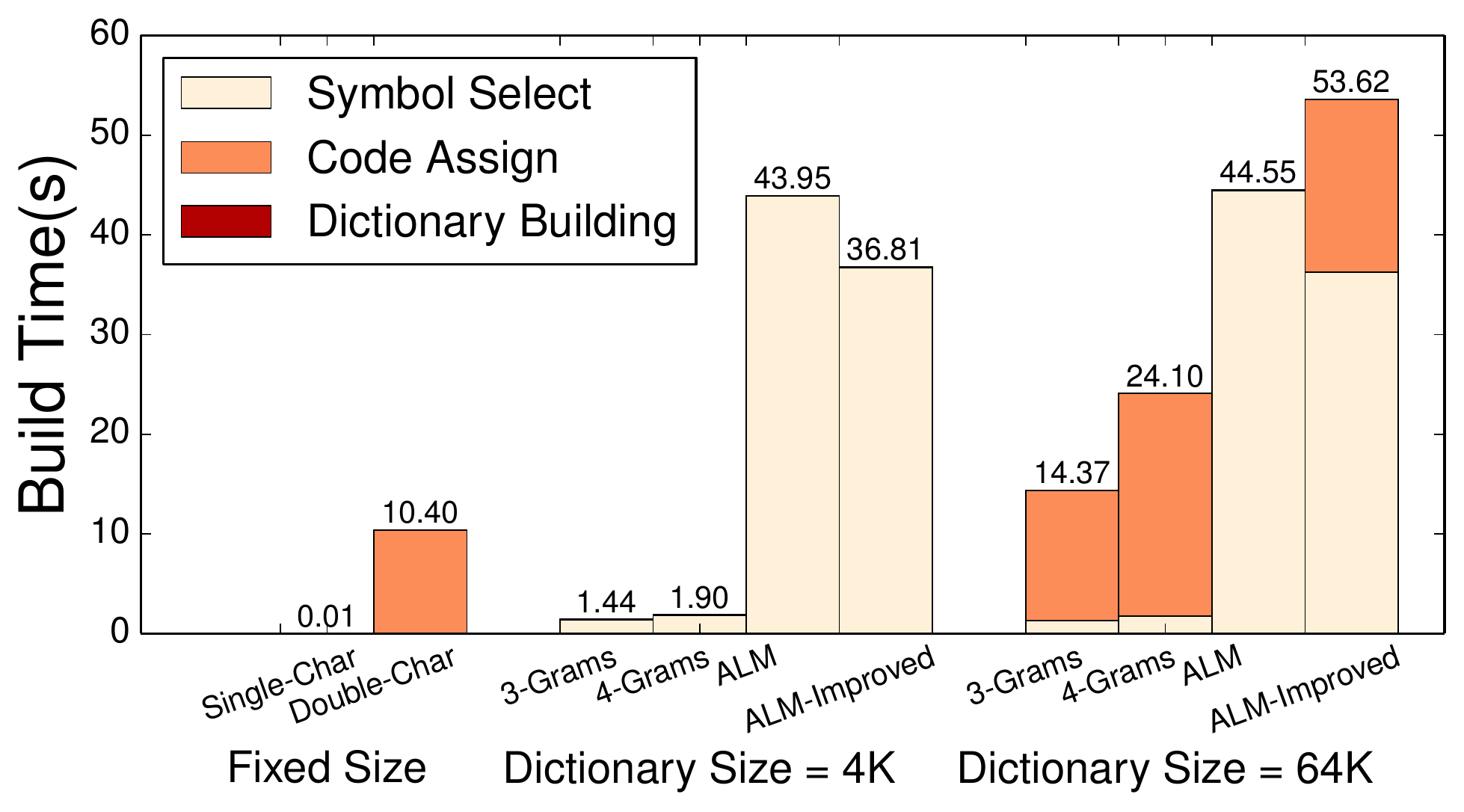}
    \caption{
        \textbf{Dictionary Build Time} --
        A breakdown of the time it takes for \ope to build dictionaries on a 1\% sample of email 
        keys.
    }
    \label{fig:build-time}
\end{figure}

%% ---------------------------------------------------------------
%% Dictionary Build Time
%% ---------------------------------------------------------------
\subsection{Dictionary Build Time}
\label{sec:micro-eval:build}
We next measure how long \ope takes to construct the dictionary using
each of the six compression schemes. We record the time \ope spends in
the modules from \cref{sec:impl:impl} when building a dictionary:
(1) \compSelector, (2) \compAssigner, and (3) \compDictionary.
The last step is the time required to populate the dictionary from the key samples.
We present only the Email dataset for this experiment; the results for the other datasets produce 
similar results and thus we omit them. For the variable-length-interval schemes, we perform the 
experiments using two dictionary sizes ($2^{12}$, $2^{16}$).

\cref{fig:build-time} shows the time breakdown %of the three steps
of building the dictionary in each scheme.
%The most prominent result is that
First,
the \compSelector dominants the cost for \schemeALM and 
\schemeALMImproved because these schemes collect statistics for
substrings of all lengths, which has a super-linear cost relative to the number of keys.
For the other schemes, the \compSelector's time grows linearly
with the number of keys.
Second, the time used by the \compAssigner rises dramatically
as the dictionary size increases
because the Hu-Tucker algorithm has quadratic time complexity.
Finally, the \compDictionary build time is negligible
compared to the \compSelector and \compAssigner modules.

%% ---------------------------------------------------------------
%% Batch Encoding
%% ---------------------------------------------------------------
%% \subsection{Batch Encoding}
%% \label{sec:micro-eval:batch}
%% We also evaluate the batching optimization described
%% in \cref{sec:impl:impl}.
%% In this experiment, we sort the email dataset and then
%% encode the keys with varying batches sizes (1, 2, 32).
%% %In this experiment, we sort the input keys before encoding them.
%% %\todo{Need more explaination of this experiment's setup.}

%% As shown in \cref{fig:batch},
%% batch encoding significantly improves encoding performance
%% because it encodes the common prefix of a batch only once
%% to avoid redundant work.
%% \schemeALM and \schemeALMImproved schemes do not benefit from batch encoding.
%% Since these schemes have dictionary symbols of arbitrary lengths,
%% we cannot determine a priori a common prefix that is aligned with
%% the dictionary symbols for a batch without encoding them.

\section{Search Tree Evaluation}
\label{sec:tree-eval}

\begin{figure*}[t]
    \centering
    \begin{adjustbox}{width=1.0\textwidth,totalheight=\textheight,keepaspectratio}

\begin{tabular}{cccc}

  \multicolumn{4}{c}
  {\includegraphics[width=\textwidth]{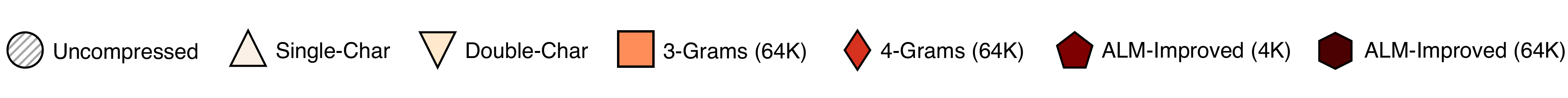}} \\

  & {\Large \textbf{Email}}
  & {\Large \textbf{Wiki}}
  & {\Large \textbf{URL}} \\

  %% ----------------------------------------------------------------

  {\rotatebox{90}{\hskip 1.6em \Large \textbf{Point vs. Memory}}}

  & \includegraphics[width=0.33\textwidth]{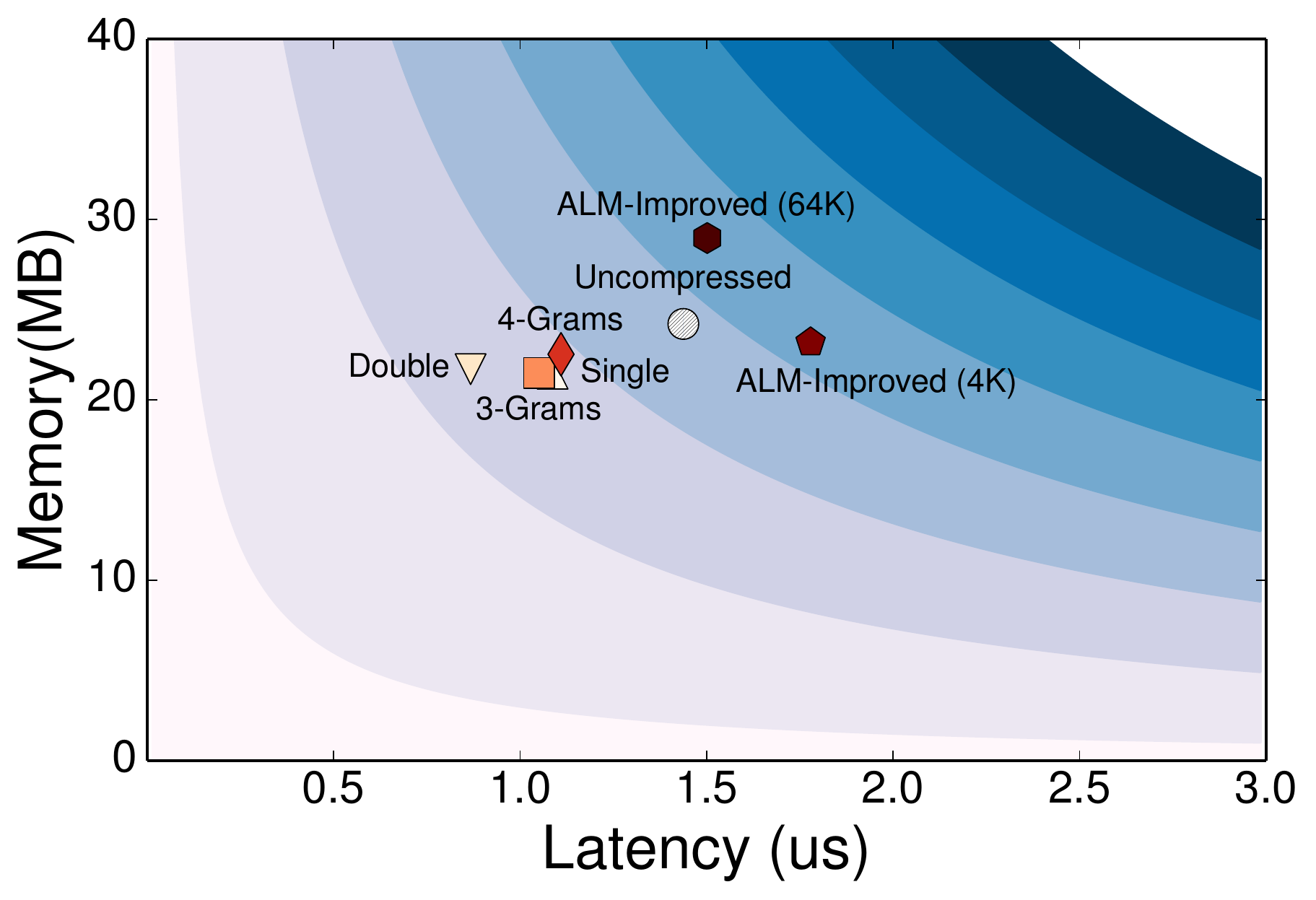}

  & \includegraphics[width=0.33\textwidth]{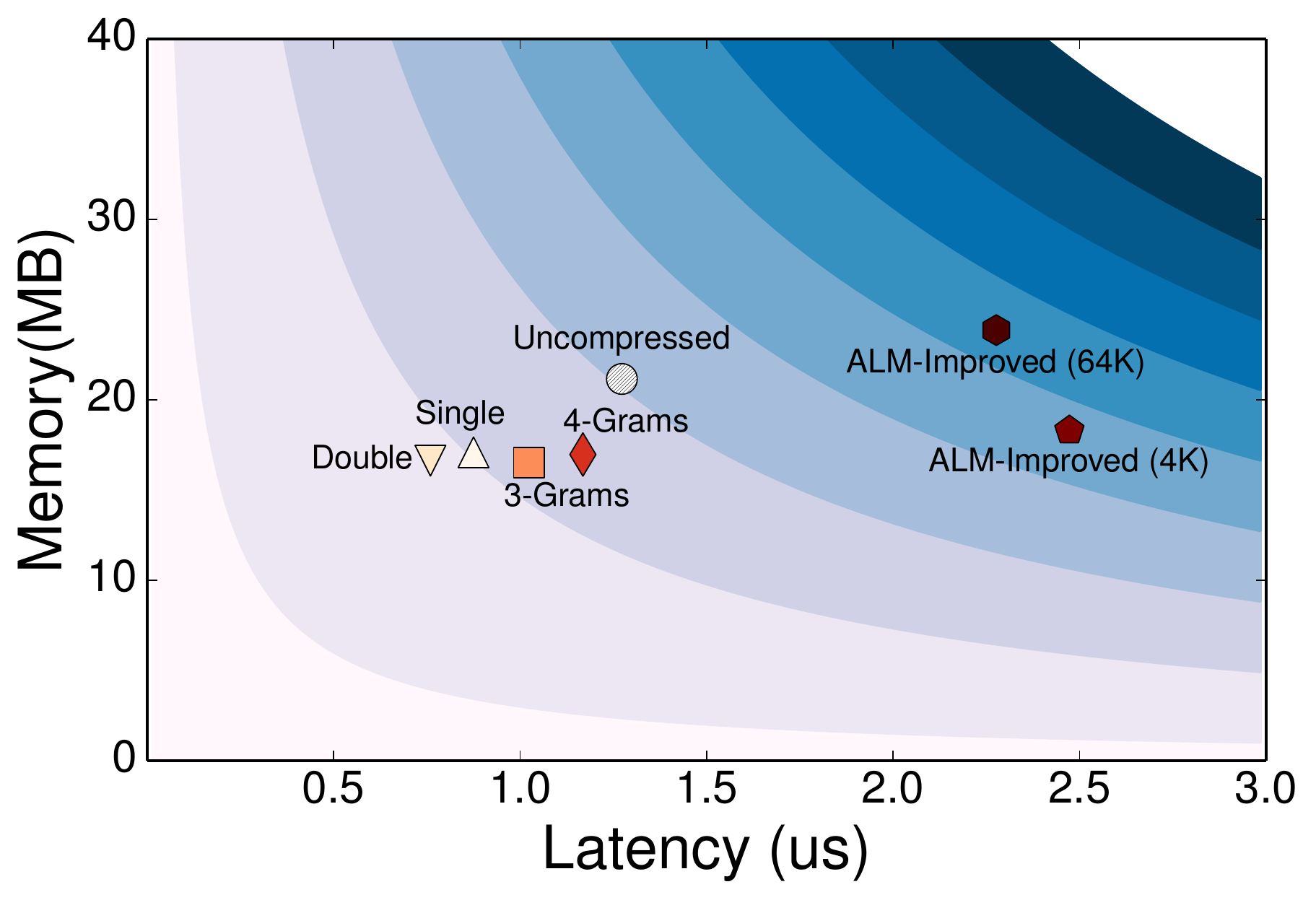}

  & \includegraphics[width=0.33\textwidth]{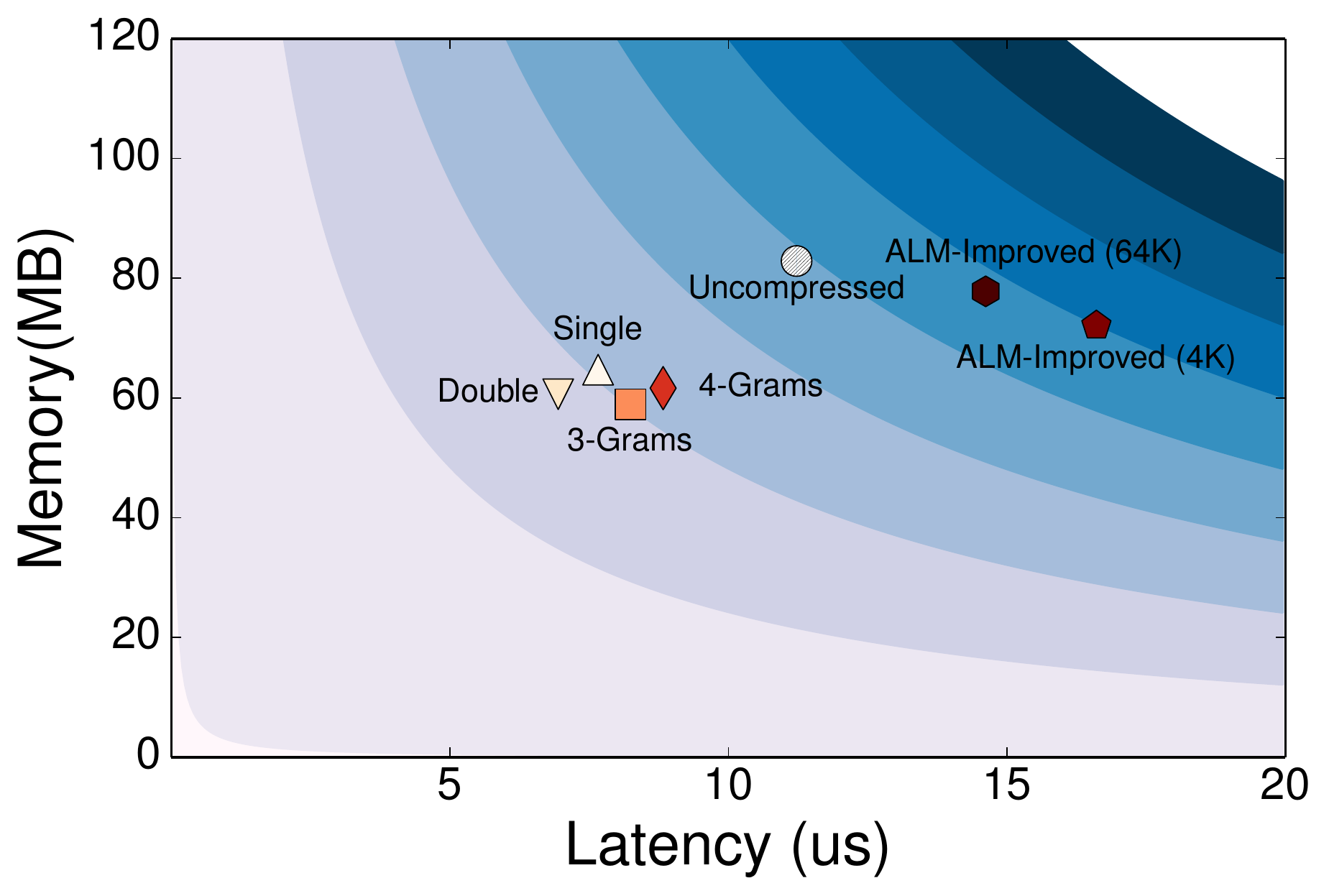} \\

  %& \includegraphics[width=0.33\textwidth]{figures/SuRF_uniform/point/lat_mem_email_surf_point.pdf}

  %& \includegraphics[width=0.33\textwidth]{figures/SuRF_uniform/point/lat_mem_wiki_surf_point.pdf}

  %& \includegraphics[width=0.33\textwidth]{figures/SuRF_uniform/point/lat_mem_url_surf_point.pdf} \\

  %% ----------------------------------------------------------------

  {\rotatebox{90}{\hskip 1.6em \Large \textbf{Range \& Build}}}

  & \includegraphics[width=0.33\textwidth]{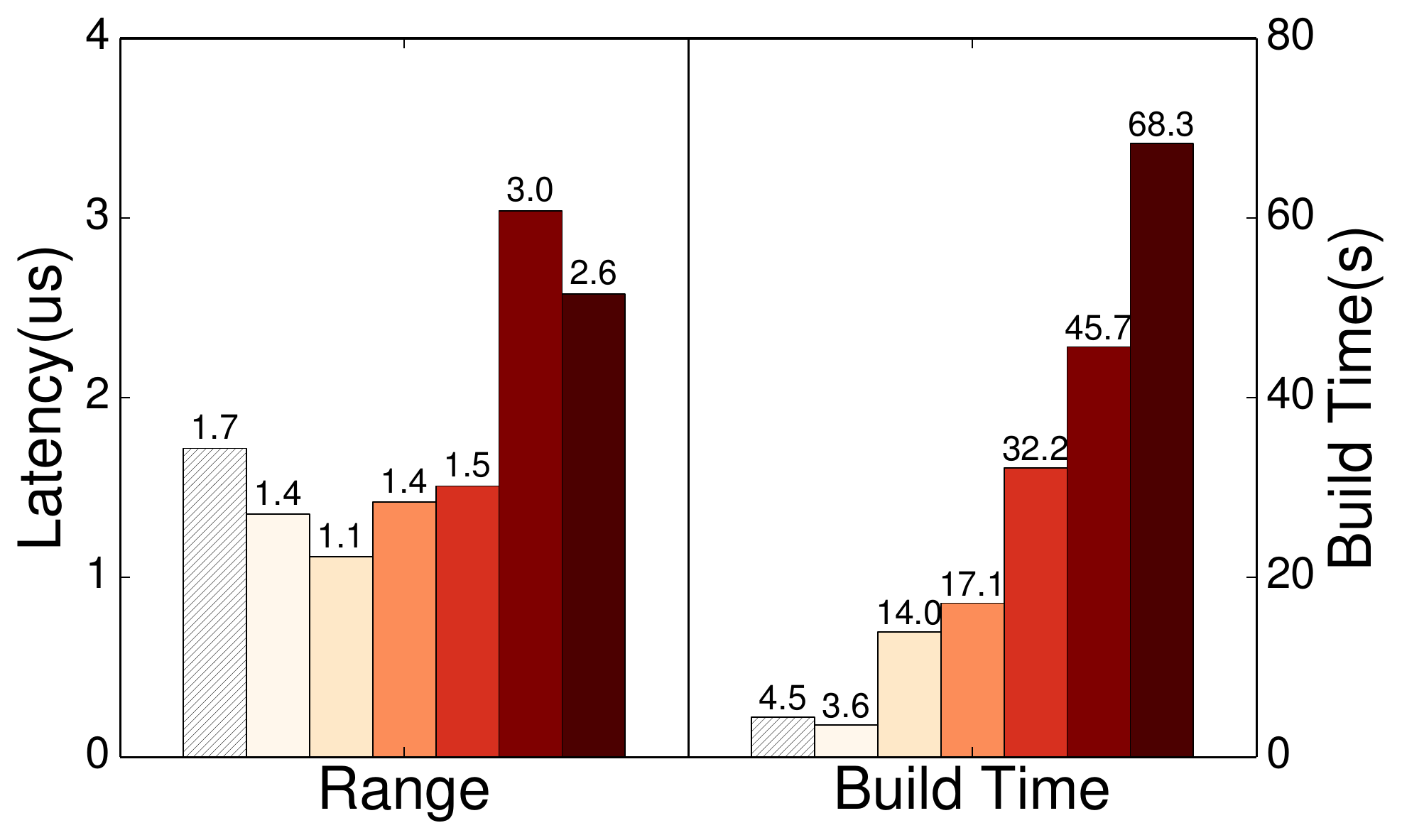}

  & \includegraphics[width=0.33\textwidth]{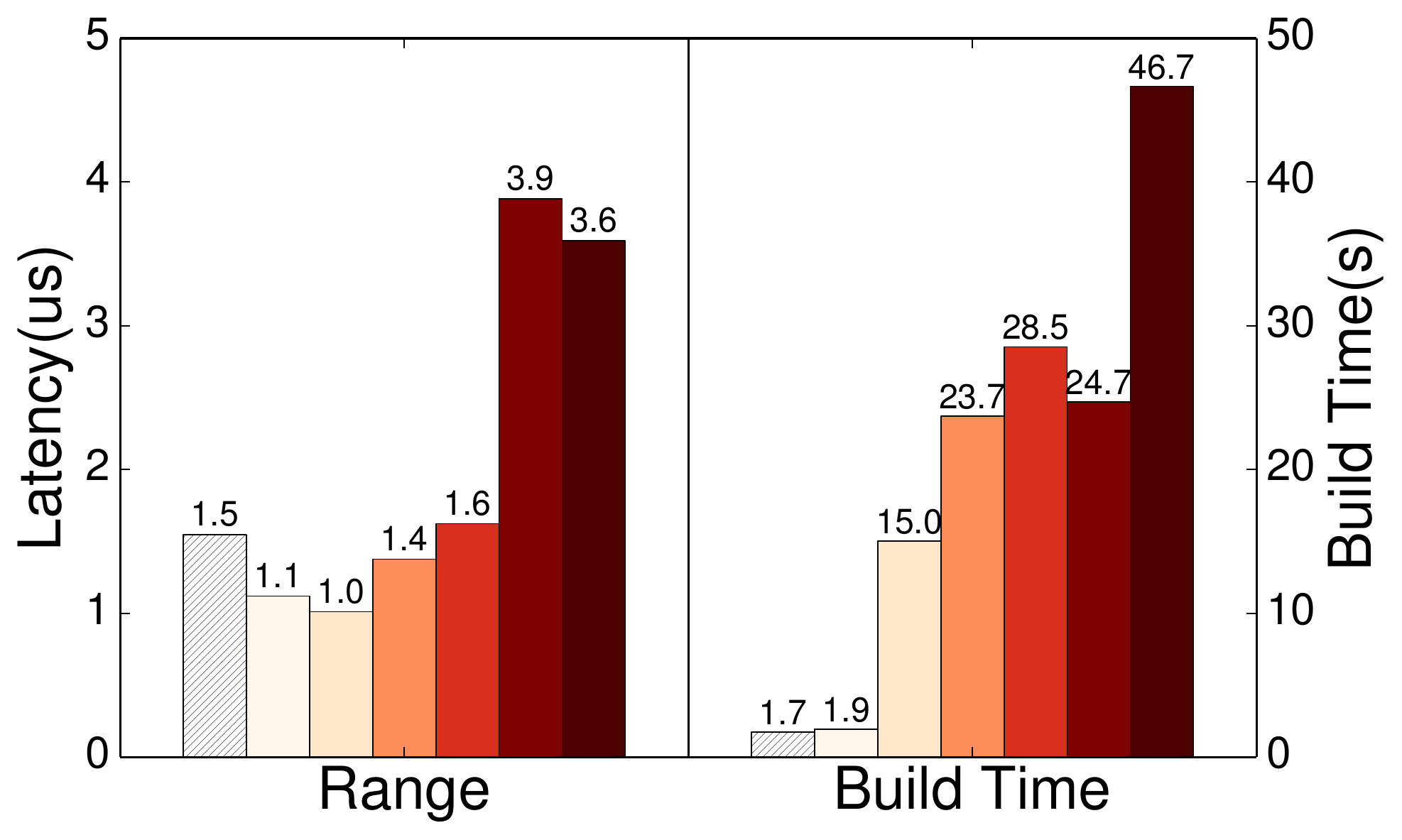}

  & \includegraphics[width=0.33\textwidth]{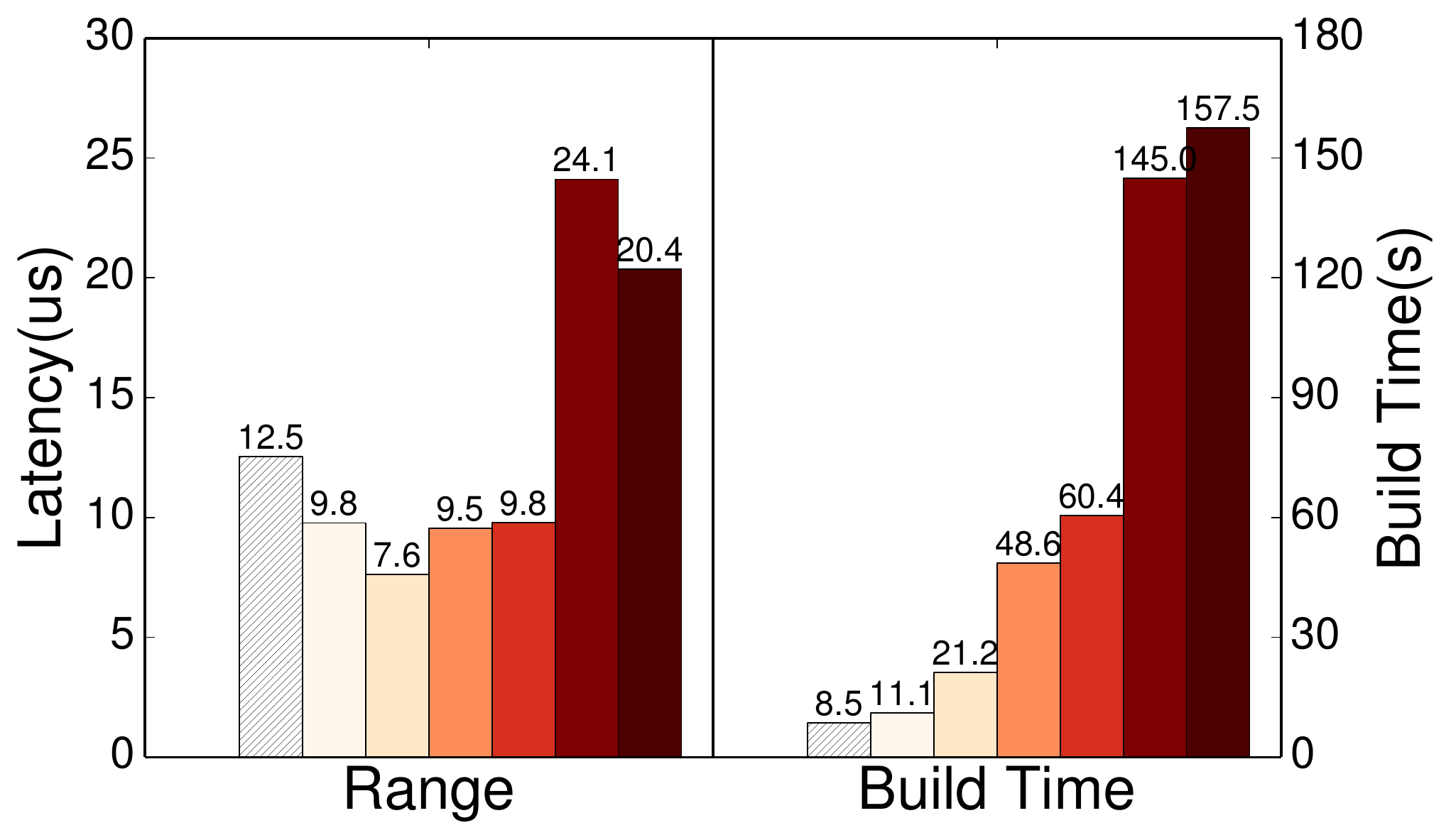} \\

  {\rotatebox{90}{\hskip 1.8em \Large \textbf{Trie Height}}}

  & \includegraphics[width=0.32\textwidth]{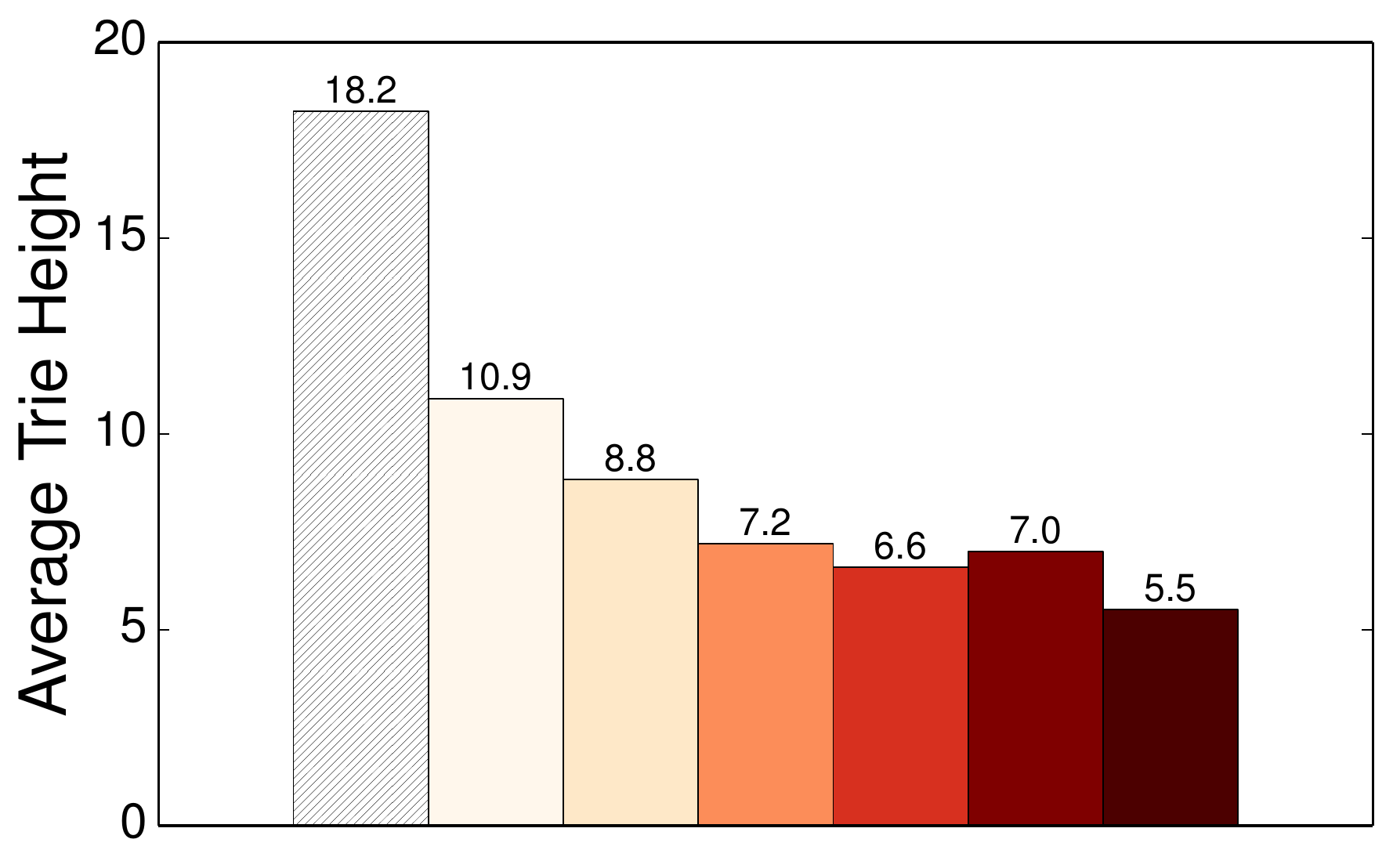}

  & \includegraphics[width=0.32\textwidth]{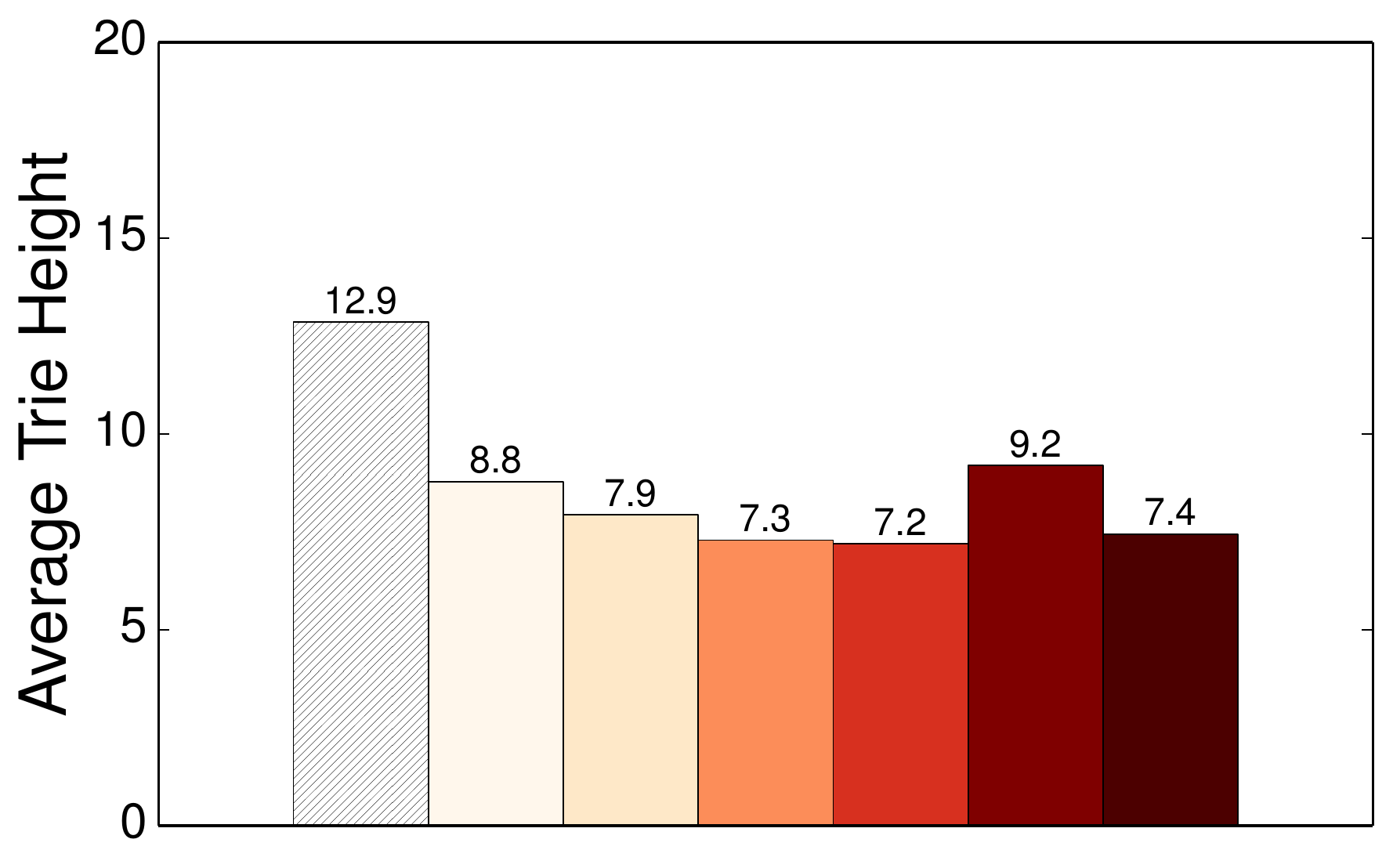}

  & \includegraphics[width=0.32\textwidth]{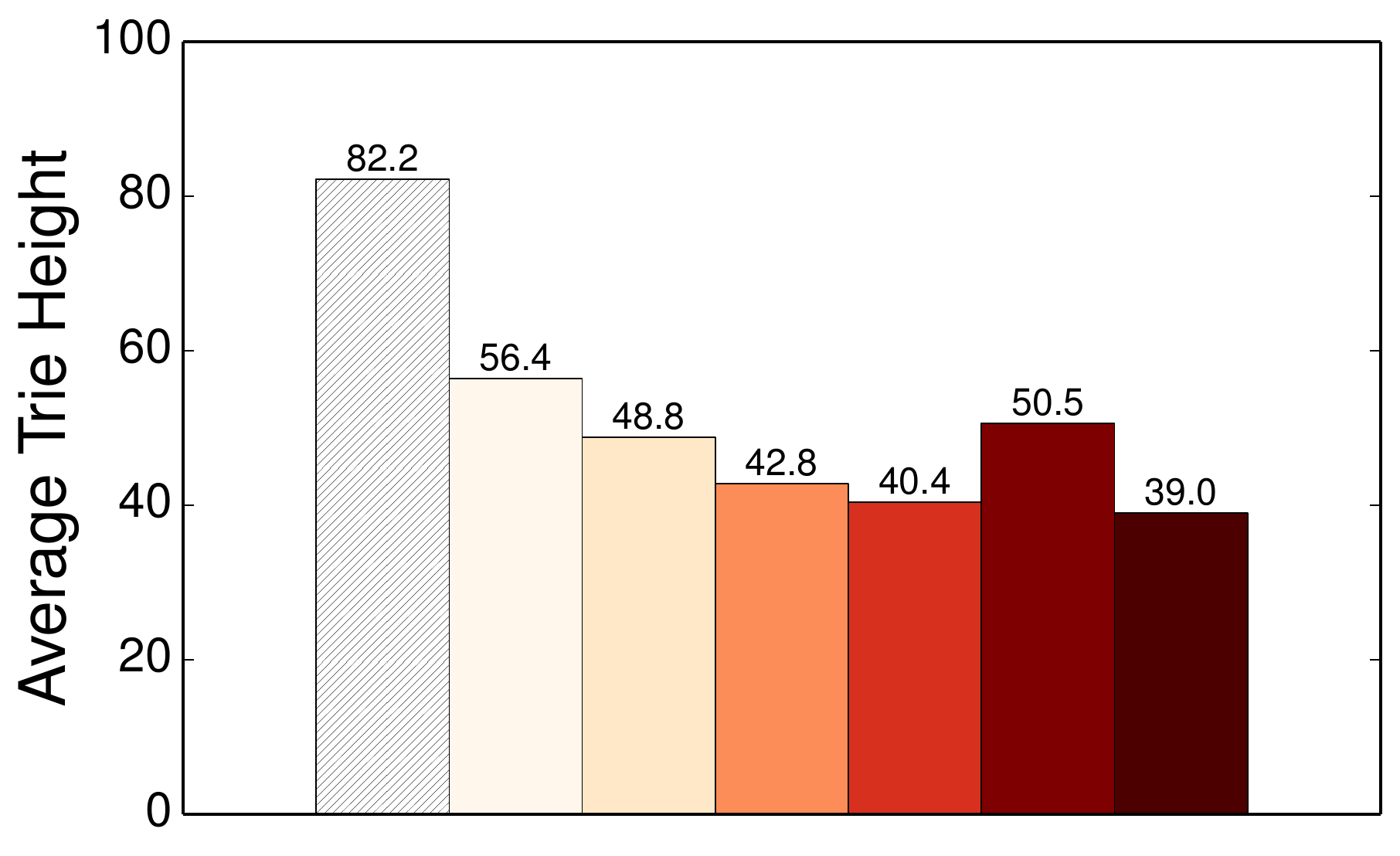} \\

\end{tabular}

\end{adjustbox}

    \caption{
        \textbf{\surf YCSB Evaluation} --
        Runtime measurements for executing YCSB workloads on \ope-optimized \surf with three 
        datasets. The trie height is the average height of each leaf node after loading all keys.
    }
    \label{fig:surf-eval}
\end{figure*}

\begin{figure}[t]
  \centering
  \includegraphics[width=0.8\columnwidth]{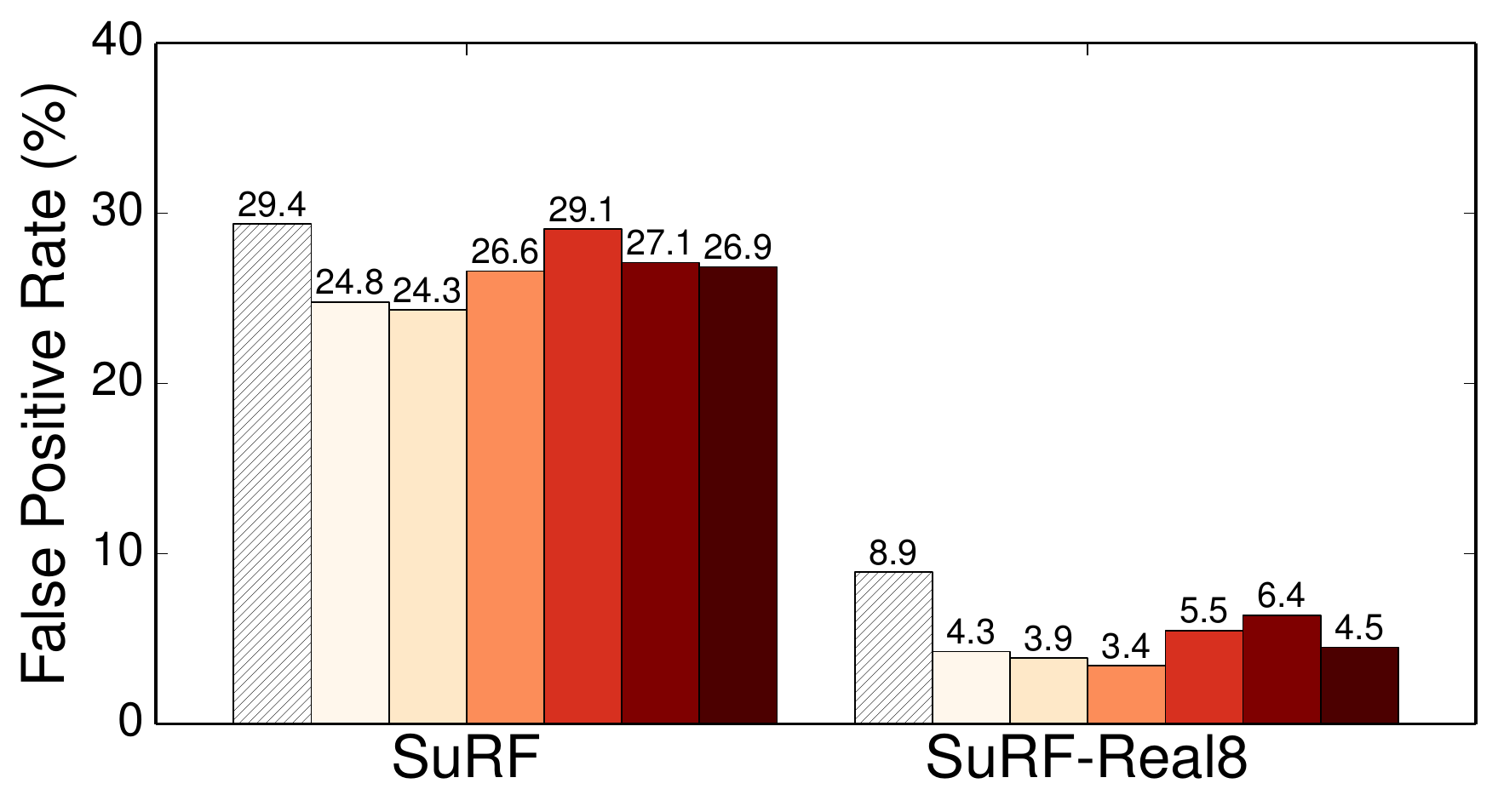}
  \caption{
    \textbf{\surf False Positive Rate} --
    Point queries on email keys.
    \surf-Real8 means it uses 8-bit real suffixes.
  }
  \label{fig:surf-fpr}
\end{figure}

%% \begin{figure*}[t]
%%     \centering
%%     \input{art-eval.tex}
%%     \caption{
%%       \textbf{\art YCSB Evaluation} --
%%       {Integrate HOPE into ART. Measure memory use, point and range query latency, and insert latency on the email,
%%       wiki, and url datasets. Note that the figures have different Y-axis scales.} 
%%     }
%%     \label{fig:art-eval}
%% \end{figure*}

\begin{figure*}[t!]
  \centering
  %\scalebox{0.965}{\input{art-hot-btree-eval.tex}}
  \begin{adjustbox}{width=\textwidth,totalheight=\textheight,keepaspectratio}

\newcommand{\mysize}{0.34}

\begin{tabular}{cccc}

  \multicolumn{4}{c}
  {\includegraphics[width=\textwidth]{figures/ppt/legend_new2.pdf}} \\

  & {\LARGE \textbf{Email}}
  & {\LARGE \textbf{Wiki}}
  & {\LARGE \textbf{URL}} \\

  %% ----------------------------------------------------------------

  \rotatebox{90}{\hskip 5em \LARGE \textsc{\textbf{\art}}}

  & \includegraphics[width=\mysize\textwidth]{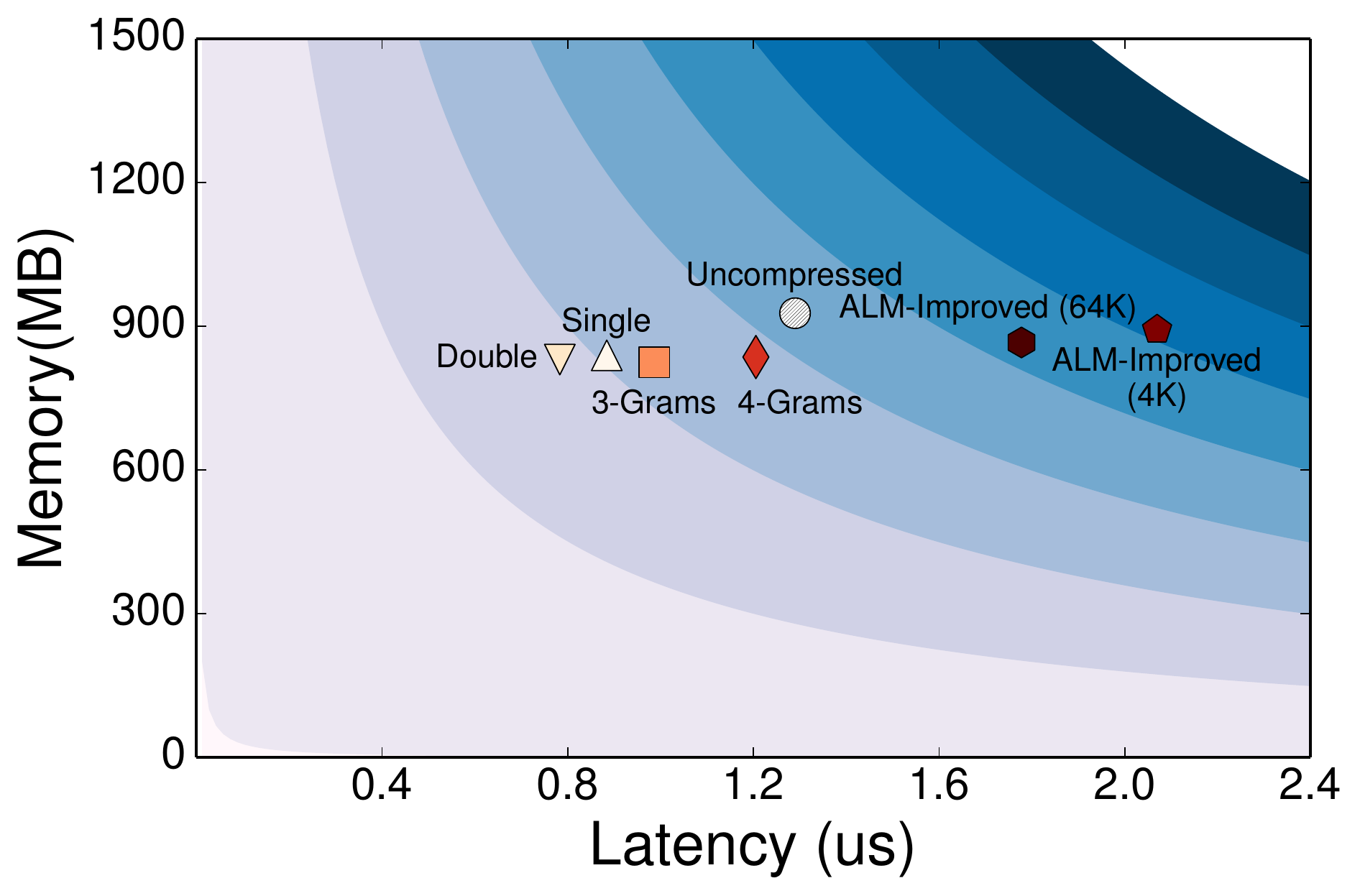}

  & \includegraphics[width=\mysize\textwidth]{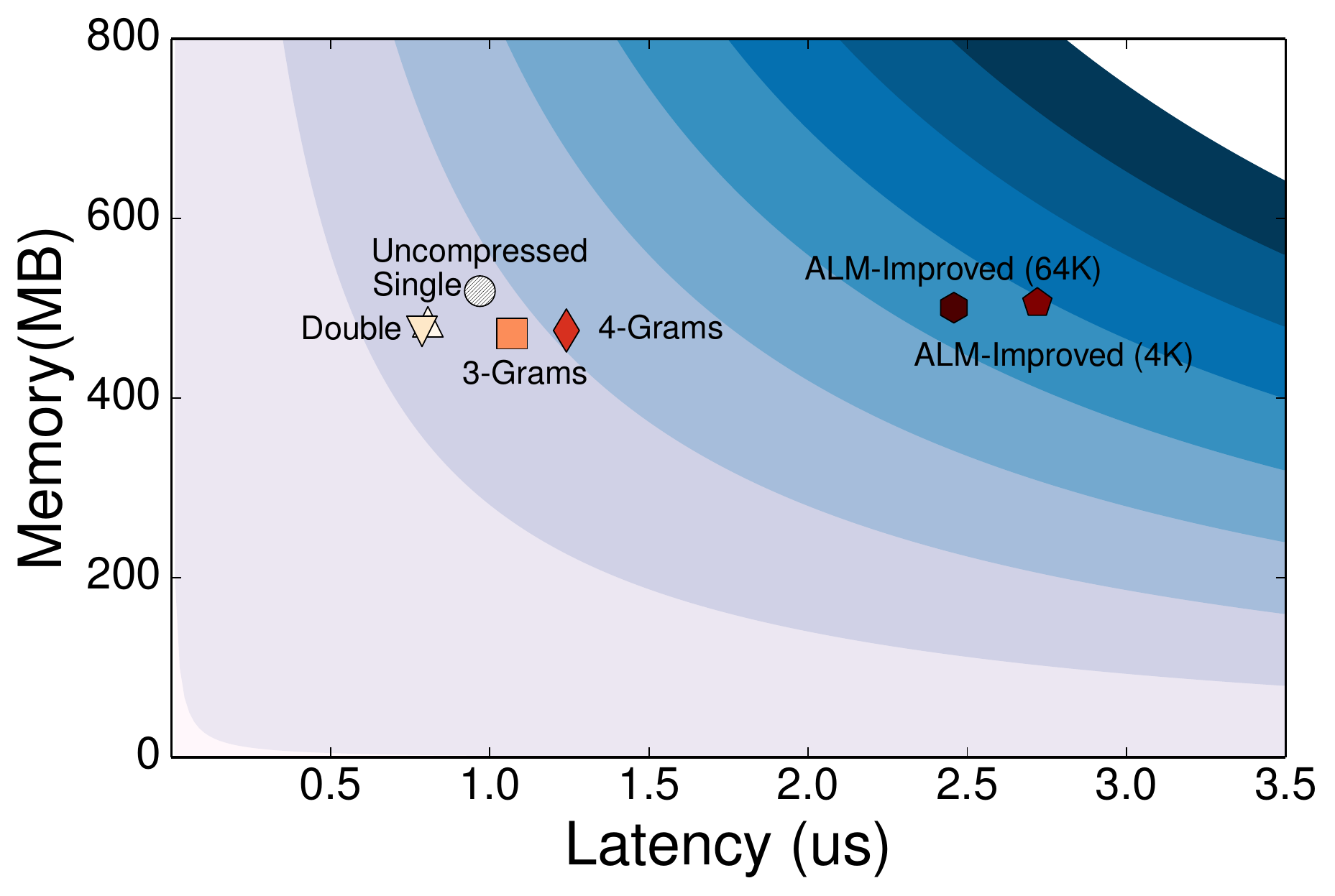}

  & \includegraphics[width=\mysize\textwidth]{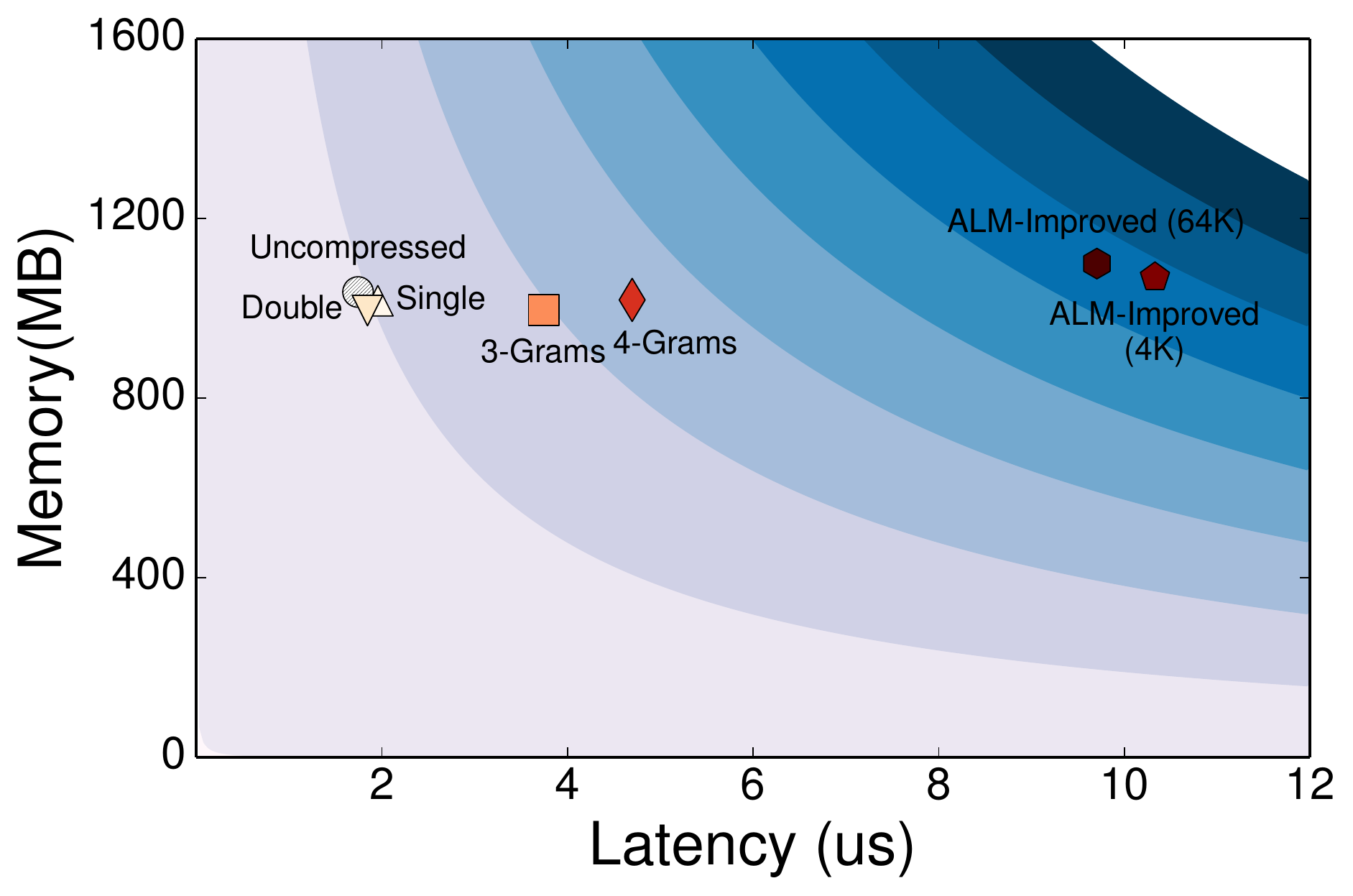} \\

  %\hlineB{2} \\

  %% ----------------------------------------------------------------

  \rotatebox{90}{\hskip 5em \LARGE \textsc{\textbf{\hot}}}

  & \includegraphics[width=\mysize\textwidth]{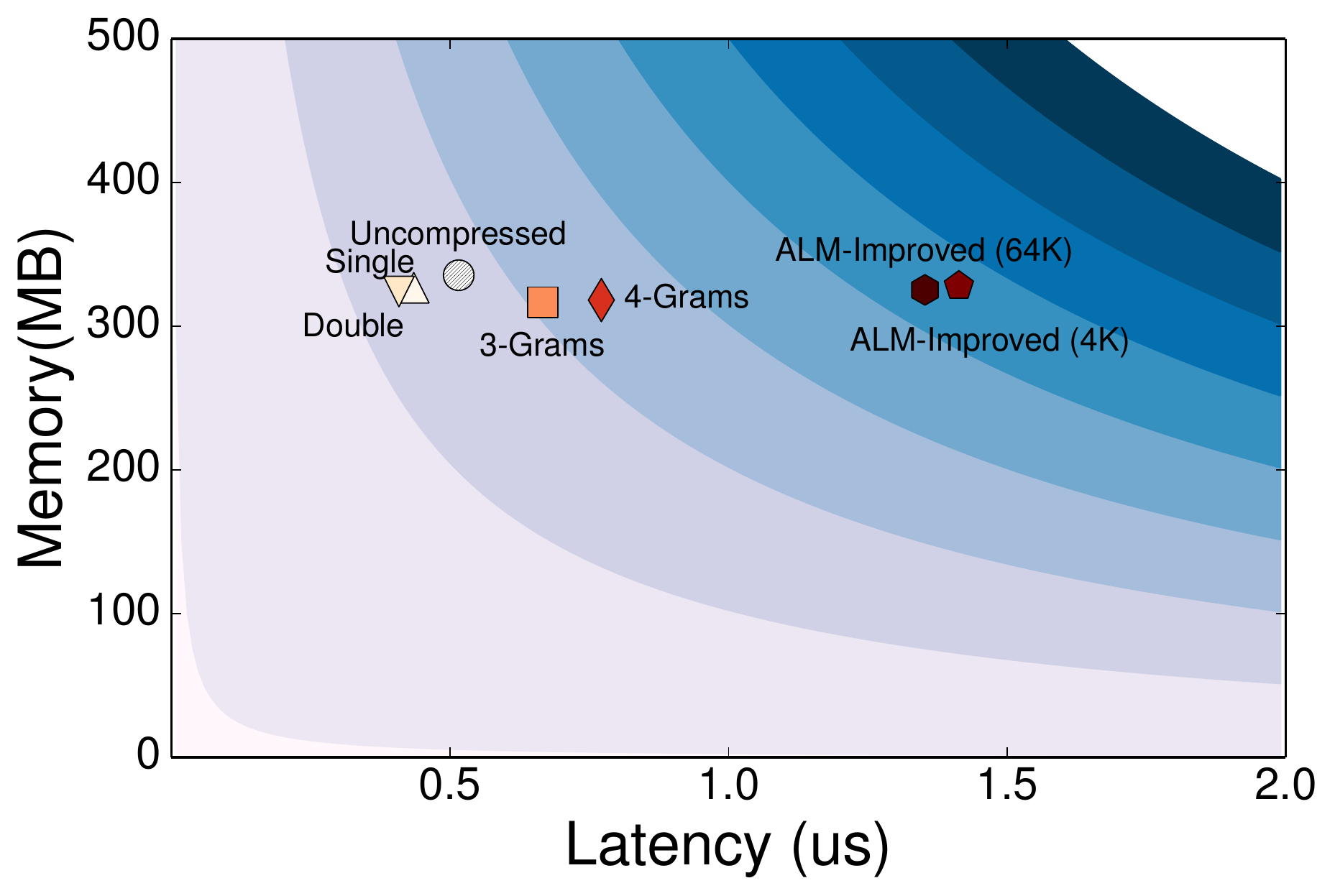}

  & \includegraphics[width=\mysize\textwidth]{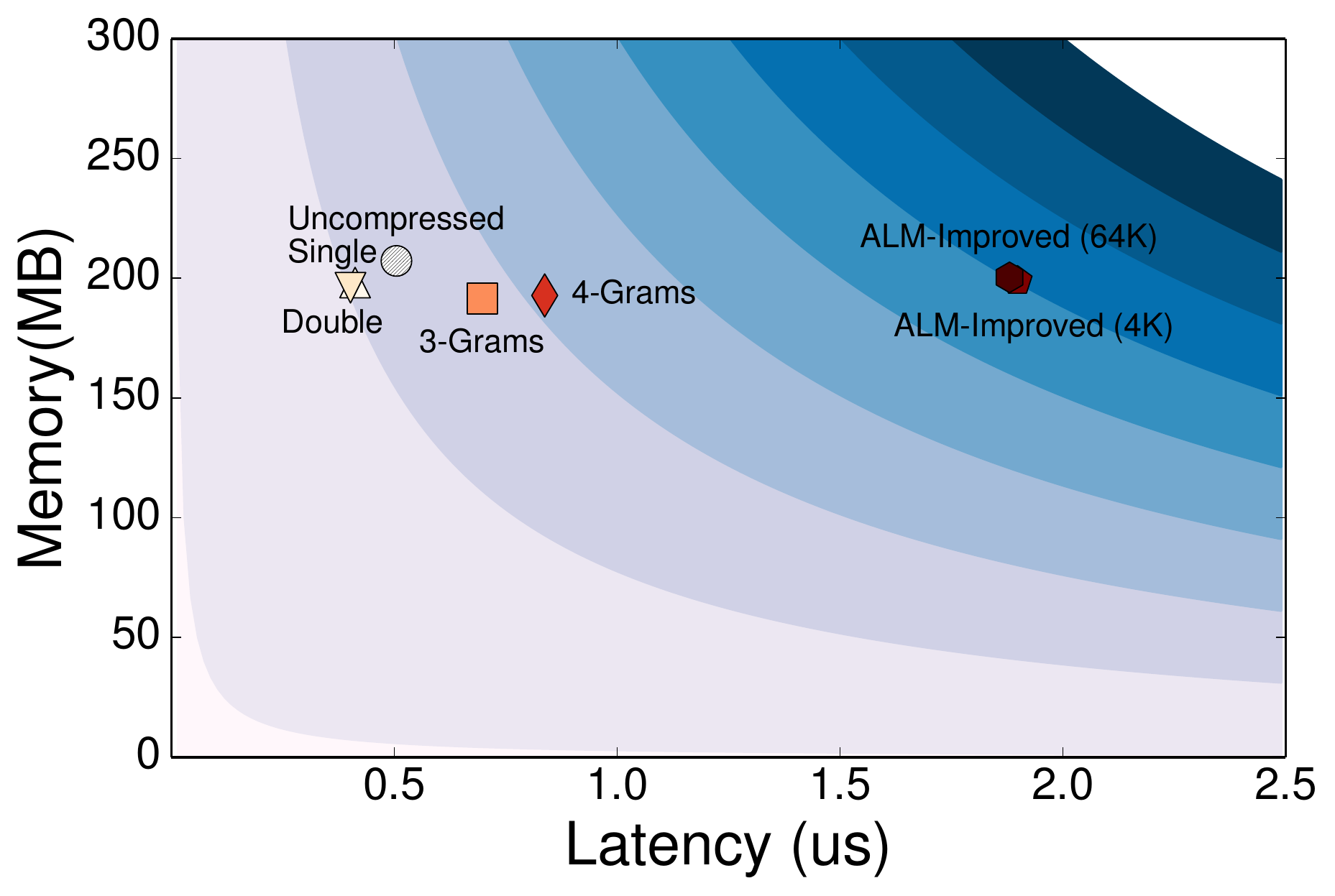}

  & \includegraphics[width=\mysize\textwidth]{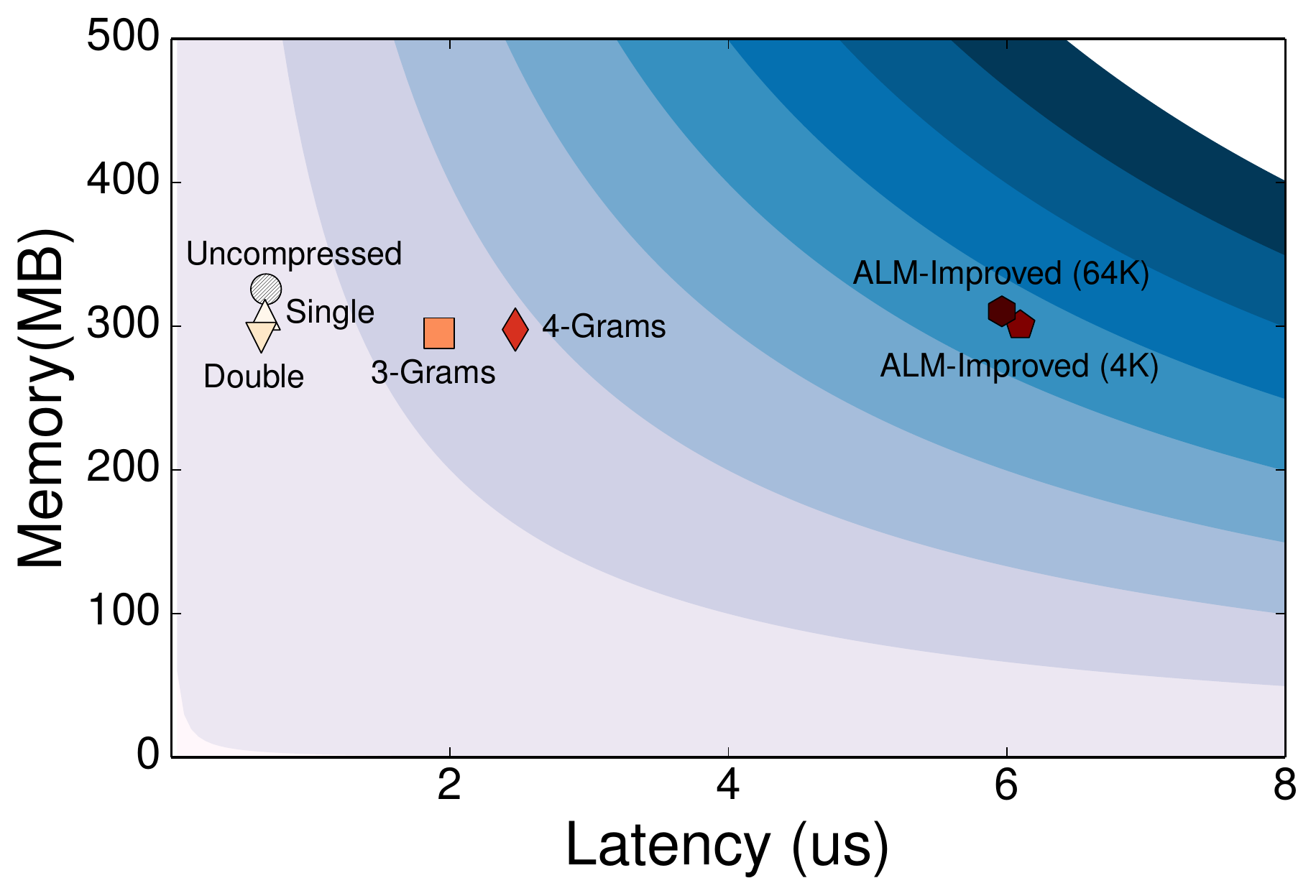} \\

  %\hlineB{2} \\

  %% ----------------------------------------------------------------

  \rotatebox{90}{\hskip 4em \LARGE \textsc{\textbf{\bplustree}}}

  & \includegraphics[width=\mysize\textwidth]{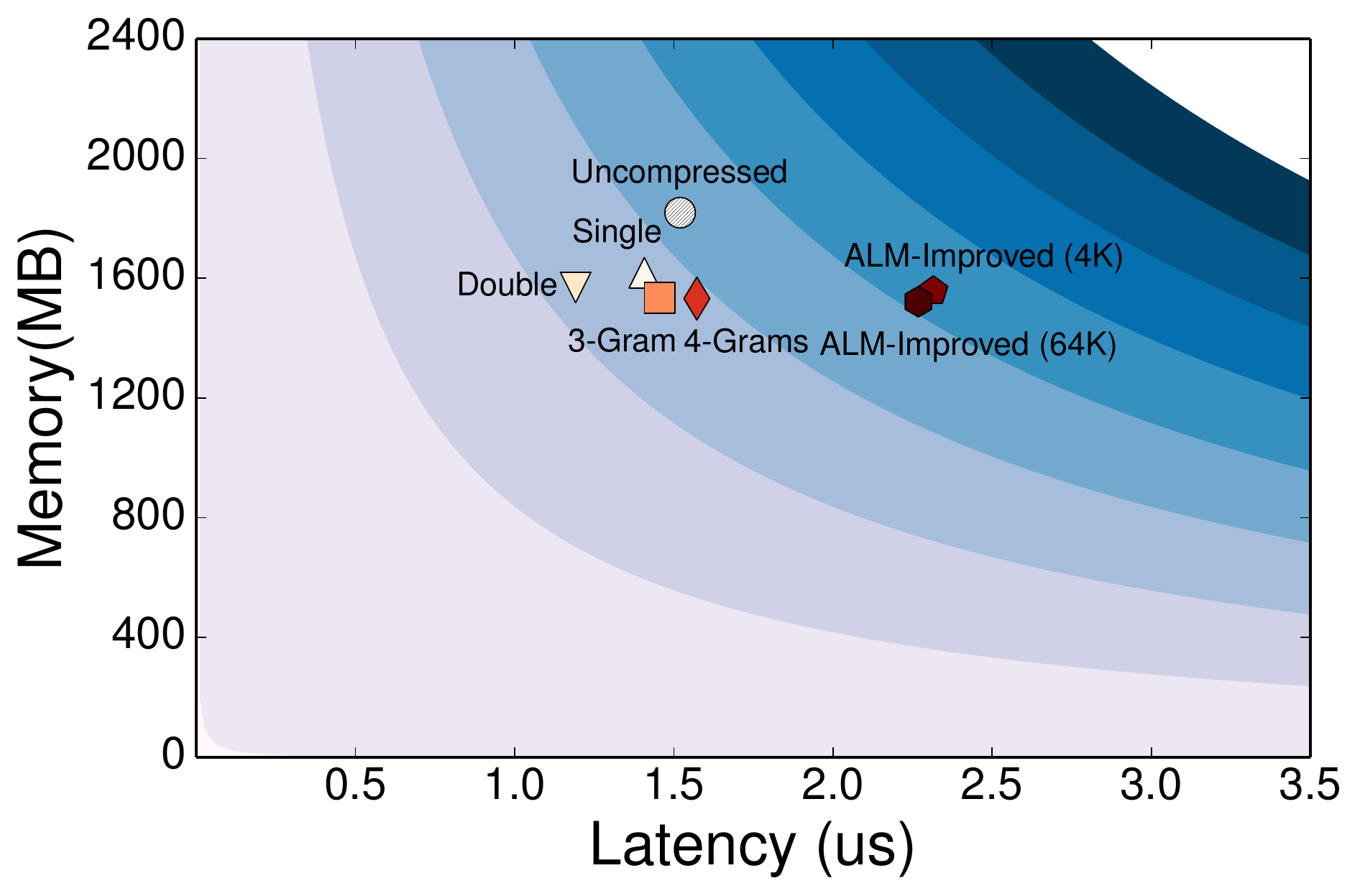}

  & \includegraphics[width=\mysize\textwidth]{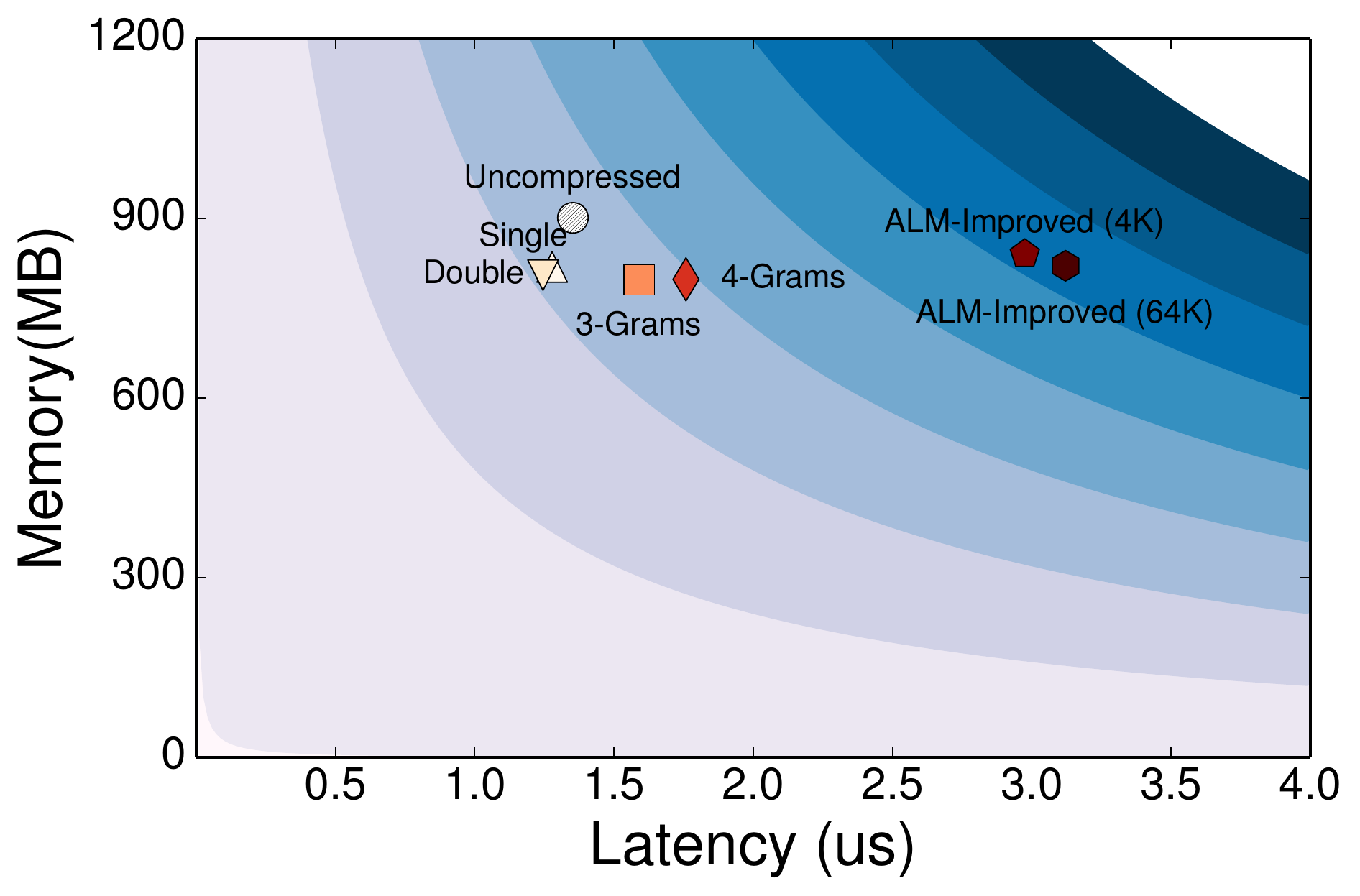}

  & \includegraphics[width=\mysize\textwidth]{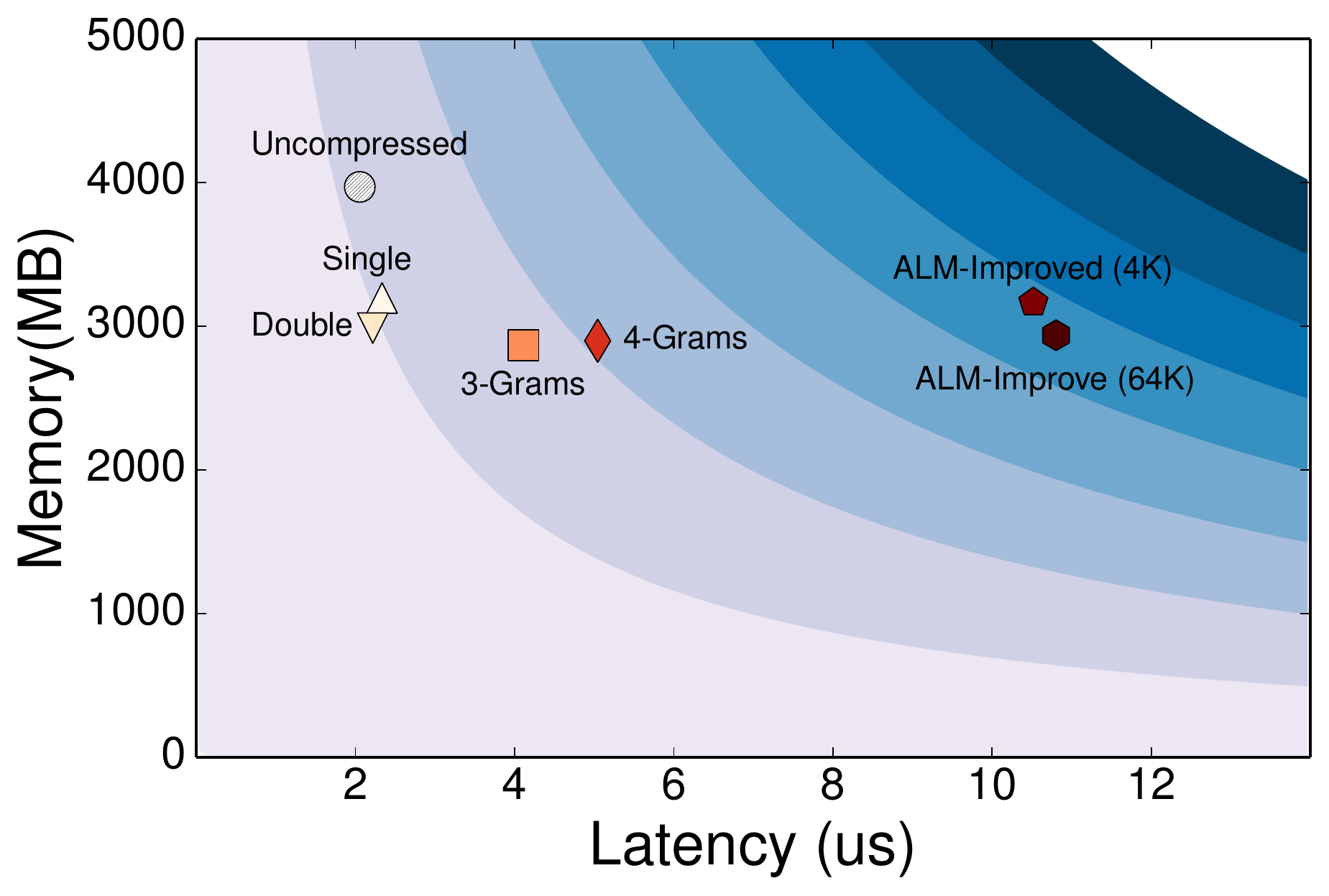} \\

  %\hlineB{2} \\

  %% ----------------------------------------------------------------

  \rotatebox{90}{\hskip 1.5em \LARGE \rev{\textsc{\textbf{\pbtree}}}}

  & \includegraphics[width=\mysize\textwidth]{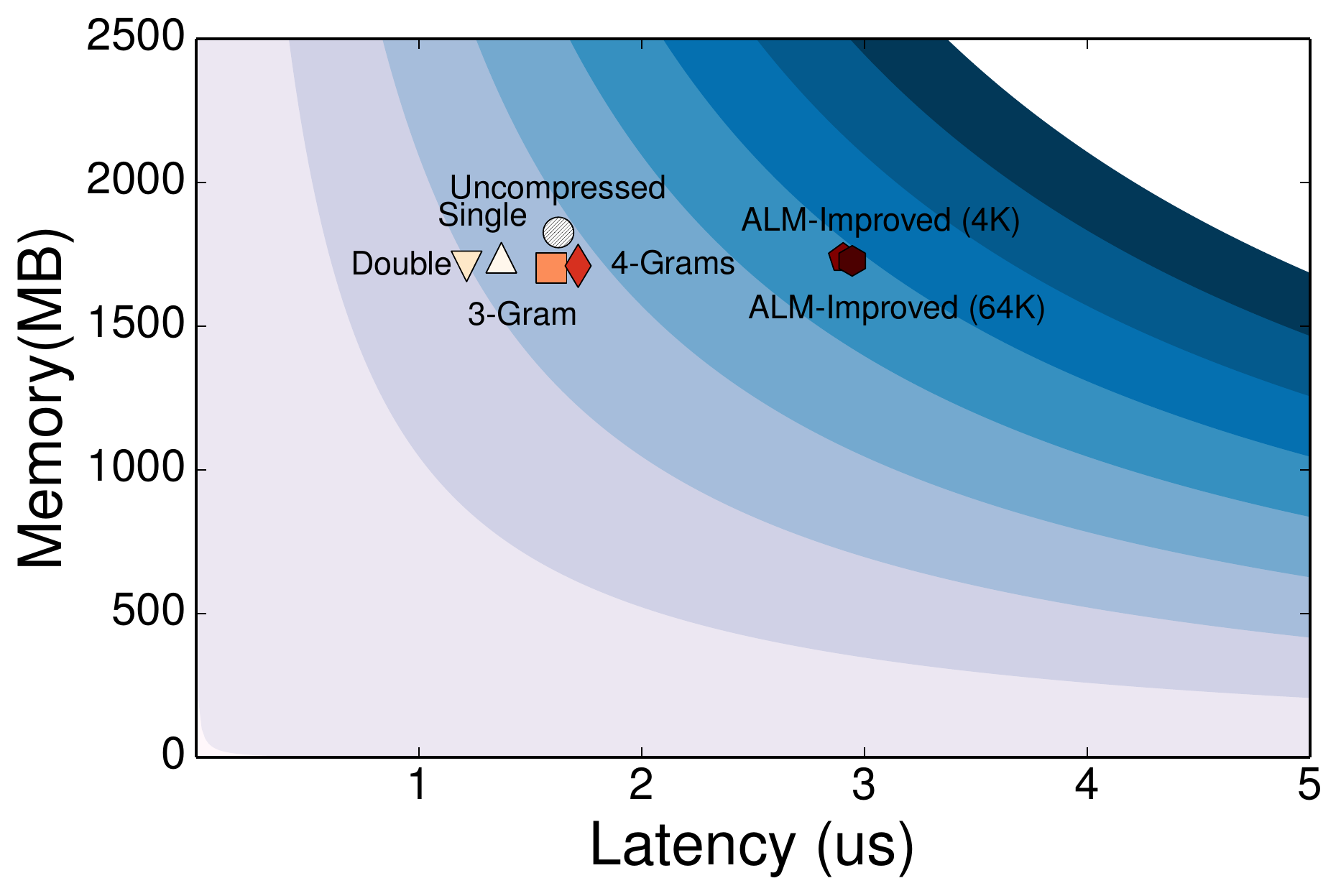}

  & \includegraphics[width=\mysize\textwidth]{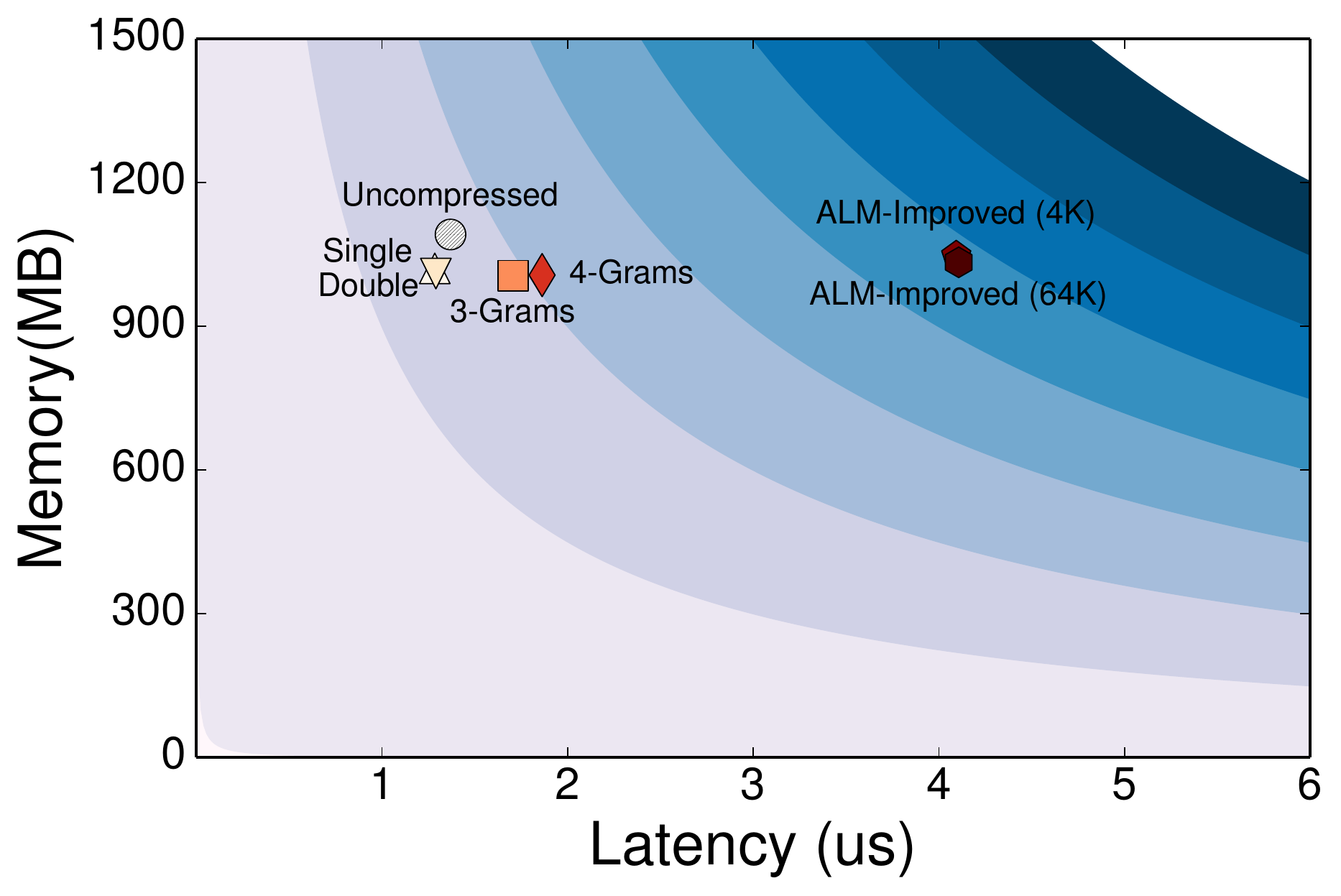}

  & \includegraphics[width=\mysize\textwidth]{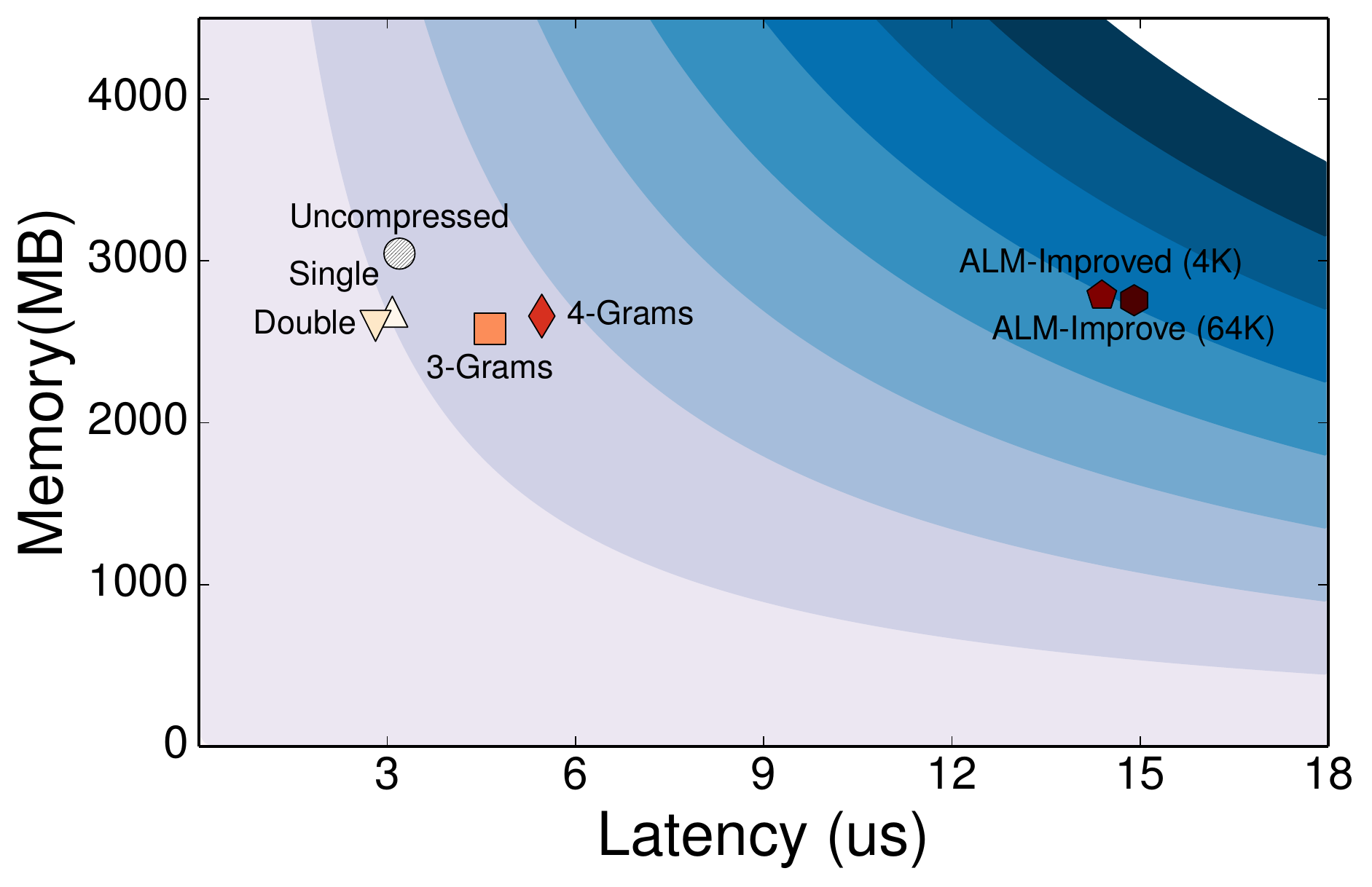} \\

  %\hlineB{2} \\

  %% ----------------------------------------------------------------

%%   \rotatebox{90}{\hskip 1.5em \LARGE \textsc{\textbf{\pbtree}}}

%%   & \includegraphics[width=\mysize\textwidth]{figures/cpsbtree_32_suffix/point/lat_mem_email_btree_point.pdf}

%%   & \includegraphics[width=\mysize\textwidth]{figures/cpsbtree_32_suffix/point/lat_mem_wiki_btree_point.pdf}

%%   & \includegraphics[width=\mysize\textwidth]{figures/cpsbtree_32_suffix/point/lat_mem_url_btree_point.pdf} \\

\end{tabular}

\end{adjustbox}

  \caption{
    %\textbf{YCSB Evaluation on \art, \hot, and \bplustree} --
    %Measurements for executing YCSB workloads on \ope-optimized indexes.
    \rev{\textbf{YCSB Point Query Evaluation on \art, \hot, \bplustree, and \pbtree}
      -- Measurements for executing YCSB point query workloads on
      \ope-optimized indexes.}
  }
  \label{fig:art-hot-btree-eval}
\end{figure*}

To experimentally evaluate the benefits and trade-offs of applying
\ope to in-memory search trees,
%we integrated \ope into four data structures:
%\surf, \art, \hot, and \bplustree
\rev{we integrated \ope into five data structures:
\surf, \art, \hot, \bplustree, and \pbtree}
(as described in \cref{sec:integration}).
Based on the microbenchmark results in \cref{sec:micro-eval},
we evaluate six \ope configurations for each search tree:
(1) \schemeSingleChar, (2) \schemeDoubleChar,
(3) \schemeThreeGrams with 64K ($2^{16}$) dictionary entries,
(4) \schemeFourGrams with 64K dictionary entries,
(5) \schemeALMImproved with 4K ($2^{12}$) dictionary entries,
and (6) \schemeALMImproved with 64K dictionary entries.
\rev{We include the original search trees as baselines
(labeled as ``Uncompressed'').}
We choose 64K for \schemeThreeGrams, \schemeFourGrams, and \schemeALMImproved
so that they have the same dictionary size as \schemeDoubleChar.
We evaluate an additional \schemeALMImproved configuration with 4K dictionary size
because it has a similar dictionary memory as \schemeDoubleChar,
\schemeThreeGrams (64K), and \schemeFourGrams (64K).
We exclude the original \schemeALM scheme because it is always
worse than the others.

%% ---------------------------------------------------------------
%% Workload
%% ---------------------------------------------------------------
\subsection{Workload}
\label{sec:tree-eval:workload}

We use the YCSB-based~\cite{cooper2010} index-benchmark framework
proposed in the Hybrid Index~\cite{zhang2016}
and later used by, for example, \hot~\cite{binna2018} and \surf~\cite{zhang2018}.
We use the YCSB workloads \textbf{C} and \textbf{E} with a Zipf distribution
to generate \textbf{point} and \textbf{range} queries.
Point queries are the same for all trees.
Each range query for \art, \hot, \bplustree, and \pbtree
is a start key followed by a scan length.
Because \surf is a filter, its range query
is a start key and end key pair, where the end key is a copy of the start 
key with the last character increased by one 
(e.g., {\small [``\texttt{com.gmail@foo}'', ``\texttt{com.gmail@fo\underline{p}}'']}).
% 
%% These workloads are composed of three types of queries:
% 
%% \begin{itemize}[leftmargin=*]
%% \item[$\bullet$] \textbf{Insert:}
%%   For \art, \hot, and \bplustree, the queries are individual inserts.
%%   For \surf, it means bulk-loading and filter construction.
% 
%%     \item[$\bullet$] \textbf{Point:}
%%     Retrieve a single key. It is the same for all trees.
%     
%%     \item[$\bullet$] \textbf{Range:}
%%     Retrieve a sequence of keys. Each range query for \art, \hot, and \bplustree is
%%     a start key followed by a scan length. Since \surf does not support iterators, a range 
%%     query is comprised of a start key and end key pair, where the end key is a copy of the start 
%%     key with the last character increased by one 
%%     (e.g., \small{[``\texttt{com.gmail@foo}'', ``\texttt{com.gmail@fo\underline{p}}'']}).    
%% \end{itemize}
% 
We replace the original YCSB keys with the keys in our email, wiki and URL datasets.
We create one-to-one mappings between the YCSB keys and our keys
during the replacement to preserve the Zipf distribution.\footnote{We omit
the results for other query distributions
(e.g., uniform) because they demonstrate similar performance
gains/losses as in the Zipf case.}

%% ---------------------------------------------------------------
%% YCSB Evaluation
%% ---------------------------------------------------------------
\subsection{YCSB Evaluation}
\label{sec:tree-eval:ycsb}
We start each experiment with the building phase
% where we create the target scheme in \ope
using the first $1\%$ of the dataset's keys.
Next, in the loading phase, we insert the keys one-by-one into the tree
(except for \surf because it only supports batch loading).
% Since we are constrained by the dataset sizes, we initialize each tree
% with 25M email keys, 14M wiki keys, and 25M URL keys.
Finally, we execute 10M queries on the compressed keys with a single thread
using a combination of point and range queries according to the workload.
We obtain the point, range, and insert query latencies by
dividing the corresponding execution time by the number of queries.
We measure memory consumption (\ope size included) after the loading phase.

\rev{\cref{fig:surf-eval,fig:surf-fpr,fig:art-hot-btree-eval} show
the benchmark results.
%\rev{The results for the benchmarks are shown in 
%\cref{fig:surf-eval,fig:surf-fpr,fig:art-hot-btree-eval,fig:art-hot-btree-eval-range}.
%\cref{fig:surf-eval,fig:surf-fpr,fig:art-hot-btree-eval}.
Range query and insert results for \art, \hot, \bplustree, and \pbtree
are included in~\cref{sec:tree-eval-range} because the performance results
are similar to the corresponding point query cases for similar reasons.}
% In all figures, lower is better.
We first summarize the high-level observations and
then discuss the results in more detail for each tree.
\\ \vspace{-0.1in}

%% ---------------------------------------------------------------
%% Observations
%% ---------------------------------------------------------------
\noindent \textbf{High-Level Observations:}
%There are several high-level observations from these experiments.
First, in most cases, multiple schemes in \ope provide a \emph{Pareto improvement}
to the search tree's performance and memory-efficiency.
%\todo{$\leftarrow$ need to explain what this means.}
\rev{
Second, the simpler \encodingFIVC schemes, especially \schemeDoubleChar,
stand out to provide the best trade-off between
query latency and memory-efficiency for the search trees.
Third, more sophisticated \encodingVIVC schemes produce the lowest
search tree memory in some cases.
Compared to \schemeDoubleChar, however,
their small additional memory reduction does not justify the significant
performance loss in general.
}
%Finally, the results also show that \ope reduces
%the false positive rate in \surf with the same configuration.
\\ \vspace{-0.1in}

%% \begin{figure*}[t]
%%     \centering
%%     \input{hot-eval.tex}
%%     \caption{
%%         \textbf{\hot YCSB Evaluation} --
%%         {Integrate HOPE into HOT. Measure memory use, point and range query latency, and insert 
%% latency on the email,
%%         wiki, and url datasets. Note that the figures have different Y-axis scales.} 
%%     }
%%     \label{fig:hot-eval}
%% \end{figure*}

%% \begin{figure*}[t]
%%     \centering
%%     \input{btree-eval.tex}
%%     \caption{
%%         \textbf{\bplustree YCSB Evaluation} --
%%         {Integrate HOPE into \bplustree. Measure memory use, point and range query latency, and 
%% insert latency on the email,
%%          wiki, and url datasets. Note that the figures have different Y-axis scales.} 
%%     }
%%     \label{fig:btree-eval}
%% \end{figure*}

%% -----------------------
%% SURF
%% -----------------------
\noindent \textbf{\surf:}
The heatmaps in the first row of \cref{fig:surf-eval} show the
point query latency vs. memory 
trade-offs made by \surf{s} with different \ope configurations.
% We refer to~\cite{zhang2018} for such heatmap illustration.
We define a cost function $C = L \times M$,
where $L$ represents latency, and $M$ represents memory.
This cost function assumes a balanced performance-memory trade-off.
We draw the equi-cost curves (as heatmaps) where
points on the same curve have the same cost.
%This cost function indicates that \surf achieves a balanced performance vs. memory configuration.
%The curves in the heatmaps are equi-cost
%(i.e., the points on the same curve have the same cost).

\ope reduces \surf's query latencies by up to 41\% in all workloads
with non-ALM encoders.
%\schemeSingleChar, \schemeDoubleChar, \schemeThreeGrams, and \schemeFourGrams encoders.
This is because compressed keys generate shorter tries,
as shown in the third row of \cref{fig:surf-eval}.
According to our analysis in \cref{sec:integration},
the performance gained by fewer levels in the trie
outweighs the key encoding overhead.
Although \surf with \schemeALMImproved (64K) has the lowest trie height,
it suffers high query latency because encoding is slow for
\schemeALMImproved schemes.
%(refer to \cref{fig:cpr-lat-mem}).

Although the six \ope schemes under test achieve compression rates
of 1.5--2.5$\times$ in the microbenchmarks,
they only provide $\sim$30\% memory savings to \surf.
The reason is that compressing keys only reduces the number
of internal nodes in a trie (i.e., shorter paths to the leaf nodes).
The number of leaf nodes, which is often the majority of the storage cost, stays the same.
\surf with \schemeALMImproved (64K) consumes more memory than others
because of its larger dictionary.

%The results for \surf with \schemeALMImproved (4K) are interesting.
\cref{sec:micro-eval:perf} showed that \schemeALMImproved (4K) achieves
a better compression rate than \schemeDoubleChar for email keys
with a similar-sized dictionary.
When we apply this scheme to \surf, however, the memory saving is smaller
than  \schemeDoubleChar even though it produces a shorter trie.
%Although this seems counterintuitive, it
This is because \schemeALMImproved
favors long symbols in the dictionary.
%allows dictionary symbols to have arbitrary lengths and it favors long symbols.
Encoding long symbols one-at-a-time can prevent prefix sharing.
As an example, \schemeALMImproved  may treat the keys ``\texttt{com.gmail@c}'' and 
``\texttt{com.gmail@s}'' as two separate symbols and thus have completely different codes.
% (e.g., ``\small{\texttt{01110101}}'' and ``\small{\texttt{100001}}'').
% Such encodings greatly compromise the inherent prefix compression
% in tries.

All schemes, except for \schemeSingleChar, add computational overhead in building \surf.
The dictionary build time grows quadratically
with the number of entries because of the Hu-Tucker algorithm.
One can reduce this overhead by shrinking the
dictionary size, but this diminishes performance
and memory-efficiency gains.

Finally, the \ope-optimized \surf achieves lower false positive rate
under the same suffix-bit configurations,
as shown in \cref{fig:surf-fpr}.
This is because each bit in the compressed keys
carries more information and is, thus, more distinguishable
than a bit in the uncompressed keys.
\\ \vspace{-0.1in}

%% -----------------------
%% ART + HOT
%% -----------------------
\noindent \textbf{\art, \hot:}
%\cref{fig:art-eval,fig:hot-eval}
%show the results for \art and \hot, respectively.
\cref{fig:art-hot-btree-eval}
shows that \ope improves \art and \hot's performance and memory-efficiency for
similar reasons as for \surf because they are also trie-based data structures.
%%%%%Due to space limitations, we omit the trie-height results\footnote{These additional 
%%%%%results are available here \cite{www-more-results}.}.
Compared to \surf, however, the amount of improvement
for \art and \hot is less.
%the improvements with \ope decrease
%\todo{$\leftarrow$ relative to what?}
%in the \art and \hot experiments.
This is for two reasons.
First, \art and \hot include a 64-bit value pointer for each key,
which dilutes the memory savings from the key compression.
More importantly, as described in \cref{sec:impl:impl,sec:integration},
\art and \hot only store partial keys using \textit{optimistic common prefix skipping} (OCPS).
\hot is more optimistic than \art as it only stores the \emph{branching points} in a trie
(i.e., the minimum-length partial keys needed to uniquely map a key to a value).
Although OCPS can incur false positives, the DBMS will verify the match when it retrieves the 
tuple. Therefore, since \art and \hot store partial keys, they do not take full advantage of 
key compression.
The portion of the URL keys skipped is large because they share long
prefixes. Nevertheless, our results show that \ope still provides some benefit and thus are 
worth applying to both data structures.
% \todo{MK: Feels a little like ART and HOT don't need key compression.  Someone
% could ask well then why don't I just use one of those indexes and not HOPE?}
\\ \vspace{-0.1in}

%% \begin{figure}[t]
%%   \centering
%%   \includegraphics[width=0.85\columnwidth]{figures/hot/point/lat_mem_email_hot.pdf}
%%   \caption{
%%     \textbf{Evaluating \ope schemes on \hot} --
%%     The experiment is point queries on email keys.
%%   }
%%   \label{fig:hot-eval}
%% \end{figure}

%% \begin{figure}[t]
%%   \centering
%%   \begin{minipage}{.49\columnwidth}
%%     \centering
%%     \includegraphics[width=\linewidth]{figures/SuRF/point/fpr_email_surf_point.pdf}
%%     \caption{
%%       \textbf{\ope reduces false positive rate in \surf}
%%     }
%%   \label{fig:surf-fpr}
%%   \end{minipage}
%%   \begin{minipage}{.49\columnwidth}
%%     \centering
%%     \includegraphics[width=\linewidth]{figures/hot/point/lat_mem_email_hot.pdf}
%%     \caption{
%%       \textbf{Evaluating \ope on \hot}
%%     }
%%     \label{fig:hot-eval}
%%   \end{minipage}
%% \end{figure}

\noindent \textbf{\bplustree, \pbtree:}
The results in \cref{fig:art-hot-btree-eval} show that
\ope is beneficial to search trees beyond tries.
Because the \bplustrees %\tlx 
use reference pointers to store
variable-length string keys outside of each node,
compressing the keys does not change the tree structure.
In addition to memory savings, the more lightweight \ope schemes (\schemeSingleChar and 
\schemeDoubleChar) also improve the \bplustree's query performance because of faster string 
comparisons and better cache locality.
We validate this by running the point-query workload on email keys and
measuring cache misses using \texttt{cachegrind}~\cite{www-cachegrind}.
%To validate this assumption, we re-ran the point-query workload on email keys and used 
%\texttt{cachegrind}~\cite{www-cachegrind} to measure cache misses.
We found that \schemeDoubleChar on \tlx
reduces the L1 and last-level cache misses by 34\% and 41\%, respectively.

\rev{Compared to plain \bplustrees, we observe smaller memory saving
  percentages when using \ope on \pbtrees. This is because prefix compression
  reduces the storage size for the keys, and thus making the
  structural components of the \bplustree (e.g., pointers) relatively larger.
  Although \ope provides similar compression rates when applied to
  a \pbtree and a plain \bplustree, the percentages of space reduction
  brought by \ope-compressed keys in a \pbtree is smaller with respect
  to the entire data structure size.}

\rev{As a final remark, \ope still improves the performance
  and memory for highly-compressed trees such as \surf.
  It shows that \ope is orthogonal to many other compression
  techniques and can benefit a wide range of data structures.}

\section{Conclusions}
\label{sec:concl}

We introduced \ope, a dictionary-based entropy-encoding compressor.
\ope compresses keys for in-memory search trees
in an order-preserving way with low performance and memory overhead.
To help understand the solution space for key compression, we developed a theoretical model 
and then show that one can implement multiple compression schemes in \ope based on this model.
Our experimental results showed that
using \ope to compress the keys for five in-memory search trees improves both their 
runtime performance and memory-efficiency for many workloads.

\vspace{0.3cm}

\noindent
\textbf{Acknowledgements.}
This work was supported %(in part) by funding from the U.S.
by the National Science Foundation under awards
\href{https://www.nsf.gov/awardsearch/showAward?AWD_ID=1700521}{SDI-CSCS-1700521},
\href{https://www.nsf.gov/awardsearch/showAward?AWD_ID=1846158}{IIS-1846158},
and \href{https://www.nsf.gov/awardsearch/showAward?AWD_ID=1822933}{SPX-1822933},
Google Research Grants, and the
\href{https://sloan.org/grant-detail/8638}{Alfred P. Sloan Research Fellowship} program.

%\begin{acks}
%To Robert, for the bagels and explaining CMYK and color spaces.
%\end{acks}

%% ==================================================================
%% BIBLIOGRAPHY
%% ==================================================================
\newpage
%\balance
\bibliographystyle{ACM-Reference-Format}
\bibliography{paper}

%% ==================================================================
%% APPENDIX
%% ==================================================================

\clearpage
\appendix
\balance
%% ---------------------------------------------------------------
%% Sample Size Sensitivity Test
%% ---------------------------------------------------------------
\section{Sample Size Sensitivity Test}
\label{sec:sample-size}

\begin{figure}[t]
  \centering
  \subfloat[Email]{
    \includegraphics[width=\columnwidth]{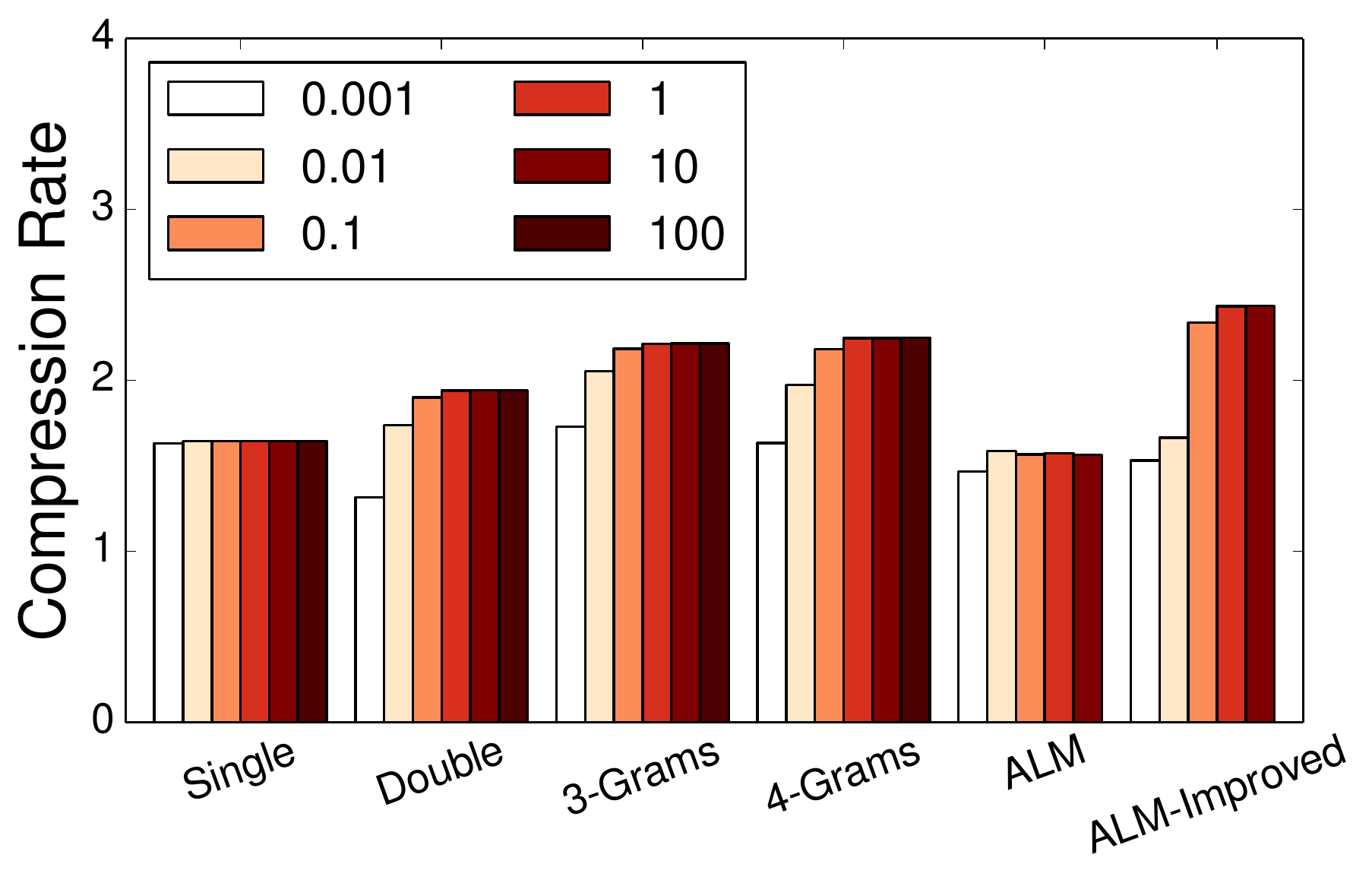}
    \label{fig:sample-email}
  }
  \\
  \subfloat[Wiki]{
    \includegraphics[width=\columnwidth]{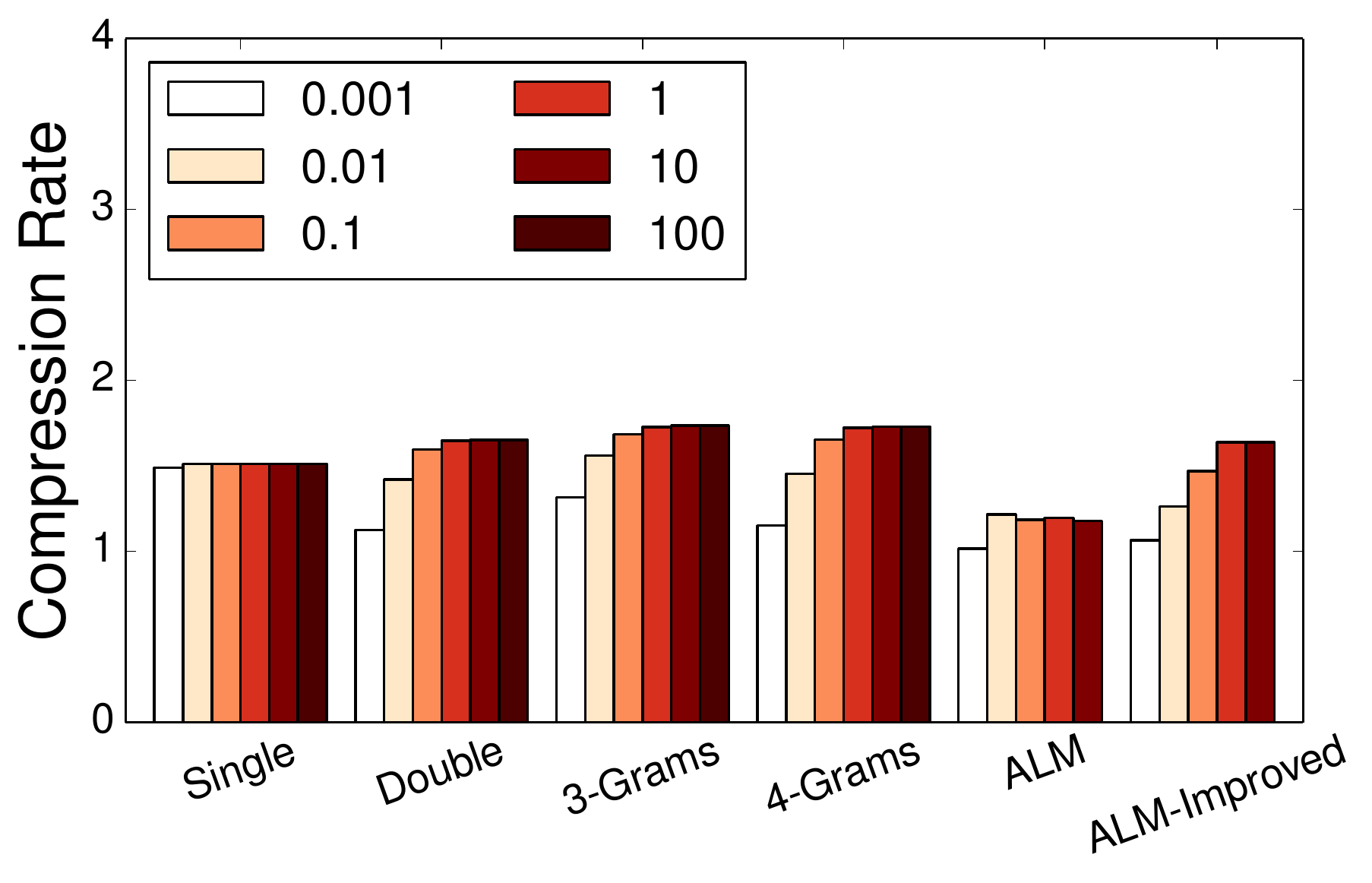}
    \label{fig:sample-wiki}
  }
  \\
  \subfloat[URL]{
    \includegraphics[width=\columnwidth]{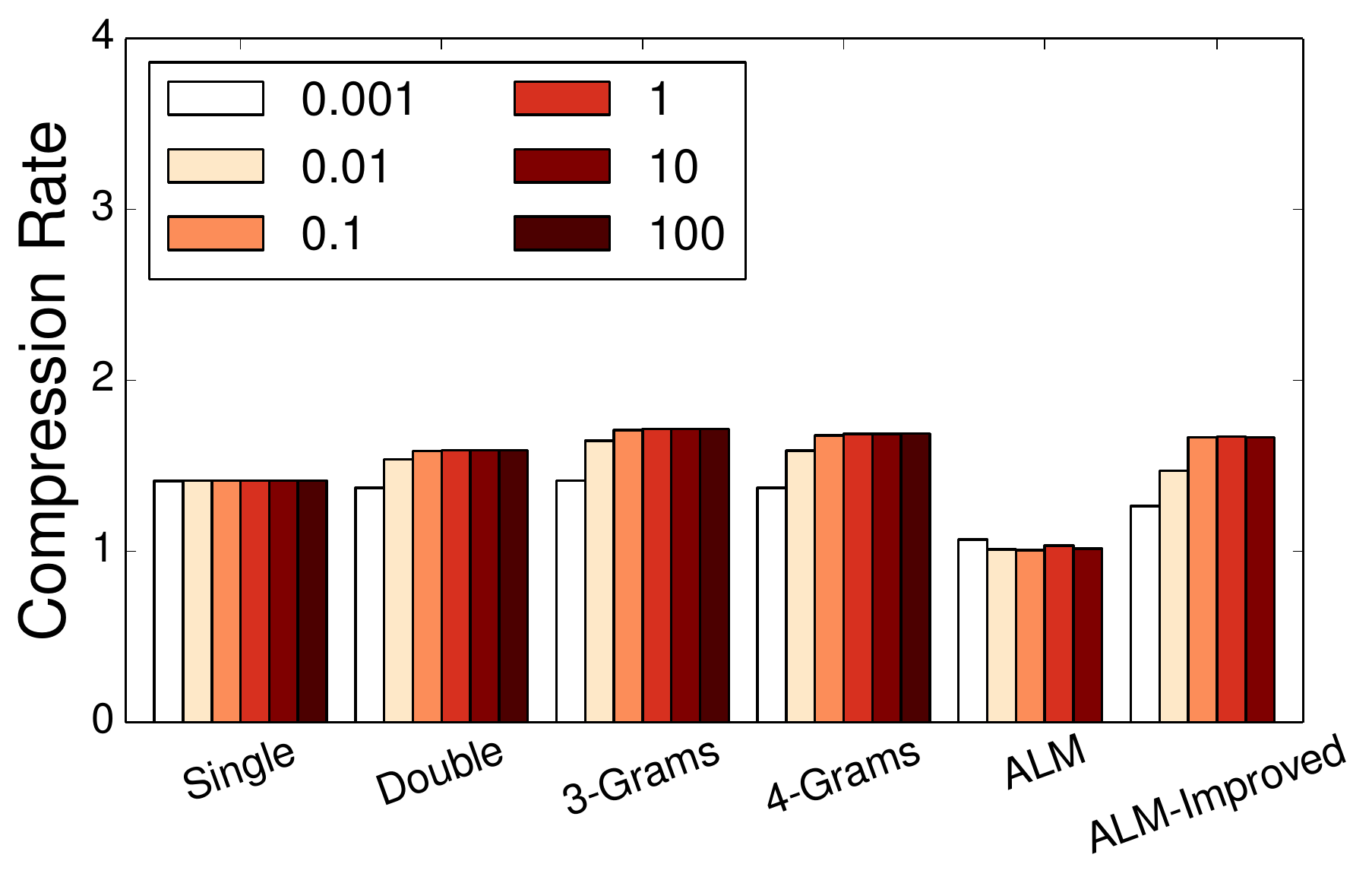}
    \label{fig:sample-url}
  }
  \caption{
    \textbf{Sample Size Sensitivity Test} --
    Compression rate measured under varying sample sizes for all
    schemes in \ope. The dictionary size limit is set to
    $2^{16}$ (64K) entries.
  }
  \label{fig:sample-size}
\end{figure}

In this section, we perform a sensitivity test on
how the size of the sampled key list affects \ope's compression rate.
We use the three datasets (i.e., Email, Wiki, and URL) introduced
in \cref{sec:micro-eval}.
We first randomly shuffle the dataset and then select the first $x$\% of
the entries as the sampled keys for \ope.
We set $x$ to 0.001, 0.01, 0.1, 1, 10, and 100, which translates to
250, 2.5K, 25K, 250K, 2.5M, and 25M samples for the Email and URL datasets,
and 140, 1.4K, 14K, 140K, 1.4M, and 14M samples for the Wiki dataset.
We measure the compression rate for each scheme in \ope for each $x$.
We set the dictionary size limit to $2^{16}$ (64K) entries.
Note that for $x = 0.001, 0.01$, schemes such as \schemeThreeGrams
do not have enough samples to construct the dictionary of the limit size.

\cref{fig:sample-size} shows the results. Note that for $x = 100$,
the numbers are missing for \schemeALM and \schemeALMImproved because
the experiments did not finish in a reasonable amount of time
due to their complex symbol select algorithms.
From the figures, we observe that a sample size of 1\% of the dataset
(i.e., 250K for Email and URL, 140K for Wiki) is large enough for
all schemes to reach their maximum compression rates.
1\% is thus the sample size percentage used in the experiments in
\cref{sec:micro-eval,sec:tree-eval}.
We also notice that the compression rates for schemes that exploit
higher-order entropies are more sensitive to the sample size
because these schemes require more context information to achieve
better compression.
As a general guideline, a sample size between 10K and 100K
is good enough for all schemes, and we can use a much smaller sample
for simpler schemes such as \schemeSingleChar.

%% ---------------------------------------------------------------
%% Batch Encoding
%% ---------------------------------------------------------------
\section{Batch Encoding}
\label{sec:batch}
We evaluate the batching optimization described in \cref{sec:impl:impl}.
In this experiment, we sort the email dataset and then
encode the keys with varying batch sizes (1, 2, 32).

\begin{figure}[t]
  \centering
  \includegraphics[width=\columnwidth]{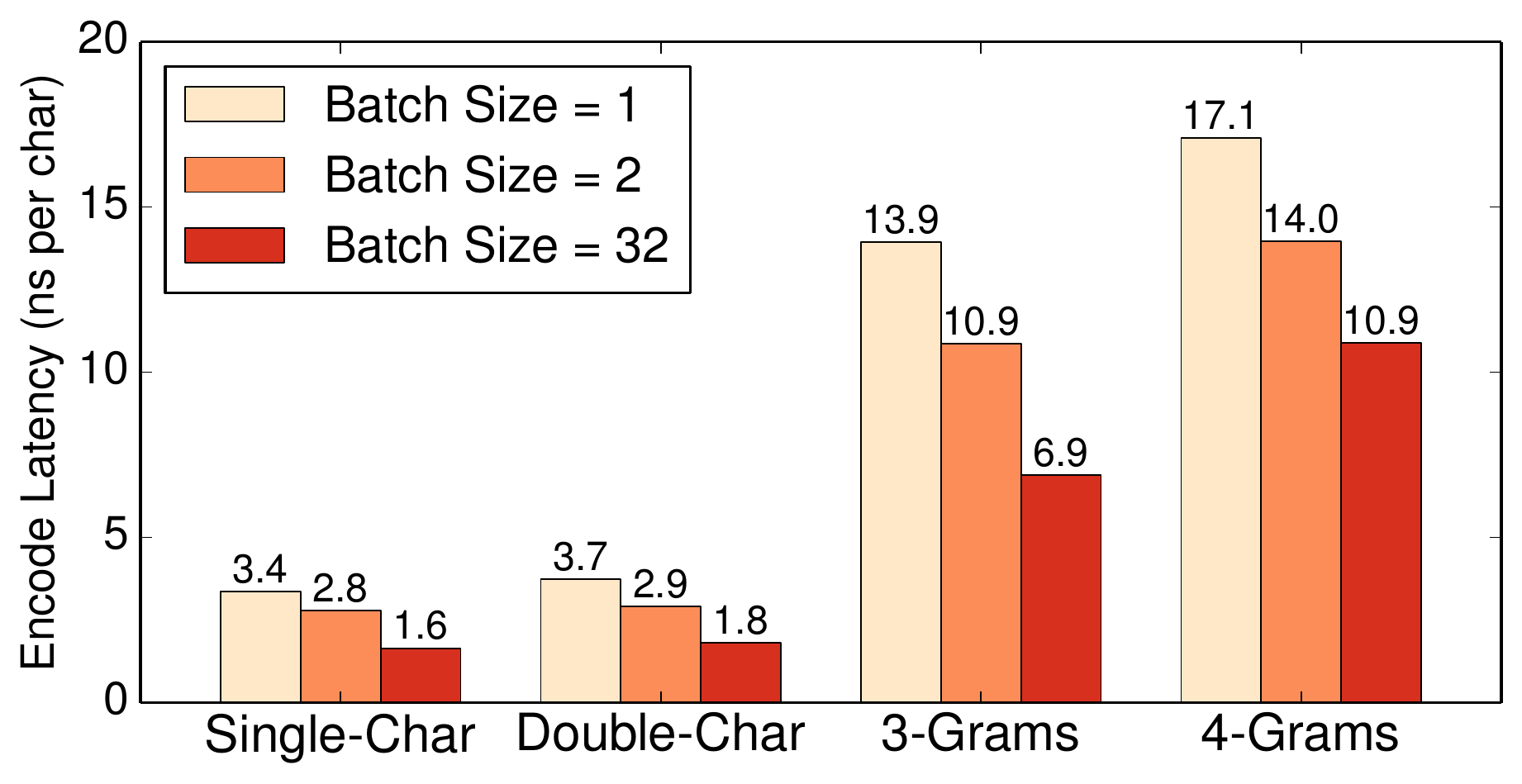}
  \caption{
    \textbf{Batch Encoding} --
    Encoding latency measured under varying batch sizes on a pre-sorted 1\% sample of email 
    keys. The dictionary size is $2^{16}$ (64K) entries for \schemeThreeGrams and \schemeFourGrams. 
  }
  \label{fig:batch}
\end{figure}

As shown in \cref{fig:batch},
batch encoding improves encoding performance significantly
because it encodes the common prefix of a batch only once
to avoid redundant work.
\schemeALM and \schemeALMImproved schemes do not benefit from batch encoding.
Because these schemes have dictionary symbols of arbitrary lengths,
we cannot determine a priori a common prefix that is aligned with
the dictionary symbols for a batch without encoding them.

%% ---------------------------------------------------------------
%% Updates and Key Distribution Changes
%% ---------------------------------------------------------------
\section{Updates and Key Distribution Changes}
\label{sec:updates}

  As discussed in \cref{sec:impl,sec:integration},
  \ope can support key updates without modifying the dictionary
  because the completeness and order-preserving properties
  of the String Axis Model (refer to \cref{sec:model:axis}) guarantee
  that any \ope dictionary can encode arbitrary input keys while
  preserving the original key ordering.
  However, a dramatic change in the key distribution may hurt \ope's
  compression rate.

\begin{figure}[H]
  \centering
  \includegraphics[width=\columnwidth]{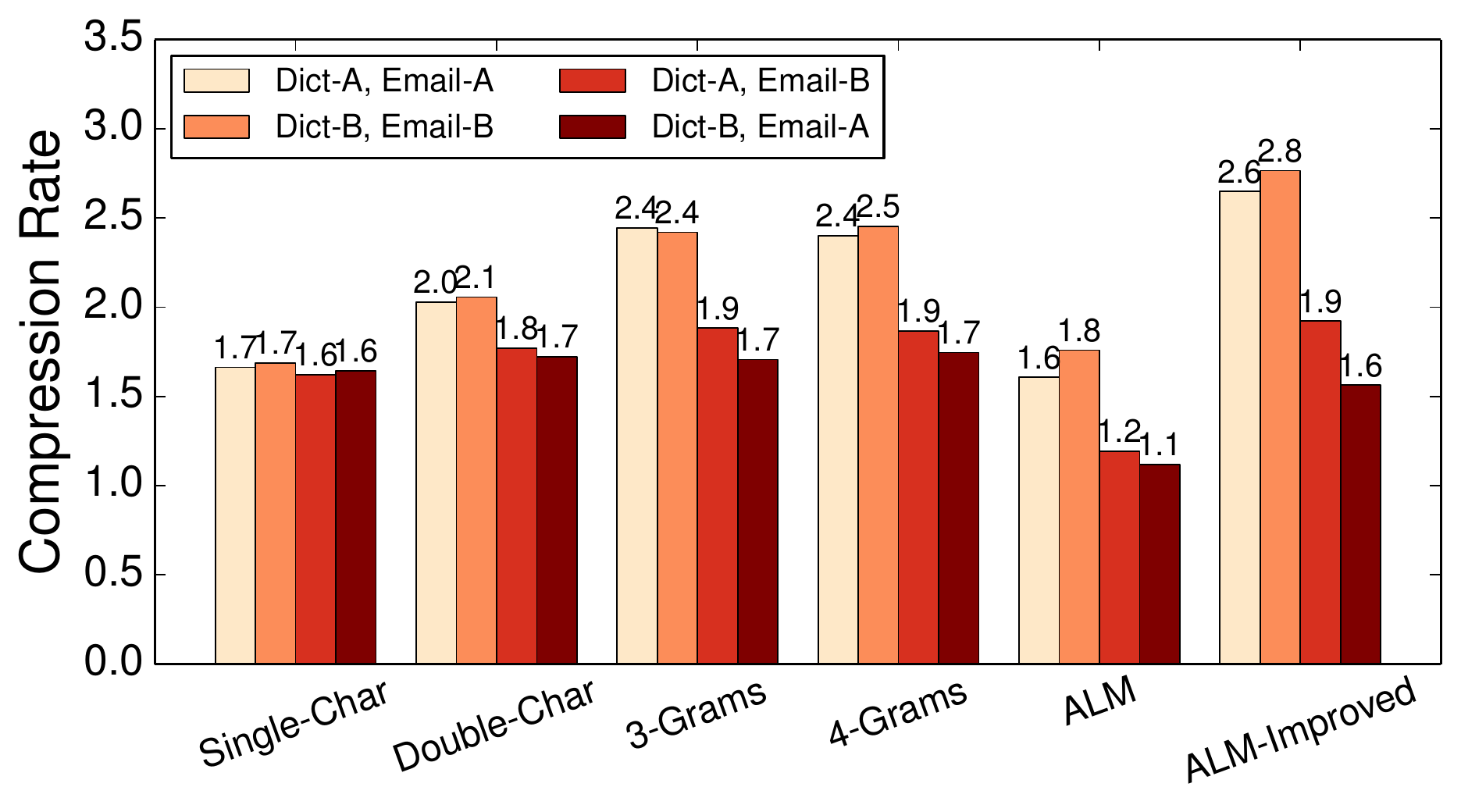}
  \caption{
    \textbf{Key Distribution Changes} --
    Compression rate measurements under stable key distributions
    and sudden key pattern changes.
  }
  \label{fig:distribution-change}
\end{figure}

  To simulate a sudden key distribution change, we divide our email dataset
  into two subsets (roughly the same size): Email-A and Email-B.
  Email-A contains all the \texttt{gmail} and \texttt{yahoo} accounts
  while Email-B has the rest, including accounts from \texttt{outlook},
  \texttt{hotmail}, and so on.
  In the experiments, we build two dictionaries (i.e., Dict-A and Dict-B)
  using samples from Email-A and Email-B, respectively for each
  compression scheme in \ope.
  We use the different dictionaries to compress the keys in the different
  datasets and then measure the compression rates.

  \cref{fig:distribution-change} shows the results.
  ``Dict-A, Email-A'' and ``Dict-B, Email-B'' represent cases where
  key distributions are stable,
  while ``Dict-A, Email-B'' and ``Dict-B, Email-A'' simulate dramatic
  changes in the key patterns.
  From the figure, we can see that \ope's compression rate decreases
  in the ``Dict-A, Email-B'' and ``Dict-B, Email-A'' cases.
  This result is expected because the dictionary built based on earlier
  samples cannot capture the new common patterns in the new distribution
  for better compression.
  We also observe that simpler schemes (i.e., schemes that exploit
  lower-order entropy) such as \schemeSingleChar are less affected
  by the workload changes.
  We note that a compression rate drop does not mean that we must rebuild
  the \ope-integrated search tree immediately because \ope still guarantees
  query correctness.
  A system can monitor \ope's compression rate to
  detect a key distribution change and then schedule an index rebuild
  to recover the compression rate if necessary.

%% ---------------------------------------------------------------
%% Range and Insert Queries
%% ---------------------------------------------------------------
\section{Range Queries and Inserts}
\label{sec:tree-eval-range}

\begin{figure*}[t]
  \centering
  \begin{adjustbox}{width=\textwidth,totalheight=\textheight,keepaspectratio}

\newcommand{\mysize}{0.34}

\begin{tabular}{cccc}

  \multicolumn{4}{c}
  {\includegraphics[width=\textwidth]{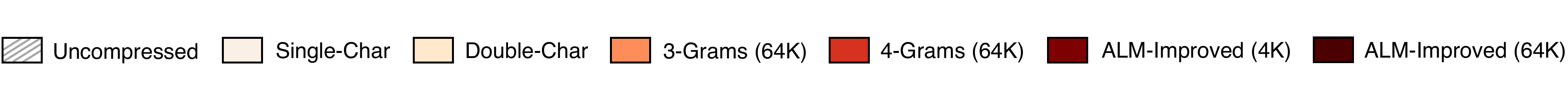}} \\

  & {\LARGE \textbf{Email}}
  & {\LARGE \textbf{Wiki}}
  & {\LARGE \textbf{URL}} \\

  %% ----------------------------------------------------------------

  \rotatebox{90}{\hskip 5em \LARGE \textsc{\textbf{\art}}}

  & \includegraphics[width=\mysize\textwidth]{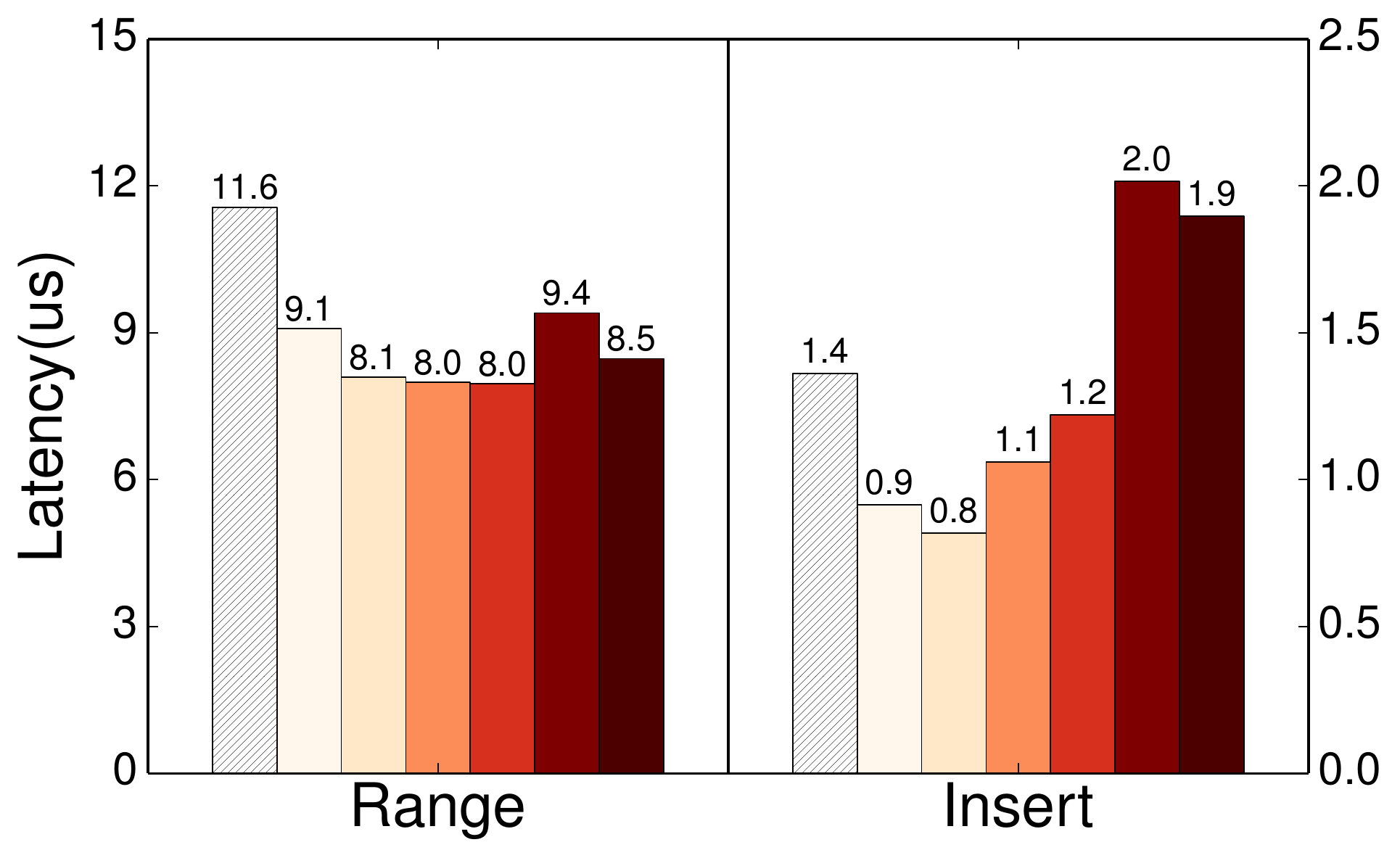}

  & \includegraphics[width=\mysize\textwidth]{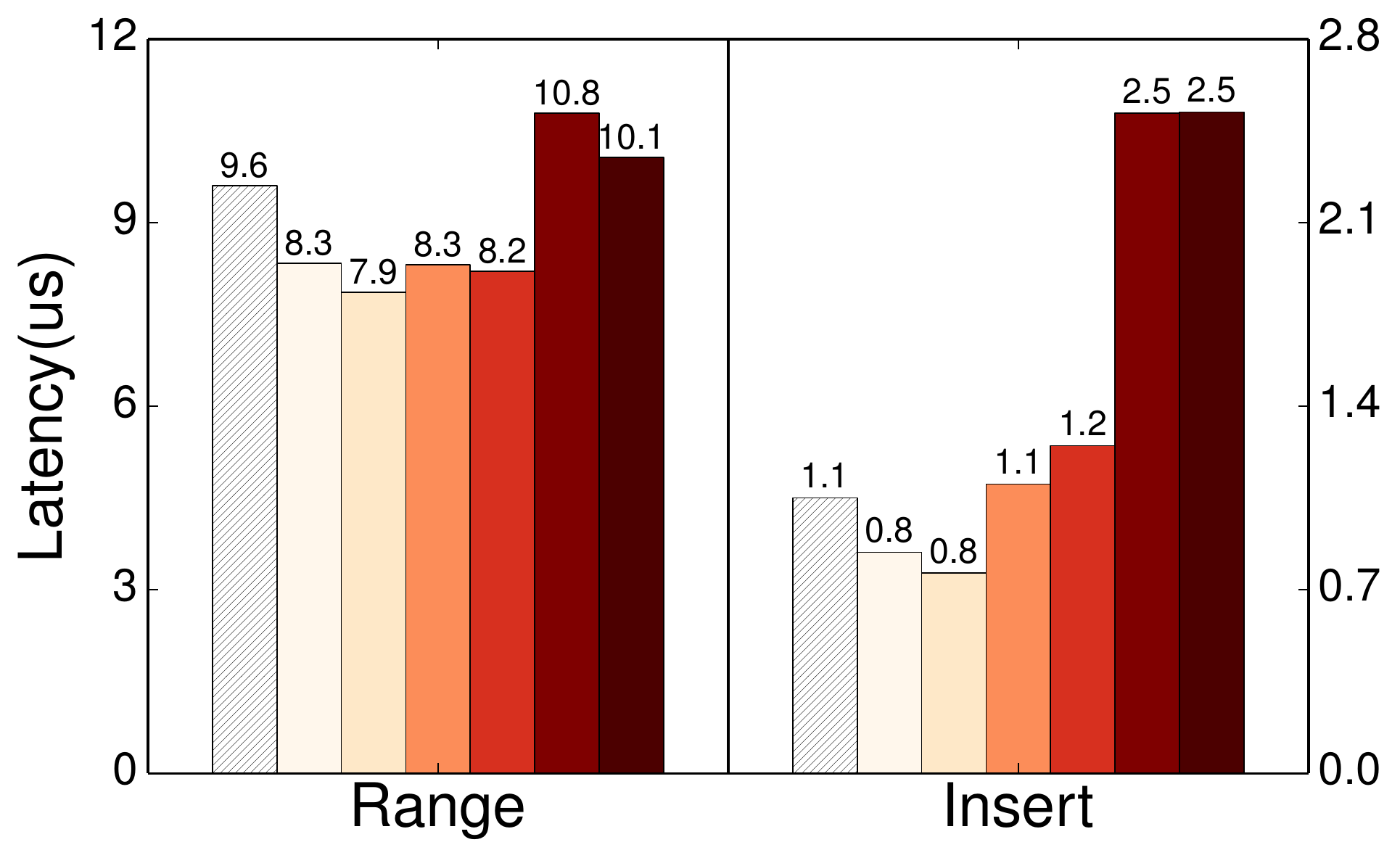}

  & \includegraphics[width=\mysize\textwidth]{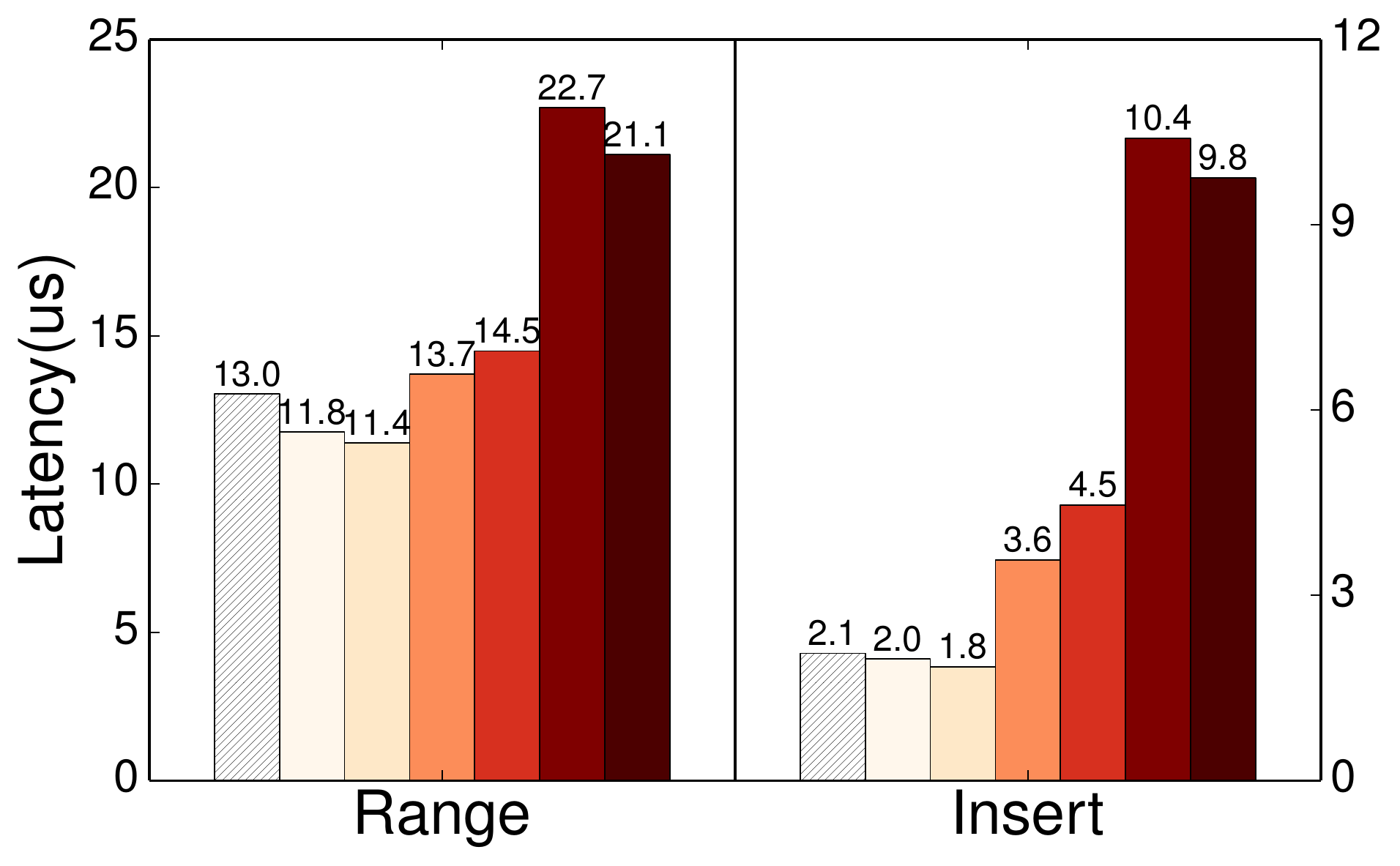} \\

  %\hlineB{2} \\

  %% ----------------------------------------------------------------

  \rotatebox{90}{\hskip 5em \LARGE \textsc{\textbf{\hot}}}

  & \includegraphics[width=\mysize\textwidth]{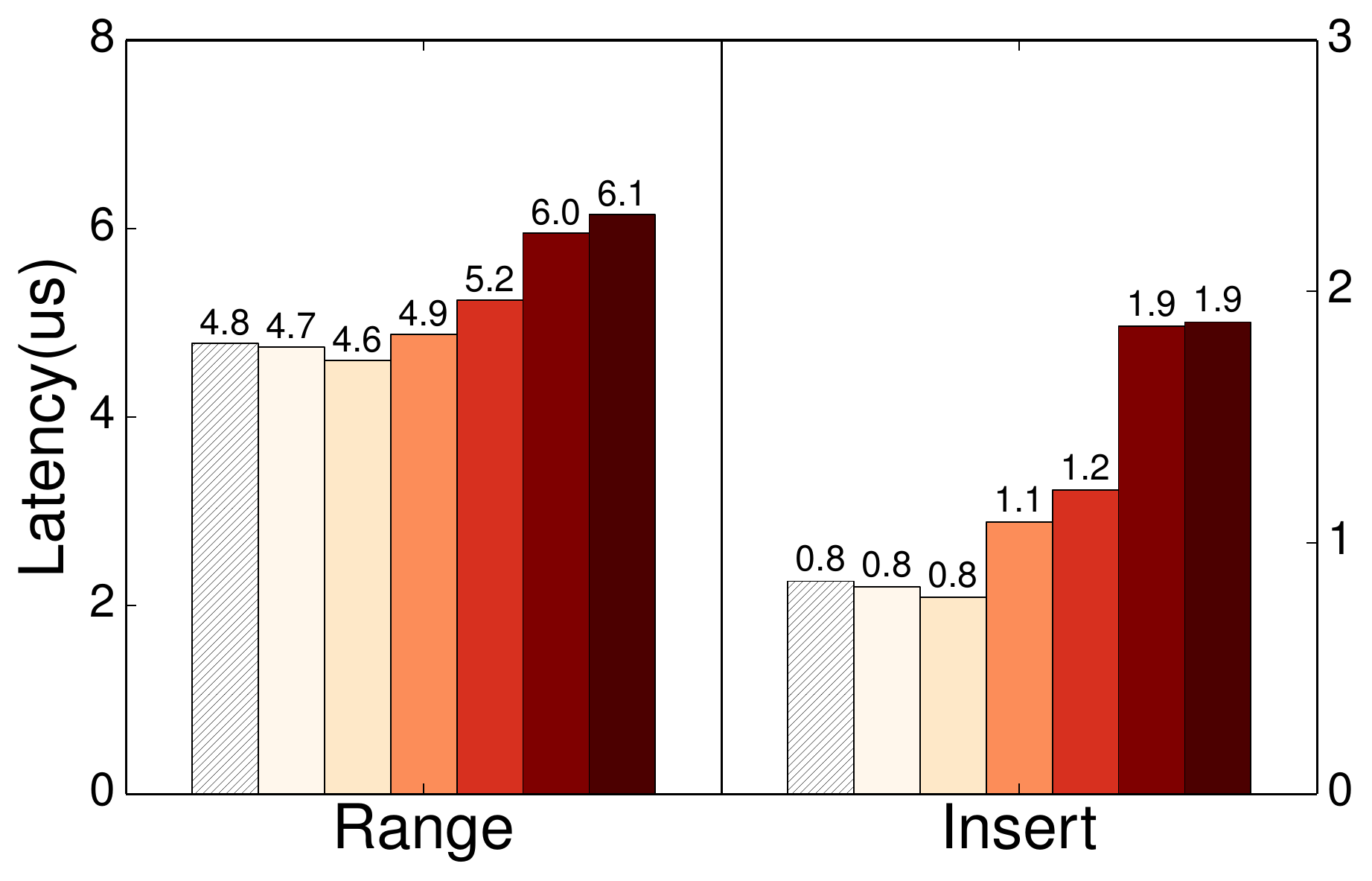}

  & \includegraphics[width=\mysize\textwidth]{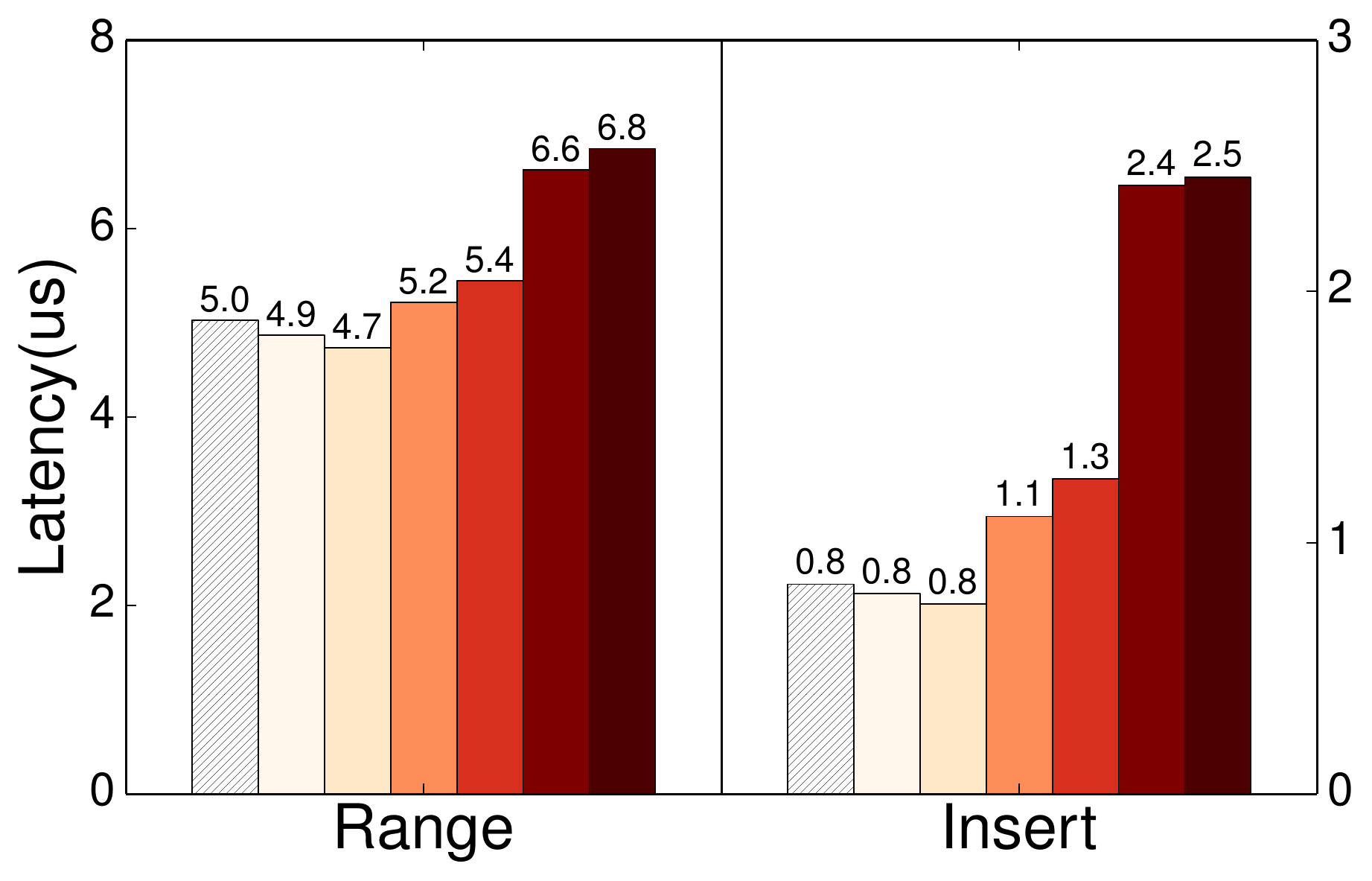}

  & \includegraphics[width=\mysize\textwidth]{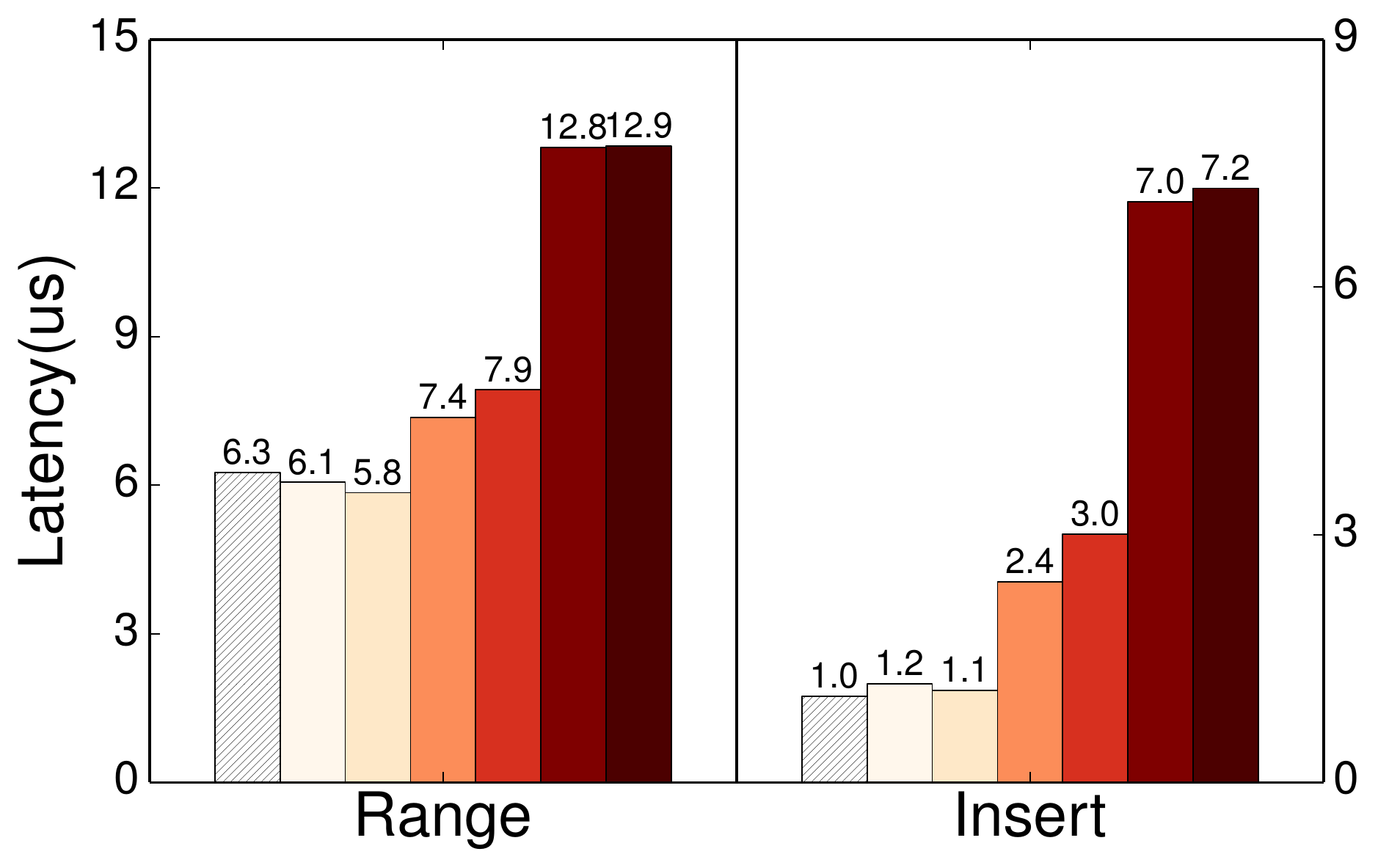} \\

  %\hlineB{2} \\

  %% ----------------------------------------------------------------

  \rotatebox{90}{\hskip 4em \LARGE \textsc{\textbf{\bplustree}}}

  & \includegraphics[width=\mysize\textwidth]{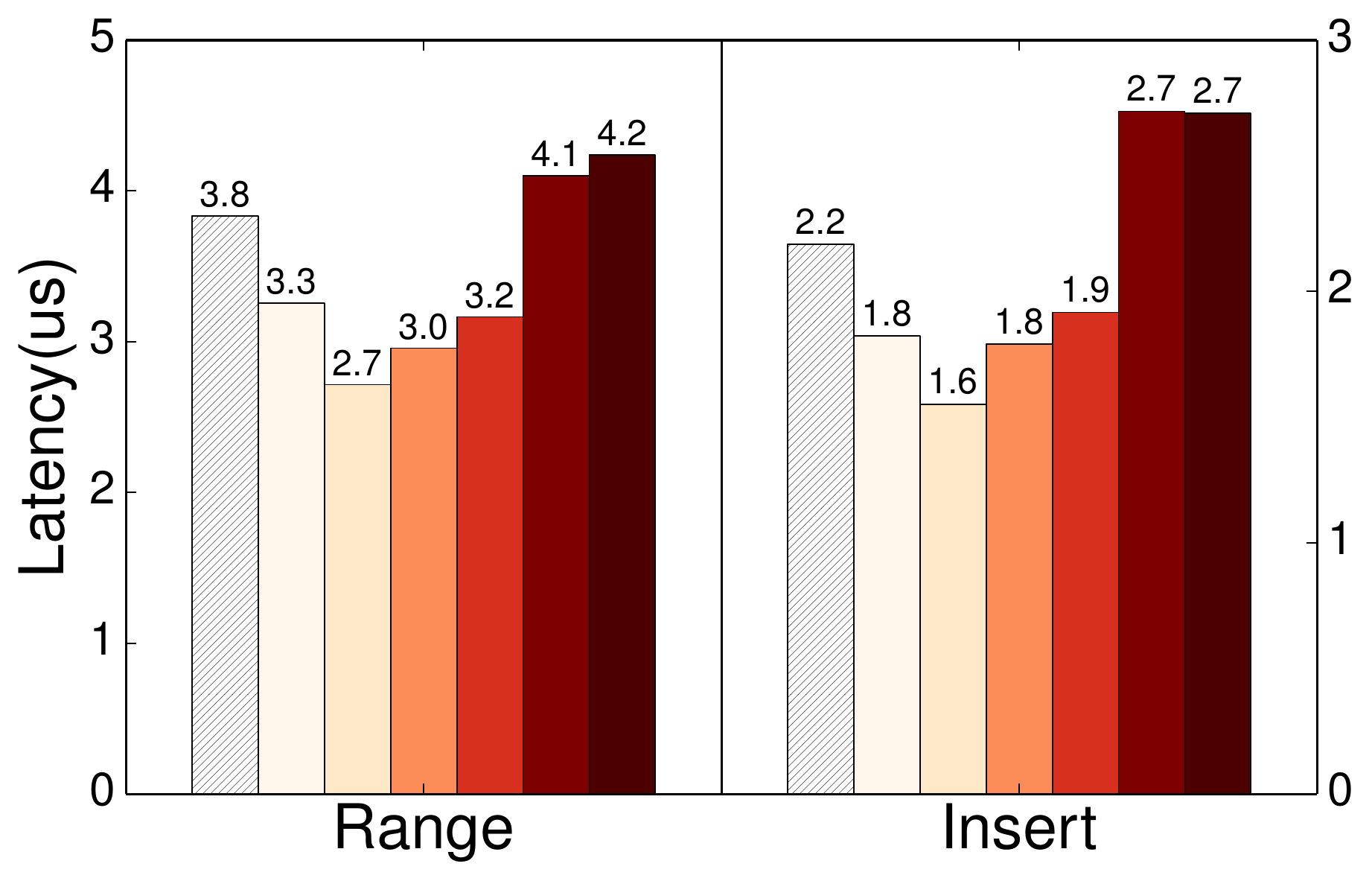}

  & \includegraphics[width=\mysize\textwidth]{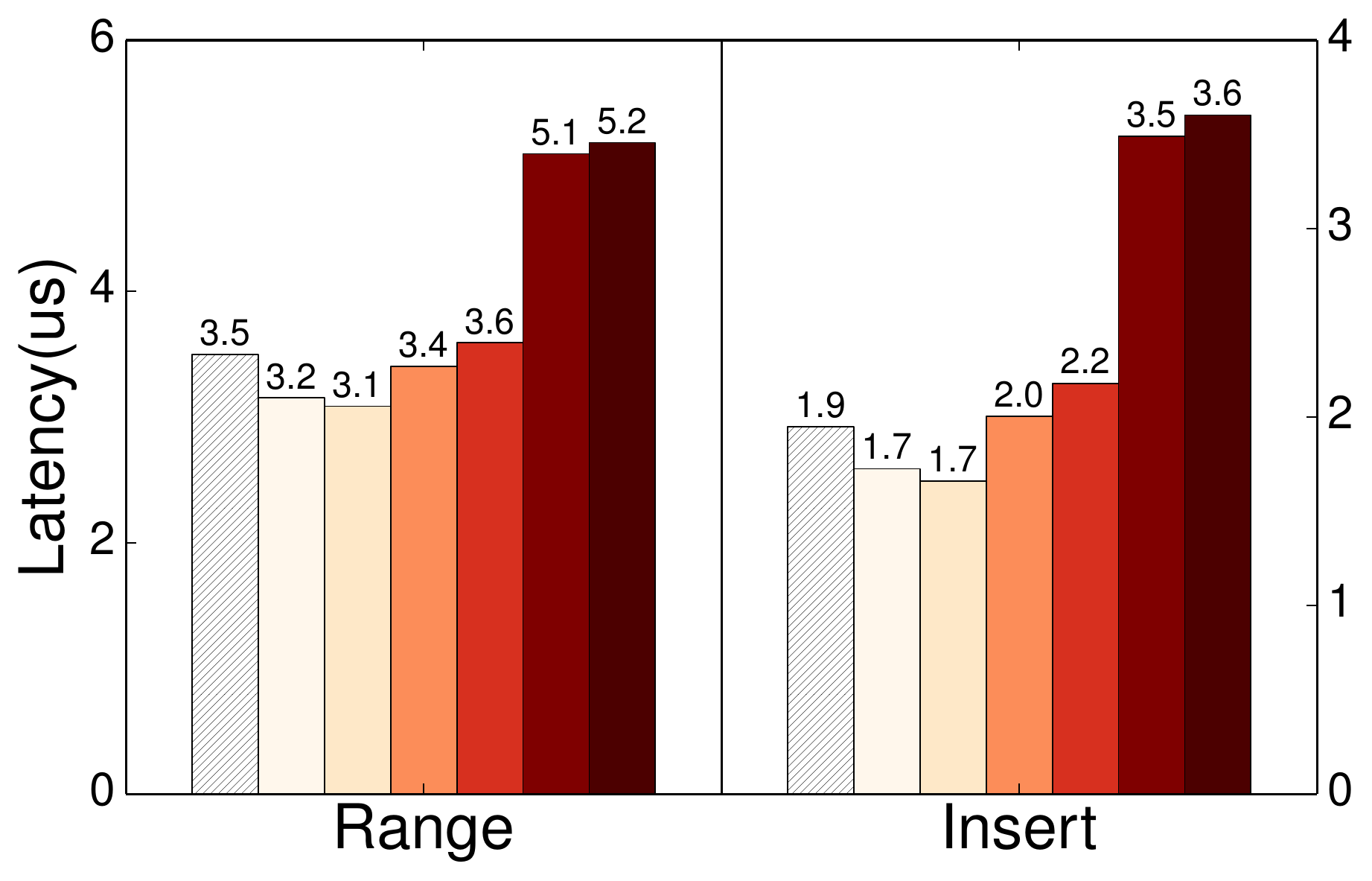}

  & \includegraphics[width=\mysize\textwidth]{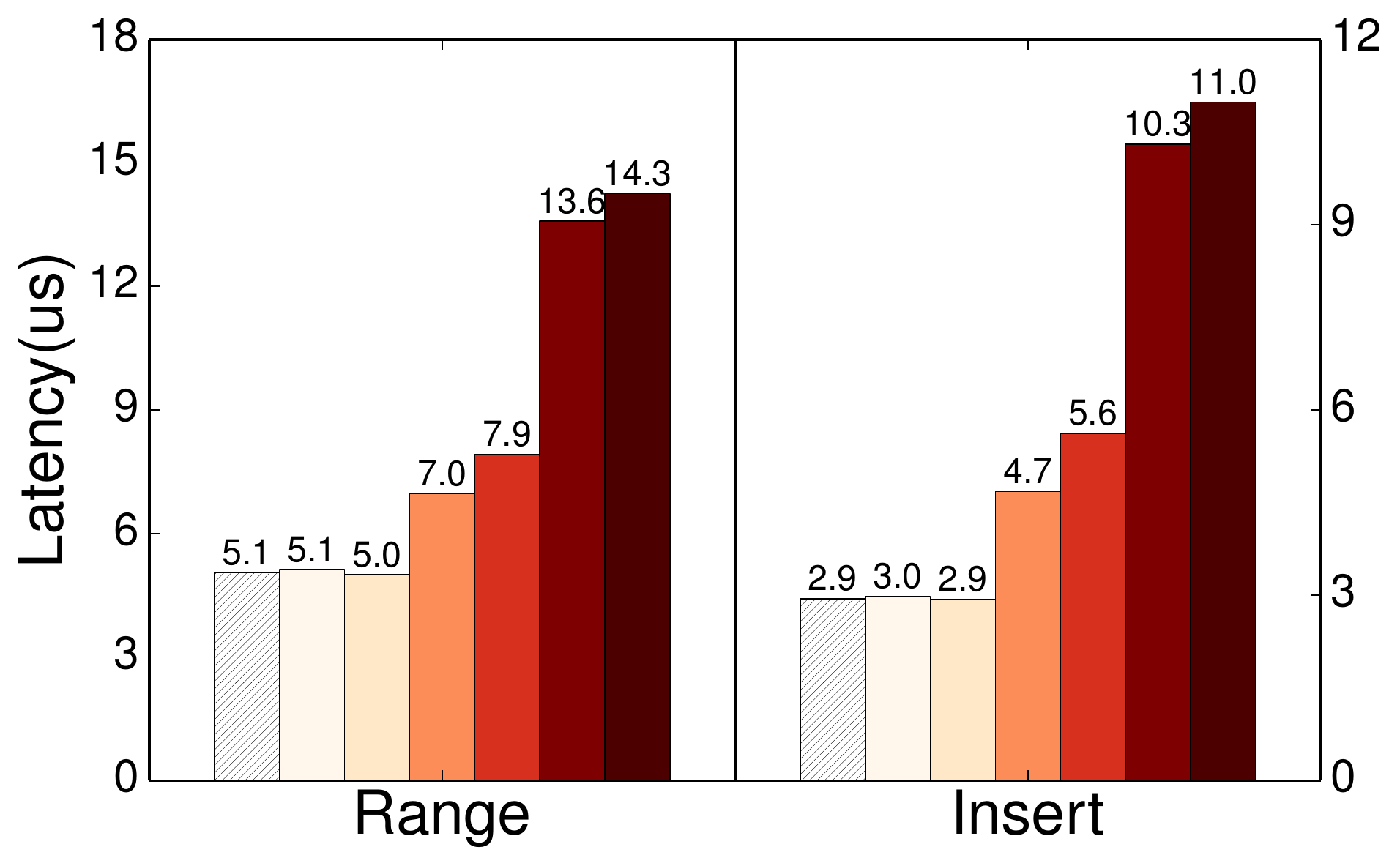} \\

  %\hlineB{2} \\

  %% ----------------------------------------------------------------

  \rotatebox{90}{\hskip 1em \LARGE \textsc{\textbf{\pbtree}}}

  & \includegraphics[width=\mysize\textwidth]{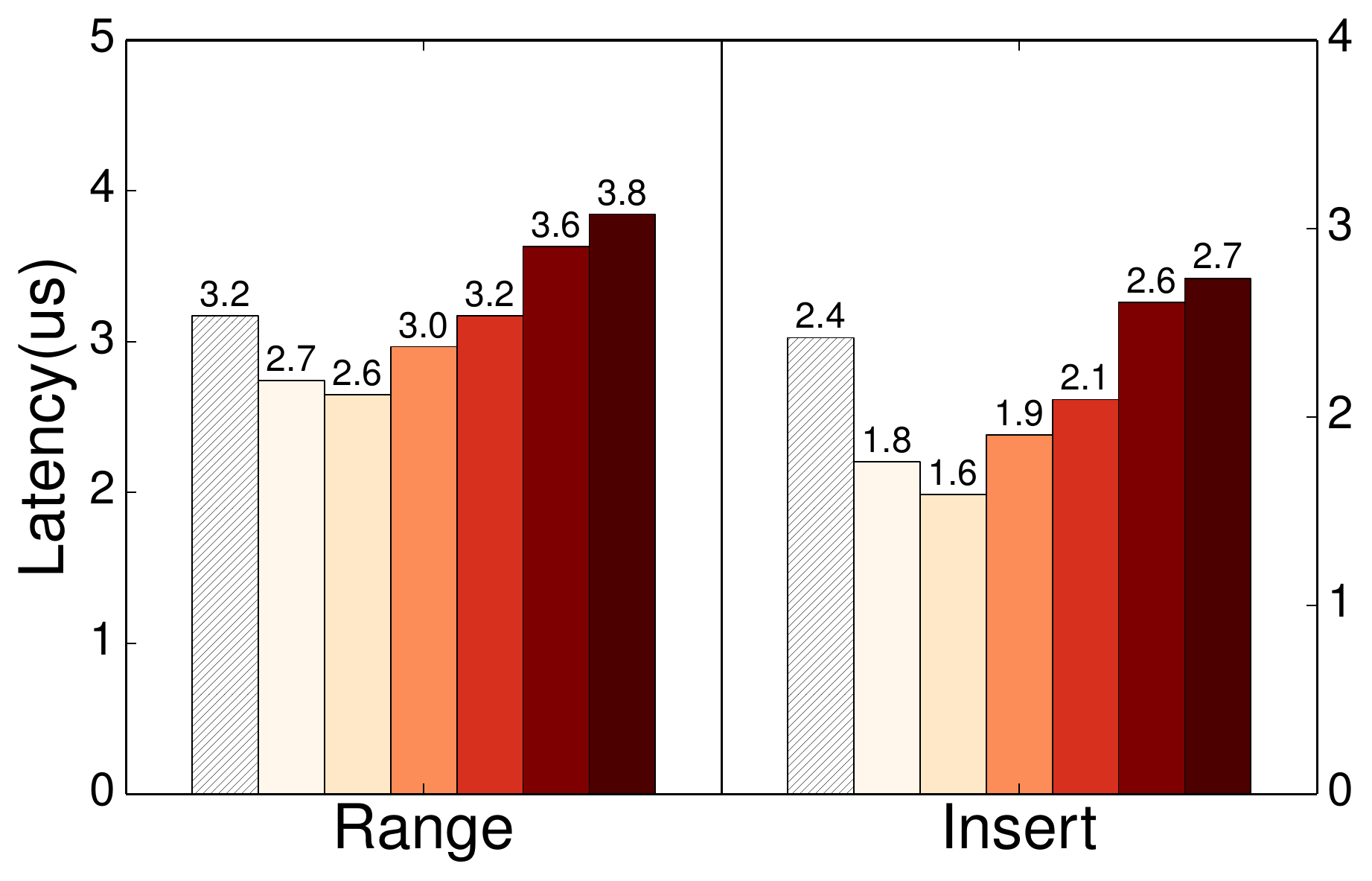}

  & \includegraphics[width=\mysize\textwidth]{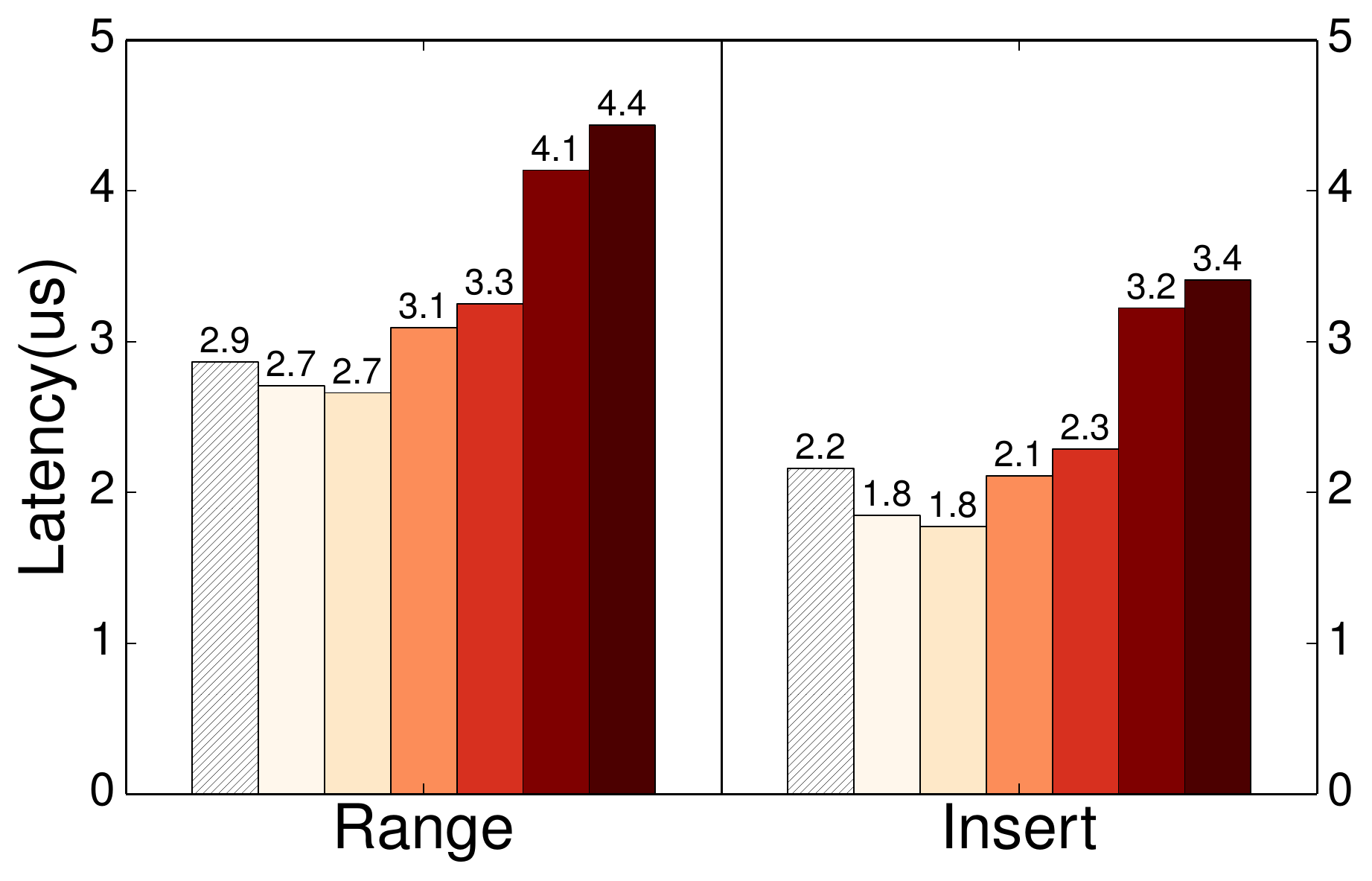}

  & \includegraphics[width=\mysize\textwidth]{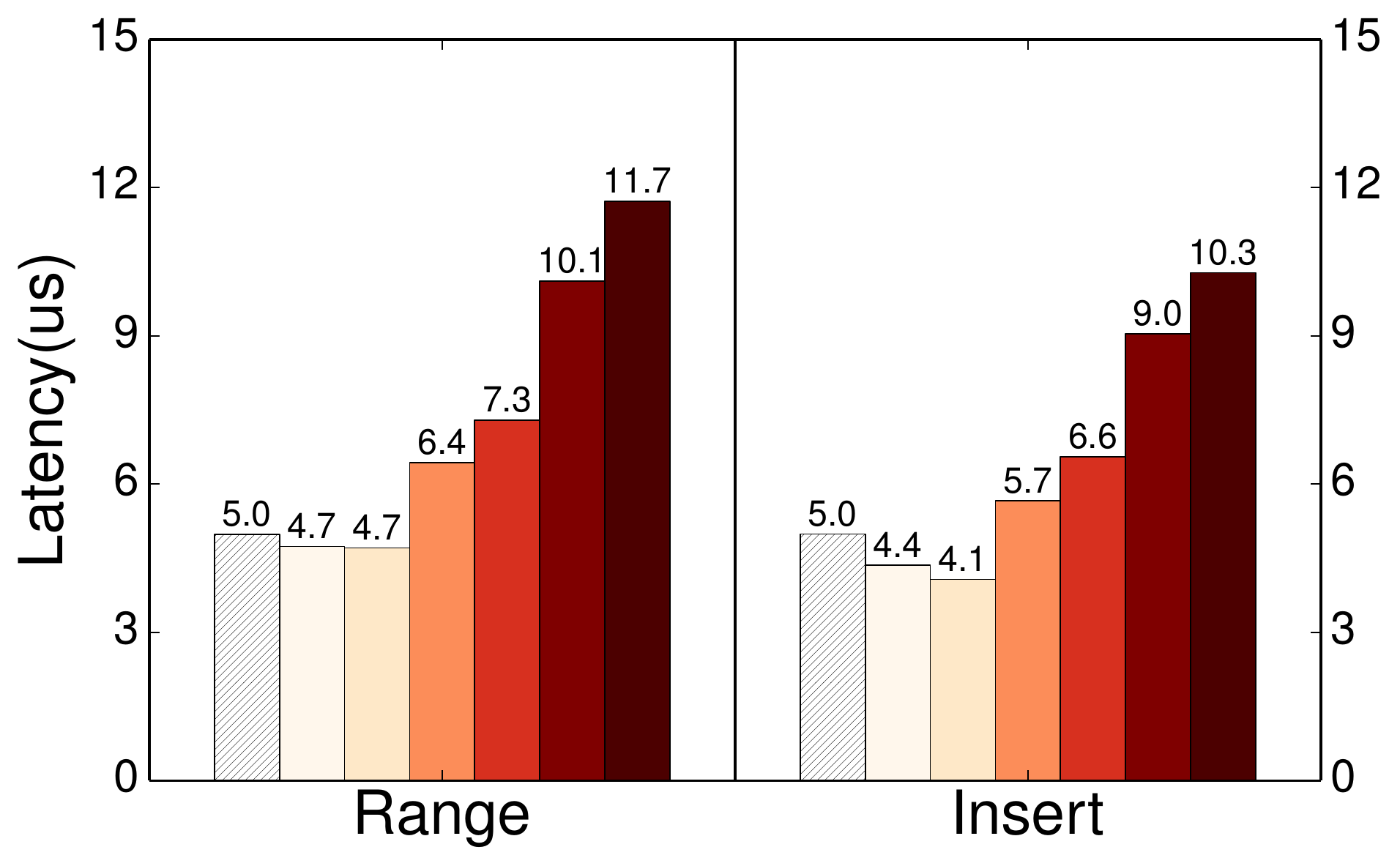} \\

\end{tabular}

\end{adjustbox}

  \caption{
    \textbf{YCSB Range and Insert Query Evaluation on \art, \hot, \bplustree, and \pbtree}
    -- Measurements on range queries and inserts for executing YCSB workloads
    on \ope-optimized indexes.
  }
  \label{fig:art-hot-btree-eval-range}
\end{figure*}

  This section presents the results of the YCSB evaluation for
  range queries and inserts.
  The experiment settings are described in \cref{sec:tree-eval:ycsb}.
  As shown in \cref{fig:art-hot-btree-eval-range} (next page), \ope improves
  (for some schemes, hurts) range query and insert performance
  for similar reasons as in the corresponding point query experiments
  in \cref{sec:tree-eval:ycsb}.
  For range queries, \ope applies the batch encoding (i.e., batch size = 2)
  optimization
  (introduced in \cref{sec:impl:impl} and evaluated in \cref{sec:batch})
  to reduce the latency of encoding both boundary keys in the query.

%% ---------------------------------------------------------------
%% Acknowledgements
%% ---------------------------------------------------------------
%% \section{Acknowledgements}
%% \label{sec:ack}

%% This work was supported (in part) by the U.S. National Science
%% Foundation under award
%% \href{https://www.nsf.gov/awardsearch/showAward?AWD_ID=1700521}{SDI-CSCS-1700521},
%% \href{https://www.nsf.gov/awardsearch/showAward?AWD_ID=1846158}{IIS-1846158},
%% and \href{https://www.nsf.gov/awardsearch/showAward?AWD_ID=1822933}{SPX-1822933},
%% Google Research Grants, and the
%% \href{https://sloan.org/grant-detail/8638}{Alfred P. Sloan Research Fellowship} program.

\end{document}